\documentclass[aps,pra,reprint,groupedaddress]{revtex4-2}
\usepackage{amsmath}
\usepackage[utf8]{inputenc}
\usepackage{graphicx}
\usepackage[breaklinks=true,colorlinks=true,linkcolor=blue,urlcolor=blue,citecolor=blue]{hyperref}

\usepackage{orcidlink}

\begin{document}


\title{Double-$K$-hole resonances in single photoionization of He-like B$^{\bf{3+}}$ ions}


\author{A.~M\"{u}ller\,\orcidlink{0000-0002-0030-6929}}
\email[]{Alfred.Mueller@physik.jlug.de}
\author{ P.-M.~Hillenbrand\,\orcidlink{0000-0003-0166-2666}
}
\author{ S.-X.~Wang\,\orcidlink{0000-0002-6305-3762}
}
\author{ S.~Schippers\,\orcidlink{0000-0002-6166-7138}
}
\affiliation{I. Physikalisches Institut, Justus-Liebig-Universit\"{a}t Gie{\ss}en,
35392 Giessen, Germany}

\author{ E. Lindroth\,\orcidlink{0000-0003-3444-1317}
}
\affiliation{Department of Physics, Stockholm University, Alba Nova University Center, 106 91 Stockholm, Sweden}

\author{ F. Trinter\,\orcidlink{0000-0002-0891-9180}
}
\affiliation{Molecular Physics, Fritz-Haber-Institut der Max-Planck-Gesellschaft,
14195 Berlin, Germany}

\author{ J. Seltmann\,\orcidlink{0009-0009-7545-2101}
}
\affiliation{Deutsches Elektronen-Synchrotron (DESY),
 22607 Hamburg, Germany}

\author{ S. Reinwardt\,\orcidlink{0000-0001-8106-7124}
}
\author{ M. Martins\,\orcidlink{0000-0002-1228-5029}
}
\affiliation{Institut f\"{u}r Experimentalphysik, Universit\"{a}t Hamburg, 
22761 Hamburg, Germany}

\author{ A.~S.~Kheifets\,\orcidlink{0000-0001-8318-9408}
}
\affiliation{Research School of Physics, The Australian National University, Canberra, ACT 2601, Australia}

\author{ I. Bray\,\orcidlink{0000-0003-0166-2666}
}
\affiliation{Department of Physics and Astronomy, Curtin University, 
Perth, WA 6845, Australia}


\date{\today}

\begin{abstract}
Within a joint experimental and theoretical research project, single photoionization of He-like B$^{3+}$ ions was investigated in the energy range from approximately 250 to 1200~eV. With the parent-ion beam in the experiment containing both $1s^2~^1S$ ground-state and $1s2s~^3S$ metastable B$^{3+}$ ions, double-core-hole resonances could be studied. Two series of hollow resonant states were observed, one populated by $K$-shell double excitation $1s^2~^1S \to 2\ell n\ell'~^1P$ ($\ell=s,p$; $\ell'=p,s$; $n=2,3,..,6$) at photon energies up to about 510~eV, the other by $K$-shell single excitation $1s2s~^3S \to 2\ell n\ell'~^3P$ ($\ell=s,p$; $\ell'=p,s$; $n=2,3,..,6$) at energies up to about 310~eV. High resolving powers up to approximately 29000 were achieved.
The relativistic many-body perturbation theory was employed to determine
level-to-level cross sections for $K$-shell excitation with subsequent
autoionization. The resonance energies were calculated with inclusion
of electron correlation and radiative contributions. The energy
uncertainties of the most prominent resonances are estimated to be
below $\pm 1$ meV. Convergent close coupling (CCC) calculations provided single-photoionization cross sections $\sigma_{34}$ for B$^{3+}$ including the resonant and non-resonant channels. Apart from the resonances, $\sigma_{34}$ is dominated by direct ionization in the investigated energy range. The contribution $\sigma_{34}^{\mathrm{dir}}$ of the latter process to $\sigma_{34}$ was separately determined by using the random-phase approximation with exchange and relativistic Hartree-Fock calculations which agree very well with previous calculations. Direct ionization of one electron accompanied by excitation of the remaining electron was treated by the CCC theory and found to be a minor contribution to $\sigma_{34}$.

\end{abstract}


\maketitle

\section{\label{intro} Introduction}

Double-core-hole (DCH) states are of outstanding importance both from a fundamental physics point of view and with respect to applications, e.g., in molecular structure and dynamics studies. DCH states with very short lifetimes enabling detailed time-resolved spectroscopy in molecules~\cite{Cederbaum1986,Santra2009} have received increasing attention  with the advent of x-ray free-electron lasers \cite{Berrah2011}. These light sources can provide x-ray beams with very high brightness as well as ultrashort pulse durations  and, therefore, DCH states can be produced with relatively high yields by single or sequential double photoabsorption and used to investigate the properties of molecules (see, for example, Refs.~\cite{Nakano2013,Ismail2023,Trinter2024} and references therein).

Due to their exotic nature and, in particular, the delicate effects of electron-electron correlation, double-$K$-shell vacancies in atoms have been a topic of high interest over many decades. Correlation effects govern the birth, life, and death of DCH states. Therefore, understanding the mechanisms for their production, the exact excitation energies, as well as the decay processes and associated lifetimes provides stimulating challenges for experiment and theory. Few-electron systems are particularly interesting because theoretical approaches can produce very accurate results~\cite{Drake1988,Madsen2003,Indelicato2019,Zaytsev2023} that often surpass experimental precision.

The most fundamental DCH states are produced by double excitations in He-like systems. Pioneering work by Madden and Codling~\cite{Madden1963} on double excitation of neutral helium by a single photon  producing autoionizing empty-$K$-shell states opened the door to an unmatched flow of high-impact experimental and theoretical publications on doubly excited helium atoms and DCH states of atoms in general~\cite{Rost1997,Wuilleumier2000,Tanner2000a,Wehlitz2010,Hoszowska2009,Hoszowska2013,Zaytsev2019,Lyashchenko2021,Wu2023,Zaytsev2023} identifying just a few examples here with numerous references to previous high-quality investigations.

Double-$K$-shell vacancies in He-like systems can be prepared in several different ways. Madden and Codling used synchrotron light for the one-photon double excitation of helium atoms~\cite{Madden1963}. Different from this very controlled resonant production of autoionizing two-electron levels, the same states can be populated by charged-particle impact on He-like atoms~\cite{Becker1964,Westerveld1979,Padhy1990,Mikhailov2013}, by ion-atom or ion-molecule collisions~\cite{Woods1975a,Rodbro1979,Mack1989,Barat1992,Benis2004a} as well as by multiple collisions in beam-foil experiments (see, e.g., Refs.~\cite{Bruch1975,Kasthurirangan2013}) or in plasma environments~\cite{Kennedy2004a}.

 In contrast to the extensive investigations on neutral He, experimental studies on the controlled production of He-like ions with empty $K$ shells under single-collision conditions are less numerous. Experimental access to doubly excited states in He-like ions with an outstanding potential of spectroscopic precision is possible via dielectronic recombination in collisions of H-like ions with electrons~\cite{Kilgus1990,Dewitt1995,Bernhardt2011,Wang2024} and electron-impact $K$-shell excitation of $1s2s$ metastable He-like ions~\cite{Mueller2014g,Mueller2018b}. The electron-energy spread in such experiments is limited by electron space-charge and kinematic effects in beams of fast particles resulting in resolving powers $E/\Delta E$ of the order of 1000 for studies of double-$K$-vacancy levels.

 High-resolution detection of decay products from intermediate doubly excited states produced in ion-atom or ion-molecule collisions can provide detailed information, e.g., by spectroscopy of emitted Auger electrons~\cite{Rodbro1979,Mack1989,Benis2004a}, however, typical resolving powers of approximately 300 to 500 mark the inherent limitations of the early experiments. Better resolution can be achieved with optical instruments detecting photons emitted from doubly excited states. Resolving powers in the range 1000 to 2000 have been reached (see, e.g., Ref.~\cite{Kasthurirangan2013}). An exceptionally high resolution was achieved with electron spectroscopy of doubly excited states in the Li$^+$ ion which could be populated starting from neutral Li ~\cite{Diehl1999a}. The maximum resolving power reached in this experiment was of the order of 1600. Another high-resolution access to doubly excited states in He-like ions is provided by the dual-laser-plasma technique~\cite{Kennedy2004a} employing optical spectrometers. Observations of ${n\ell n'\ell'}$ states were made for Li$^+$ with resolving powers between 1500 and 1900~\cite{Carroll1977,Kiernan1994} and for Be$^{2+}$ with $E/\Delta E$ = 2200~\cite{Jannitti1984}.

 Much better resolution is achievable by excitation with narrow-bandwidth photon beams in experiments employing  merged beams of photons and ions. However, the controlled production of doubly excited states starting from He-like ions by a single photon is very challenging due to low cross sections for double excitation and low densities of He-like ions in colliding-beams experiments. So far, only two experiments on photoproduction of doubly excited He-like ions by employing the photon-ion merged-beams technique have been published. In one of the measurements, single ionization of ground-state Li$^+$ ions was investigated~\cite{Scully2006a} where doubly excited states were observed with resolving powers up to 13000. Absolute cross sections for direct photo-double-excitation of Li$^{+}(1s^{2}~^{1}S)$ were measured. In a second experiment of the same kind, cross sections for direct single excitation of the $K$ shell of metastable C$^{4+}(1s2s~^3S)$ were determined~\cite{Mueller2018d}. Resolving powers up to almost 26000 could be achieved in the spectroscopy of doubly excited He-like C$^{4+}$. This was sufficient, e.g., for separating the fine structure within the resonantly excited C$^{4+}(2s2p~^3P)$ term that decays by an Auger process to C$^{5+} + e^-$. In the experiments, the C$^{5+}$ product ions were detected as a function of the incident-photon energy.

 Beside the high photon flux available at the third-generation synchrotron light source employed in the experiment and the relatively large cross sections for single $K$-shell excitation necessary to produce doubly excited levels, one other effect helped with resolving the fine structure of doubly excited resonances in He-like carbon: by starting from  the $1s2s~^3S$ level of long-lived C$^{4+}$ that already carries a single $K$ vacancy in the initial state, the excitation of the second $K$-shell electron leads to $^3P$ DCH states. These live much longer and have much smaller lifetime widths than the $^1P$ DCH states that can be excited by one photon from the $1s^2~^1S$ ground level. In the present work, we find a lifetime of the $2s2p~^3P$ levels in B$^{3+}$ ions that is more than a factor of 10 longer than that of $2s2p~^1P$. Clearly, not all DCH states have very short lifetimes as  needed for fast time-resolved spectroscopy of molecules. Long lifetime means small natural width and thus facilitates high-resolution measurements with relatively low but monochromatized photon flux without losing too much signal.

In this paper, we present measured yields of B$^{4+}$ ions arising from the absorption of single photons by B$^{3+}$ ions
\begin{equation}
h\nu + \rm{B}^{3+} \to \rm{B}^{4+} + e^-.
\end{equation}
The ions were provided in the form of a mass-and-charge-analyzed beam. In a photon-ion merged-beams arrangement, the product ions were charge analyzed and counted as a function of photon energy while the photon flux and the ion current were monitored for normalization of the signal. The observed B$^{4+}$ final charge state of the ions can result from different reaction channels, direct, non-resonant, single photoionization with removal of one of the two electrons and an indirect process
\begin{equation}
h\nu + \rm{B}^{3+} \to [\rm{B}^{3+}]^{**} \to \rm{B}^{4+} + e^-
\end{equation}
where, in a first step, an intermediate autoionizing (double-$K$-vacancy) state is formed by photoexcitation and, in a second step, the doubly excited intermediate state decays by an Auger process in which an electron is ejected.

 The parent ions used in the experiment are produced in the relatively hot plasma of an electron cyclotron resonance ion source (ECRIS) where excited states of B$^{3+}$ are also populated. The extracted ion beam contains B$^{3+}$ ions both in the $1s^2~^1S$ ground state and in the long-lived metastable $1s2s~^3S$ $K$-shell excited level. Hence, the yield spectrum that is measured in the photon-energy range from approximately 250 to 1200~eV contains contributions from both initial levels. Similar to astrophysical observations or the spectroscopy of earth-bound radiation sources where complex mixtures of different chemical elements in atomic or molecular form and in different charge states are present, the contributions of the different components can be disentangled on the basis of their characteristic-energy fingerprints. In particular, in the present experiment two series of doubly excited states ${n\ell n'\ell'~^1P}$ and ${n\ell n'\ell'~^3P}$ with $n,n' \geq 2$ could be observed with high resolution. Their origins are clearly distinguished by their resonance energies. Resonance parameters of the strongest resonances in the single-ionization cross section were determined from the experimental spectra and by using state-of-the-art theoretical methods. Theory and experiment are in very good agreement.

The present paper is organized as follows. Next to this introduction, in Sec.~\ref{Sec:experiment}, the experimental procedures are outlined. In Sec.~\ref{Sec:theory},  the theoretical approaches are briefly described. Sec.~\ref{Sec:t-results} summarizes the theoretical results and Sec.~\ref{Sec:normalization} discusses normalization procedures based on theory by which the experimental results are put on an absolute scale. Sec.~\ref{Sec:comparison}  presents the experimental  results and compares them with theory. The paper ends with a summary~(Sec.~\ref{Sec:summary}) and acknowledgements~(Sec.~\ref{Sec:acknowledgement}).

\section{\label{Sec:experiment} Experimental Procedures}

The present experiments were conducted at the soft-x-ray beamline P04~\cite{Viefhaus2013} of the synchrotron lightsource PETRA III~\cite{Schroer2022} operated by DESY in Hamburg, Germany. The permanent endstation PIPE (Photon-Ion Spectrometer at PETRA III)~\cite{Schippers2020a} of beamline P04  provided the complete experimental setup for the measurements which employed the photon-ion merged-beams technique~\cite{Phaneuf1999}  for the determination of relative cross sections for single ionization of B$^{3+}$ ions by single photons with energies between approximately 250 and 1200~eV. The experimental procedures for the measurement of absolute cross sections at PIPE have been previously described in great detail~\cite{Schippers2014,Mueller2017}. Therefore, the present description only summarizes the experimental background of the present project and especially provides the specific details of the present experiments.

A permanent-magnet ECRIS~\cite{Broetz2001} was operated with BF$_3$ gas to produce the desired B$^{3+}$ ions. Positive ions generated in the ECRIS plasma were extracted and accelerated towards ground potential by a static acceleration voltage $U_{\rm acc} = +6000$~V applied to the plasma chamber. An ion beam was formed by suitable electrostatic focusing elements. The B$^{3+}$ ions in the beam had an energy of 18~keV and a velocity $v_{\rm ion} = 5.62 \times 10^7$~cm/s when traversing a magnetic analyzer field that filtered the $^{11}$B$^{3+}$ component from the numerous other ingredients of the ion inventory extracted from the ECRIS. The beam of mass-and-charge-selected $^{11}$B$^{3+}$ ions was collimated to a diameter of about 1~mm and merged with the counter-propagating photon beam provided by beamline P04. Photoionized product ions, B$^{4+}$, were separated from the parent-ion beam by a second magnet and transported to a single-particle detector with almost 100\% counting efficiency. The parent ions were collected by a large Faraday cup inside the second magnet's vacuum chamber and their electrical current was measured by a sensitive electrometer. The flux of the transmitted photon beam was recorded by a calibrated photodiode. Typical $^{11}$B$^{3+}$ beam currents in the interaction region were between 2 and 3~nA.

A newly installed blazed grating with 400 lines per~mm delivered a very high photon flux of more than $10^{14}$ photons per second at a photon energy $E_{\rm ph} = 450$~eV and a monochromator exit-slit width of 1000~$\mu$m corresponding to a photon-energy bandwidth of approximately 620~meV. The product-ion detector recorded 20 to 25 counts per second (off-resonance) under these conditions. The rate of dark counts was measured to be 0.017(1)~Hz where the number in parentheses is the uncertainty of the preceding last digit. The photons did not produce background. Thus, the background under the  B$^{4+}$ signal that mainly originates from electron-stripping collisions of the ions in the residual gas almost exclusively depended on the current of incident parent ions and amounted to about 0.2~Hz/nA. The background was subtracted to obtain relative apparent cross sections $\sigma_{34}^{\rm app}$ for net single photoionization of B$^{3+}$ ions (for further explanation, see the end of this section).

The magnetically filtered parent $^{11}$B$^{3+}$ ion beam consisted of two components: $^{11}$B$^{3+}(1s^2~^1S_0)$ ground-level and long-lived $^{11}$B$^{3+}(1s2s~^3S_1)$ excited-level ions with a lifetime of 149~$\mu$s~\cite{Drake1971}. Hence, ionization signals from both beam components had to be expected. The fraction $f$ of metastable ions in the parent-ion beam was not \textit{a priori} known. In a previous experiment studying electron-impact ionization of B$^{3+}$ ions from an identical ECRIS, a fraction $f=0.105$ was derived~\cite{Renwick2009a}. The fraction may vary with the operation mode of the ion source, however, the optimization of B$^{3+}$ ion output and the required high temperature of the source plasma necessary to produce multiply charged ions probably leave little room for changing the metastable fraction of these ions.

Other parameters that influence the survival of excited states in an ion beam are the flight time between production in the ECRIS and the photoionization in the photon-ion interaction region and the possible quenching by collisions with residual-gas molecules along the flight path. It is not known how long it takes a B$^{3+}$ ion to drift from the ion source plasma to the extraction electrode. This time may be tens of microseconds given the fact that in neither experiment, indications for the presence of B$^{3+}(1s2s~^1S_0)$ metastable ions with a lifetime of 10.86~$\mu$s~\cite{Derevianko1997} could be found in the parent-ion beam. The flight path between the ion source and the center of the interaction region is approximately 9~m which results in a time of flight of roughly 16~$\mu$s. Reference~\cite{Renwick2009a} stated an upper limit of 1\% for the fraction of $1s2s~^1S$ parent ions produced in an identical ECRIS. For the present experiment, a $1s2s~^3S$ metastable fraction $f=0.091(16)$ was derived. Since theory was engaged to determine $f$, the discussion of the procedure employed to quantify the fraction of metastable B$^{3+}(1s2s~^3S_1)$ in the parent-ion beam is deferred to Sec.~\ref{Sec:normalization}.

Overview measurements of B$^{4+}$ product-ion yields were conducted by scanning the photon-energy range from 250~eV to 1200~eV in steps of 5~eV with a fixed exit-slit width of 1000~$\mu$m. In finer scans covering the energy range from 248 to 310~eV, resonances associated with photoexcitation of the metastable ion-beam component and subsequent Auger processes
\begin{equation}
\label{Eq:msresonances}
h\nu + {\mathrm{B}}^{3+}(1s2s~^3S_1) \to {\mathrm{B}}^{3+}(2\ell n\ell'~^3P_{0,1,2}) \to {\mathrm{B}}^{4+} + e^-
\end{equation}
with $\ell = s,p$, $n=2,3,4,5,6$, and $\ell'=p,s$ were observed together with the threshold region of direct $K$-shell photoionization of ground-state ions
\begin{equation}
\label{Eq:gsdirect}
h\nu + {\mathrm{B}}^{3+}(1s^2~^1S_0) \to  {\mathrm{B}}^{4+}(1s~^2S_{1/2}) + e^-.
\end{equation}
Recently, the threshold $E_K^{\mathrm{gs}}$ of this process has been calculated to be 259.3744095(15)~eV~\cite{Yerokhin2022}. This result provides an extremely accurate energy-calibration reference. For easier orientation in the landscape of the numerous processes that can happen, Table~\ref{Tab:thresholds} collects theoretical threshold energies for a number of specified direct ionizations that will be used throughout this paper..

\begin{table}
	
\caption{\label{Tab:thresholds}
Threshold energies of specified ionization processes of B$^{3+}$ ions. The numbers were obtained on the basis of the state-of-the-art calculations by Yerokhin \textit{et al.}~\cite{Yerokhin2015,Yerokhin2022}. Numbers in parentheses indicate the uncertainties of the last two digits. In the upper part of the table, the initial B$^{3+}$ and ionized final B$^{4+}$ levels are given together with the calculated threshold energies in a.u. and eV, respectively. The lower portion of the table shows threshold energies for direct double ionization, i.e., the upper energy limit of the occurrence of resonances.
}

\begin{tabular}{llcr} \hline \hline
init. B$^{3+}$ &~fin. B$^{4+}$  &~threshold (a.u.) &~threshold  (eV)  \\
\hline
$1s^2~^1S_0$         &~$1s~^2S_{1/2}$   &~9.531833739(55)  &~259.3744095(15) \\
                     &~$2s~^2S_{1/2}$   &~18.908824556(76) &~514.5353284(21) \\
                     &~$2p~^2P_{1/2}$   &~18.908763075(79)  &~514.5336555(21) \\
                     &~$2p~^2P_{3/2}$   &~18.909806352(79)  &~514.5620445(22) \\

$1s2s~^1S_0$         &~$1s~^2S_{1/2}$   &~2.078882735(29)  &~56.56928105(80) \\
                     &~$2s~^2S_{1/2}$   &~~11.455873552(51)  &~311.7302000(14) \\

$1s2s~^3S_1$         &~$1s~^2S_{1/2}$   &~2.2345859715(31)  &~60.806181969(85)\\
                     &~$2s~^2S_{1/2}$   &~11.611576788(25)  &~315.96710092(67)\\
                     &~$2p~^2P_{1/2}$   &~11.611515308(27)  &~315.96542795(74) \\
                     &~$2p~^2P_{3/2}$   &~11.612558585(28)  &~315.99381696(75) \\

                  \hline
                  \hline
 initial B$^{3+}$ &~final ion \\
                   \hline
$1s^2~^1S_0$         &~B$^{5+}$   &~22.034909454(77)    &~599.6004320(21)  \\

$1s2s~^1S_0$         &~B$^{5+}$   &~14.581958450(50)    &~396.7953036(14)  \\

$1s2s~^3S_0$         &~B$^{5+}$   &~14.737661686(25)    &~401.03220450(68)  \\
                  \hline
\end{tabular}
\end{table}	

The resonances  together with the surrounding smooth part of the apparent cross section $\sigma_{34}^{\rm app}$ were scanned in a sequence of high-resolution sweeps with photon-energy steps of 0.005~eV at a constant monochromator exit-slit width of 200~$\mu$m. For the strongest resonance in the whole spectrum, associated with B$^{3+}(2s2p~^3P_{0,1,2})$ levels at about 248~eV, detailed scans were performed with a step width of 0.001~eV and an exit-slit width of 8~$\mu$m. For highest possible resolution, a grating with 1200 lines per mm  was employed. In these measurements, the photon flux was down to only $2 \times 10^8$~photons per second and the photon-energy resolution was sufficient for partly resolving the triplet structure of the scrutinized $^3P$ resonance term.

Nearly 90\% of the parent-ion beam being in the ground state, it was also possible to observe resonances associated with double-$K$-shell excitation and subsequent Auger decay. This was accomplished for transitions
\begin{equation}
\label{Eq:gsresonances}
h\nu + {\mathrm{B}}^{3+}(1s^2~^1S_0) \to {\mathrm{B}}^{3+}(2\ell n\ell'~^1P_1) \to {\mathrm{B}}^{4+} + e^-
\end{equation}
with $\ell = s,p$, $n=2,3,4,5,6$, and $\ell'=p,s$. These $^1P$ resonances are much broader than the $^3P$ resonances populated from the metastable $^3S$ level while their strengths are smaller by almost two orders of magnitude. Therefore, in order to obtain sufficient signal rates, the energy range 450 to 515 eV was scanned around the different resonances with lower resolution (400 lines per mm grating, monochromator exit-slit width 400~$\mu$m) and a wider step width of 0.02~eV. Under these conditions the photon flux at 500~eV was $6.3 \times 10^{13}$~s$^{-1}$.

With the beam of ions counterpropagating the photon beam in the present experiments, a Doppler shift of photon energies by approximately a factor of $1+v_{\rm ion}/c = 1.00188$ with the vacuum speed of light $c$ had to be corrected for. The exact treatment of the Doppler correction is provided in Ref.~\cite{Mueller2018d} together with remaining associated uncertainties in the experimental determination of the photon energy. The Doppler-corrected photon-energy axis was then calibrated to the double-$K$-hole resonances calculated within this project with exceptionally high accuracy.

The new reference energies were obtained by employing many-body perturbation theory with inclusion of electron correlation and radiative contributions~(see Sec.~\ref{MBPT}). The uncertainty of almost all of these resonance energies is estimated to be less than 1~meV which is very much smaller than the uncertainties of the often used soft-x-ray calibration reference transitions observable in the photoabsorption by neutral gases such as CO, N$_2$, or Ne (see, e.g., Ref.~\cite{Sodhi1984a}). The present experiment demonstrates the feasibility of much improved photon-energy calibrations on the basis of resonance energies in the photoionization of He-like B$^{3+}$ ions. Known level energies of heliumlike ions have been employed recently in an attempt to re-determine x-ray transition energies of Ne, CO$_2$ and SF$_6$ with improved accuracy~\cite{Stierhof2022}.

The relative apparent cross sections $\sigma_{34}^{\rm app}$ for net single ionization of B$^{3+}$ ions were obtained by normalizing the measured count rate of B$^{4+}$ product ions to the photon flux and the measured B$^{3+}$  parent-ion current. ``Net single ionization'' does not distinguish between different pathways leading to the production of B$^{4+}$ from B$^{3+}$ parent ions by absorption of a single photon. The cross section is ``apparent'' because it is measured with a mixed parent-ion beam consisting of a ground-state and a metastable-ion component. In particular, it depends on the experiment-specific fraction $f$ of metastable ions in the parent B$^{3+}$ ion beam. The cross section $\sigma_{34}^{\rm app}$ determined here is ``relative'' because it is not on an absolute scale since the overlap of the interacting photon and ion beams, the so-called form factor, has not been separately determined. By employing the present theory it is possible to determine $f$ and to put the cross-section contributions on an absolute scale. The details are described in Sec.~\ref{Sec:normalization}.

\section{\label{Sec:theory} Theoretical Approaches}
 In the following subsections, the theoretical methods applied in the present project are briefly described. Detailed accounts of the different approaches can be found in the literature. Therefore, the present descriptions are kept at a minimum.

\subsection{\label{MBPT} Many-Body Perturbation Theory}
The intermediate double-$K$-hole states observed as resonances in the single ionization of B$^{3+}$ are calculated with relativistic many-body perturbation theory (MBPT) in an all-order formulation including  single and double excitations,  as described by Salomonson and \"{O}ster~\cite{Salomonson1989a}. This means that  all types of excitations that can occur in a pure two-electron system are accounted for. The  B$^{3+}$ ion is placed in a spherical box within which a discrete radial grid is used. Diagonalization of the discretized hydrogenlike Dirac Hamiltonian gives a discrete basis set for each spin-angular symmetry (defined by $\ell_j$) that is complete on the grid chosen. The basis set is then used to construct correlated wave functions to all orders in the perturbation expansion of the  electron-electron interaction.  Here, both  the Coulomb  and the Breit interaction are accounted for. The perturbation expansion is constructed from an {\em extended model space}~\cite{Lindgren1974} whenever a state is dominated by two or more nearly degenerate configurations. An example is  the $2s2p~^3P_1$ state which in $jj$-coupling has major contributions both from the  $2s2p_{1/2}$ and  the $2s2p_{3/2}$ configurations. This is a common scenario in $jj$-coupling, but also for doubly excited states in general.

A multipole expansion of the electron-electron interaction is used, making the method applicable to many-electron atoms in general. In the present calculations, the expansion allows for inclusion of all  partial waves up to $\ell_{\rm max} =10$ for the $2\ell n\ell', n\leq 3 $, resonance groups, $\ell_{\rm max} =7$ for the $2\ell 4\ell'$ resonance group, and $\ell_{\rm max} =6$ for the higher resonances. The convergence rate with respect to $\ell_{\rm max}$
is in general considerably faster for the triplet than for the  singlet states. The rate was investigated for the $2\ell 2\ell'$ and $2\ell 3\ell'$  resonance groups. For the latter group, the calculations gave convergence rates approximately proportional to $\ell_{\rm max}^{-6}$ for the triplet and $\ell_{\rm max}^{-4}$ for the singlet states, that is, with increasing $\ell_{\rm max}$ the changes of  the corrections to the level energies become smaller proportional to $\ell_{\rm max}^{-6}$ and $\ell_{\rm max}^{-4}$, respectively. The same behavior was found for the $2\ell 2\ell'$ resonance group with the exception of the $2s2p~^1P_1$ state which converges considerably faster than all the other states. In summary, it is found that contributions from $\ell_{\rm max}>10$ affect the triplet resonances  by less than $0.1$~meV, while some of the singlet resonances are affected on the $1$~meV level.

When perturbation theory is applied to autoionizing states it is obvious that the use of a discrete basis set will cause problems close to the poles in the energy denominator. A complex scaling of the radial coordinates can, however, solve this problem.  The present treatment follows the method employed by Lindroth~\cite{Lindroth1994a} for the calculation of doubly excited levels in the helium atom, and later for a number of Be-like ions (see, e.g., Ref.~\cite{Lestinsky2008a}). The method yields complex energies for  the autoionizing states, where the imaginary part corresponds to  the half-life time (due to Coulombic decay) of the state. The  decay rates due to photon emission are calculated from the dipole matrix elements between the  doubly excited states and the $1sn\ell~^3L_J$ states with $n \leq 4$ for the resonance groups $2\ell n\ell', n\leq 4$, and $n \leq 6$ for the higher groups. For details see Ref.~\cite{Tokman2002}. The radiative decay rates of all the considered doubly excited states are completely dominated (with contributions of more than 99\%)  by photoemission events that require one-electron transitions only.
For the  underlying theory of complex rotation (CR) and MBPT the reader is referred to a review by Lindroth and Argenti~(\cite{Lindroth2012a} and references therein).

To account for the  contributions to the energies that originate from the quantization of the electromagnetic field, the tabulation of the corrections to the $n\leq 2$ states of hydrogen-like ions by Yerokhin and Shabaev~\cite{Yerokhin2015}  has been used. Since the resonances are mixtures of configurations such as $2sn\ell$, $2p_{1/2}n\ell'$, and $2p_{3/2}n\ell''$,  the correction to each such configuration  is multiplied with its fraction in the mixture to estimate the effect on the resonance energies. The  contribution to the $2s$ level is around $1.6$~meV, while the corrections to the $2p$ levels are below $0.1$~meV. It is thus only resonances that have a substantial admixture of $2sn\ell$ configurations that will be visibly affected.

The normal mass shift is taken care of by the correction factor $M/(m_\mathrm{e} + M)$ where $m_\mathrm{e}$ is the electron rest mass and $M$ the rest mass of the $^{11}$B nucleus.  The mass-polarization effect (specific mass shift) has also been considered. Calculations of this effect with the non-relativistic formula~\cite{Hughes1930} resulted in shifts of at most 0.1~meV. It is safe to say that mass-polarization shifts of the investigated levels of the heliumlike B$^{3+}$ ion are well below the 1~meV level.

The principle theoretical approach described above has been sucessfully applied previously to resonance contributions to the single photoionization of C$^{4+}(1s2s~^3S)$ ions~\cite{Mueller2018d}. The resonances and the direct-ionization continuum were calculated separately and then added according to Eq. 11 in Ref.~\cite{Mueller2018d}. There, the total energies of the initial states were taken from previous results obtained by Drake~\cite{Drake1988} who used highly correlated non-relativistic wave functions of Hylleraas type to calculate ionization energies. Relativistic and radiative corrections were subsequently treated as perturbations. The calculation is thus very different from the one presented here. Yet,  the two calculations agree with one another within $0.5$~meV.

The salient feature in the MBPT approach used here is that each calculation is done for a limited number of doubly excited states, defined by a so-called model space, and all the configuration outside this space is included through an iterative procedure. It is then possible to use a very large configuration space.  The drawback is that the iterative procedure might not converge, however, this is not really a problem in multiply charged ions with few electrons where the $n$-manifolds such as, for example, $2\ell_j n\ell'_{j'}$  are well separated. The radial basis is complete on the chosen grid, and can thus be saturated, and the numerical uncertainty associated with the grid is well below the main sources of uncertainty. The main sources of uncertainty of MBPT results are instead
1) the partial wave expansion, and the convergence when including higher angular momenta (here $\ell_{\mathrm{max}} = 10$ is used for the lowest energy resonances) and
2) physics beyond the Dirac  equation.

\subsection{\label{CCC} Convergent Close Coupling Approach}

The convergent close coupling (CCC) approach to calculating single and double photoionization of
helium and helium-like ions was developed by Kheifets and
Bray~\cite{Kheifets1996,Kheifets1998}. A broader review is available~\cite{Bray2002a},
as are some recent developments focusing on the
near-threshold behavior~\cite{Bray2017}. Briefly, upon photoabsorption
the final photoelectron-ion interaction is calculated by expanding the
ionic wave functions in an orthogonal Laguerre basis. For each orbital
angular momentum $\ell\le\ell_{\rm max}$, this basis has
the parameter basis size $N_\ell$ and exponential fall-off
$\lambda_\ell$. We rely on the completeness of the basis to ensure
that convergence, to a desired precision, is obtained by simply
increasing the basis sizes. In order to reduce the number of free
parameters, we typically set $N_\ell=N_0-\ell$ and
$\lambda_\ell=\lambda$, thereby reducing convergence testing to just
the three parameters $\ell_{\rm max}$, $\lambda$, and $N_0$. The photoelectron-ion wave function is then
obtained via the CCC method, which solves the resulting close-coupling
equations in momentum space. The photoionization amplitude is
formed by evaluating the dipole operator between the target ground
state and the CCC-calculated wave function. This is evaluated in the
three gauges of length, velocity, and acceleration. When both the
initial B$^{3+}$ target state wave function and the CCC-calculated final total  wave function of $e^-$-B$^{4+}$ are
sufficiently accurate, then the three forms yield indistinguishable results.

In the present case of $\gamma-{\rm B}^{3+}$ interaction, the hydrogen-like B$^{4+}$
wave functions are obtained by setting $\ell_{\rm max}=4$, $\lambda=3$,
and $N_0=35$. Only open states are retained as the contribution of
closed states was found to be insignificant. This means the size of
the calculations rapidly increases with photon energy. Nevertheless,
with modern computational resources the calculations take no more than
just a few minutes per photon energy. The velocity form is presented throughout,
but the others are barely distinguishable except for the direct-ionization base level of the cross section for direct removal of a $2s$ electron from B$^{3+}(1s2s~^3S)$ where differences of up to about 23\% between the three gauges are found.

The accuracy of the CCC results depends critically on the accuracy of the wave functions of the initial and final states of the interaction process under consideration. The quality of the wave functions relies on the size and suitability of the basis set employed for their construction. Since the computational efforts for calculating cross sections rapidly increase with the basis size, limitations of this size are inevitable. The description of the initial state of the target B$^{3+}$ within the existing computational constraints is more difficult for the metastable $1s2s~^3S$ than for the $1s^2~^1S$ ground level. Expanding the basis for constructing a better wave function for the metastable level of B$^{3+}$ would require an unmanageable effort. Such limitations are the main reason for inaccuracies of CCC results.

\subsection{\label{HFR} Relativistic Hartree-Fock Approach}

The B$^{3+}$ $K$-shell photoabsorption cross sections for  the $1s^2~^1S$ and  $1s2s~^3S$ initial states were modeled by \textit{ab initio} configuration-interaction (CI) calculations on the basis of the Hartree-Fock method with relativistic extensions (HFR) applying the Cowan code \cite{Cowan1981}. The theoretical approach has been described in detail previously~\cite{Martins2001}.

For describing the B$^{3+}$ initial states, the CI expansion was chosen to include the configurations $1s^2$, $1s2s$, $2s^2$, and $2p^2$. The direct photo single-ionization cross sections for the removal of one $1s$ electron from B$^{3+}(1s^2~^1S)$ was derived from the dipole matrix elements $\langle \Psi_{\epsilon\ell} | e r | \Psi_{1s^2~^1S} \rangle$ where $e r$ is the dipole operator. The wave function $\Psi_{\epsilon\ell}$ represents  the residual B$^{4+}$ ion and the ionized electron with kinetic energy $\epsilon$ and angular momentum $\ell = p$. The direct removal of the $1s$ or the $2s$ electron from B$^{3+}(1s2s~^3S)$ was treated accordingly.

\subsection{\label{RPAE} Random-Phase Approximation with Exchange}
The development of the random-phase approximation (RPA) was originally directed to describe phenomena in infinite materials~\cite{Pines2016,Co2023}. Advances in the computing technologies allowed the application of the RPA also to finite
systems such as atoms~\cite{Amusia1990}. Advanced calculations additionally consider the exchange effects resulting from the fact that electrons cannot be distinguished from one another. The associated theory is the random-phase approximation with exchange (RPAE)~\cite{Amusia1990,Amusia2004a}, which was applied here to calculate cross sections for direct $1s$ and $2s$ photoionization of the B$^{3+}(1s^2~^1S)$ and B$^{3+}(1s2s~^3S)$ ions investigated in the present experiments.

\section{\label{Sec:t-results} Theoretical Results}
 In this section, an overview of the theoretical results is provided with emphasis on their applicability to the present measurements. Comparisons with previous theoretical work are presented where suitable data are available.

 The cross section $\sigma_{34}$ for net single ionization of B$^{3+}$ comprises several reaction channels. The most straightforward concept of photoionization is the direct removal of a single electron (the photoelectric effect) from the  parent ion as described by Eq.~\ref{Eq:gsdirect}. When starting from the ground state, it is described by the cross section $\sigma_{34}^{\mathrm{gs-dir}}$. For the metastable B$^{3+}$ ion, the corresponding cross section is $\sigma_{34}^{\mathrm{ms-dir}}$.  Another direct process that may occur is the removal of one electron accompanied by excitation (shake-up) of the other electron to an excited level of the remaining B$^{4+}$ ion. In the present context, cross sections $\sigma_{34}^{{\mathrm{gs}}-n}$ and $\sigma_{34}^{{\mathrm{ms}}-n}$ for direct $K$-shell ionization plus excitation to the main shell with principal quantum number $n$ were calculated for ground-state and metastable B$^{3+}$ ions, respectively, using the CCC approach. Such two-electron processes have been investigated previously~\cite{Kheifets1998} and found to be of minor importance for the net ionization of He-like ions. The further discussion of theoretical results in this section (see subsection~\ref{gs_ion-exci}) shows that ionization-excitation (IE) can be safely neglected in the assessment of $\sigma_{34}$ also for B$^{3+}$ ions. Nevertheless, it is elucidating to see how much the cross sections $\sigma_{34}^{{\mathrm{gs}}-n}$ and $\sigma_{34}^{{\mathrm{ms}}-n}$ are influenced by DCH resonances.

As the last sentence of the preceding paragraph suggests, indirect ionization may occur in addition to the direct non-resonant processes contributing to $\sigma_{34}$. Indirect ionization proceeds via the excitation of intermediate autoionizing levels as described by Eqs.~\ref{Eq:msresonances} and \ref{Eq:gsresonances}. In the energy range of the present investigation, such resonances are populated by two-electron excitations of ground-state B$^{3+}$ or by single $K$-shell excitation of metastable B$^{3+}$. The resulting doubly excited states decay with high probability by the emission of an Auger electron. In the present paper, the associated cross sections are $\sigma_{34}^{{\mathrm{gs-res}}}$ for ground-level B$^{3+}$ and $\sigma_{34}^{{\mathrm{ms-res}}}$ for the metastable heliumlike boron ions. The cross sections for IE to highly excited $n$ shells of the hydrogenlike B$^{4+}$ ionization products are also influenced by resonances. With increasing principal quantum numbers $n$, the IE cross sections reveal increasing relative contributions of doubly excited B$^{3+}$ resonances with both electrons in high Rydberg states where the Auger decay populates highly excited B$^{4+}(n\ell)$ levels.

\subsection{\label{K-direct} Direct single photoionization}

 \begin{figure}[b]
\includegraphics[width=8.0cm]{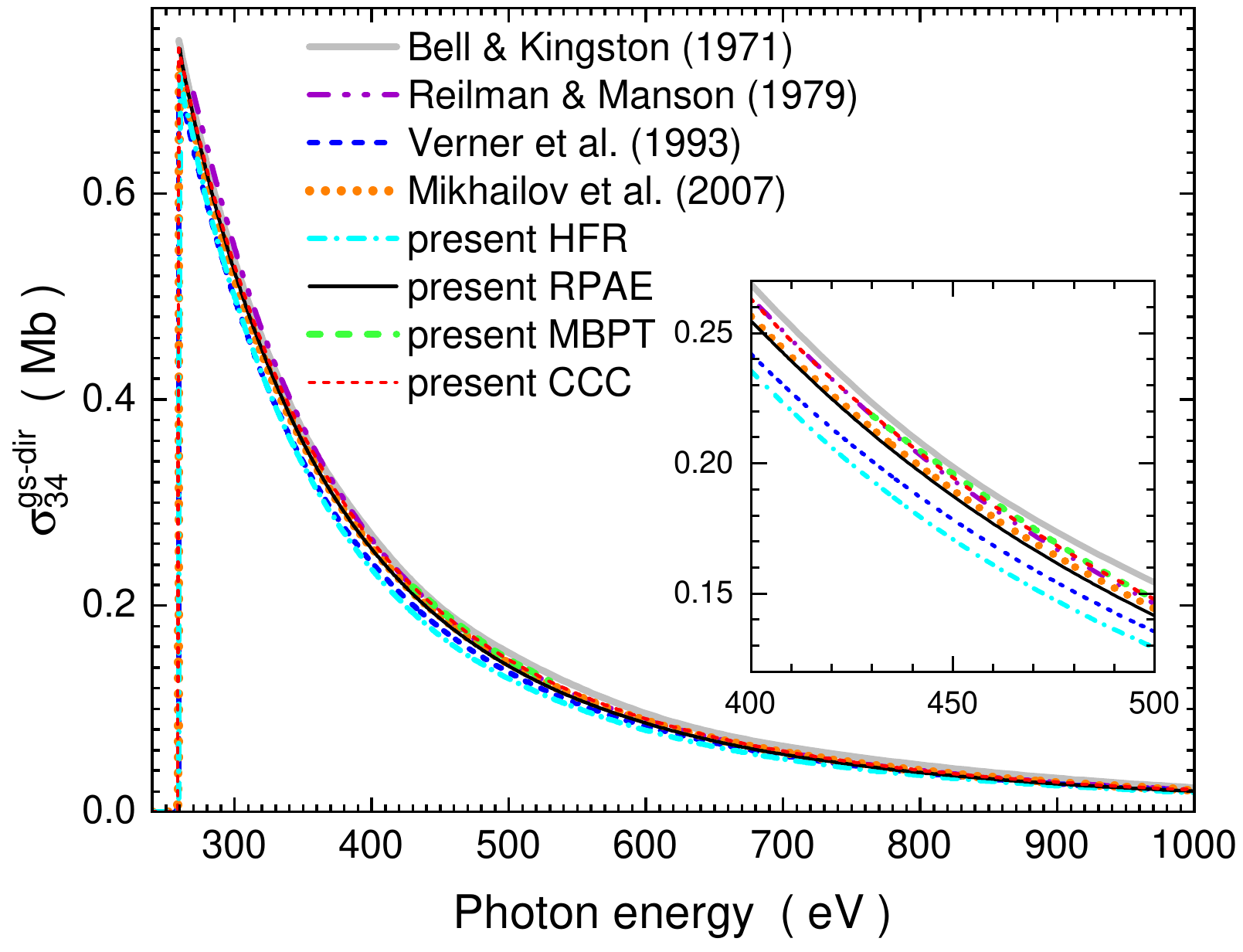}
\caption{\label{Fig:gsdirect} (color online)
Theoretical cross sections $\sigma_{34}^{\mathrm{dir}}$ for direct photoionization of B$^{3+}(1s^2~^1S)$. The light-gray solid  line is the result obtained by Bell and Kingston~\cite{Bell1971}, the magenta dash-dot-dotted line is from work of Reilman and Manson~\cite{Reilman1979}, the blue dashed line is from Verner \textit{et al.}~\cite{Verner1993a}, the orange dotted line is the calculation by Mikhailov \textit{et al.}~\cite{Mikhailov2007}, the cyan dash-dotted line is the present HFR result, and the black solid line is the result of the present RPAE theory. Direct single ionization cross sections were also obtained from the present MBPT and CCC calculations by removing the resonance contributions. The green thick dashed  line is the MBPT result and the red fine dashed curve is the CCC result. The inset zooms into the energy range 400 to 500~eV.
}
\end{figure}

\begin{figure}[t]
\includegraphics[width=\columnwidth]{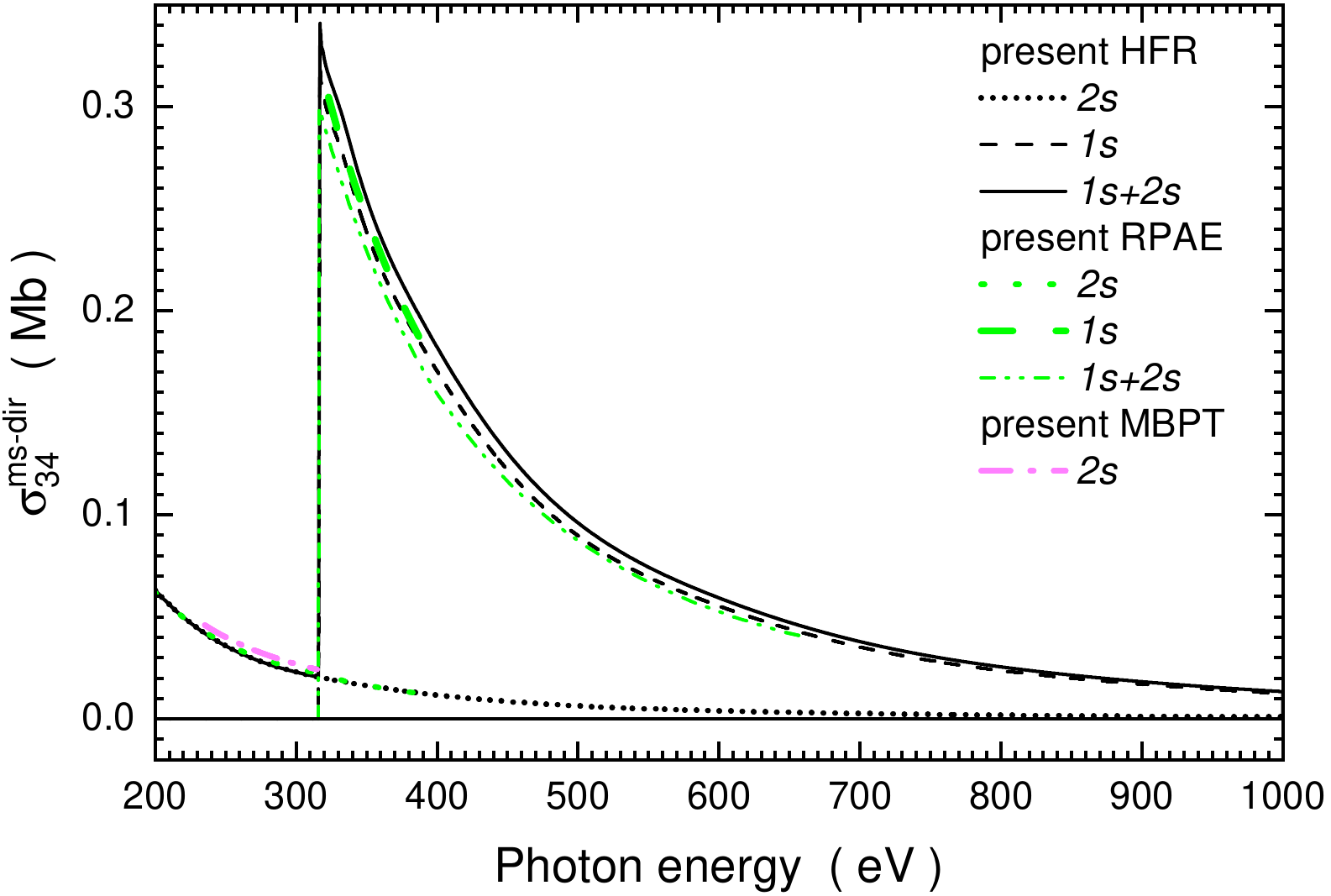}
\caption{\label{Fig:msdirect} (color online) Theoretical cross sections for direct photoionization of metastable B$^{3+}(1s2s~^3S)$. The present HFR results are shown as black lines, the RPAE results are shown as green lines, and the MBPT direct $2s$ ionization cross section is shown as a magenta dash-dotted line. The solid black curve is the total direct-ionization cross section $\sigma_{34}^{\mathrm{ms-dir}}$, i.e., the sum of the partial cross sections for $2s$ and $1s$ ionization. The black dashed line is the partial cross section $\sigma_{34}^{\mathrm{ms-dir}-1s}$ for direct removal of the $1s$ electron. The black narrow-dotted line is the partial cross section $\sigma_{34}^{\mathrm{ms-dir}-2s}$ for direct removal of the $2s$ electron. The RPAE results for direct $2s$ ionization are represented by the wide-dotted green line, for direct $1s$ ionization by the dash-dot-dotted green line, and for the sum of $1s$ and $2s$ ionization by the thick wide-dashed green line.
}
\end{figure}

In the energy range explored in this work, net single ionization of He-like B$^{3+}$ is dominated by direct $K$-shell ionization. The cross sections for direct removal of a $1s$ electron from either B$^{3+}(1s^2)$ or B$^{3+}(1s2s)$ can be calculated with good accuracy. The quality of available calculations may be judged by comparing the results of several different theoretical approaches (see Fig.~\ref{Fig:gsdirect}and its inset). The remarkable agreement among the theoretical results provides a very good basis for normalizing experimental cross sections.

Figure~\ref{Fig:gsdirect} shows cross sections $\sigma_{34}^{{\mathrm{gs}}-{\mathrm dir}}$ for direct photoionization of B$^{3+}(1s^2~^1S)$ resulting from eight different theoretical approaches. For He-like B$^{3+}$ in its ground level, $\sigma_{34}^{\mathrm{gs-dir}}$ is identical with the cross section $\sigma_{34}^{\mathrm{gs-dir}-1s}$ of direct $K$-shell single ionization by a single photon without an additional excitation of the remaining electron. These cross sections have been calculated by a number of different authors. An early result for photoionization of B$^{3+}$ was obtained by Bell and Kingston ~\cite{Bell1971} who derived the cross section from the generalized oscillator strength in the dipole approximation and suitably constructed wave functions for the ground and excited states. Their result is represented by the solid light-gray line in Fig.~\ref{Fig:gsdirect}. Extensive tables of cross sections for direct ionization of atoms and ions were generated by Reilman and Manson~\cite{Reilman1979} performing subshell-specific Hartree-Slater central-field calculations. The magenta dash-dot-dotted line shows their result for ground-state B$^{3+}$. Verner \textit{et al.}~\cite{Verner1993a} performed similarly extensive calculations and tabulated their results in the form of parameters for simple fitting functions. Their calculations are relativistic, and they are based on the self-consistent Hartree-Dirac-Slater potential. Their prediction for direct $K$-shell ionization of B$^{3+}$ is shown by the blue dashed line. Mikhailov \textit{et al.}~\cite{Mikhailov2007} generalized the early Stobbe theory~\cite{Stobbe1930} and particularly addressed the effects of electron-electron correlation. The thick-dotted orange line is their result. Within the present project, the HFR theory (see Sec.~\ref{HFR}) and the RPAE theory (see Sec.~\ref{RPAE}) were employed to calculate $\sigma_{34}^{\mathrm{gs-dir}-1s}$ for ground-level B$^{3+}$. The results are represented by the cyan dash-dotted and black solid lines, respectively. Direct single-ionization cross sections were also obtained by removing the contributions of resonances to the cross sections calculated with the unified-theory MBPT and CCC approaches. The results are displayed in  Fig.~\ref{Fig:gsdirect} with the thick dashed green line representing MBPT and the fine dashed red curve representing CCC. All these calculations give almost identical cross sections as Fig.~\ref{Fig:gsdirect} clearly demonstrates.

While direct ionization of B$^{3+}(1s^2~^1S)$ involves only a single electron shell, the cross section $\sigma_{34}^{\mathrm{ms-dir}}$ of metastable B$^{3+}(1s2s~^3S)$ comprises the two partial cross sections for the $1s$ shell and the $2s$ subshell. Figure~\ref{Fig:msdirect} displays these two partial cross sections together with their sum $\sigma_{34}^{\mathrm{ms-dir}} = \sigma_{34}^{\mathrm{ms-dir}-1s} + \sigma_{34}^{\mathrm{ms-dir}-2s}$. The partial cross sections for the $1s$ and $2s$ direct-ionization contributions were obtained by employing the present HFR, RPAE, and MBPT approaches. At this point it should be mentioned that the direct and indirect  processes, described, e.g., by Eqs.~\ref{Eq:gsdirect} and \ref{Eq:gsresonances} for ground-state B$^{3+}$ ions, cannot really be treated independently. In the vicinity of resonance contributions, the amplitudes of the two interaction channels can interfere with one another so that the reaction pathways described by Eqs.~\ref{Eq:gsdirect} and \ref{Eq:gsresonances} cannot be distinguished. This calls for a unified theoretical treatment.

\begin{table*}[ht]
\caption{\label{Tab:gs-parameters} Calculated parameters (using MBPT) of the  28 energetically lowest resonance contributions to single photoionization of ground-state B$^{3+}(1s^2~^1S)$. The most relevant excited levels are associated with $^1P_1$ states due to the selection rules for electric-dipole transitions. All levels have a total angular momentum $J = 1$. The first column provides the leading configuration and the associated term of each resonance. The entries in columns 2, 3, and 4 are results for the resonance energies relative to the B$^{3+}(1s^2~^1S)$ parent-ion initial level. $E_\mathrm{res}$ includes the most accurate results with extrapolated contributions of un-calculated high-$\ell$ partial waves. Several entries in this column are missing because extrapolations for $2\ell n\ell'~^1P$ resonances with $n \geq 4$ were not performed.  $E_\mathrm{res}^0$ represents the resonance energies obtained without extrapolation and $E_\mathrm{res}^*$ is without the resonance QED corrections (and without extrapolation). Further entries are the natural (life-time) widths $\Gamma$, the Fano $q$ parameters~\cite{Fano1961}, the Auger decay rates $A_\mathrm{a}$, the total radiative rates $A_\mathrm{r}$, the branching ratios for Auger decay $B_\mathrm{a}$, and the ionization resonance strengths $S_\mathrm{ion}$. Numbers in square brackets are powers of 10.}
\begin{ruledtabular}
\begin{tabular}{cccccccccc}
level  & $E_{\mathrm res}$ & $E_{\mathrm res}^0$ & $E_{\mathrm res}^*$ & $\Gamma$ &  $q$    &$A_\mathrm{a}$ &$A_\mathrm{r}$ &  $B_\mathrm{a}$ &  $S_\mathrm{ion}$  \\
                    &      eV       &       eV      &    eV         &   eV        &         &   s$^{-1}$    &   s$^{-1}$    &           &     Mb\,eV  \\
\hline
$	2s2p~^3P	$	& 446.8200  &	446.8200	&	446.8184	&	9.151[-3]	&	-1.331	&	1.357[13]	&	3.307[11]	&	0.9762	&	9.755[-8]	\\
$	2s2p~^1P	$	& 453.1648  &	453.1660	&	453.1644	&	8.442[-2]	&	-1.833	&	1.279[14]	&	3.540[11]	&	0.9972	&	5.958[-2]	\\
$	2s3p~^3P	$	& 487.1025  &	487.1025	&	487.1017	&	3.810[-3]	&	-1.458	&	5.566[12]	&	2.227[11]	&	0.9615	&	1.205[-7]	\\
$	2s3p~^1P	$	& 485.8058  &   485.8058	&	485.8050	&	3.599[-4]	&	-2.293	&	3.135[11]	&	2.333[11]	&	0.5733	&	1.267[-4]	\\
$	2p3s~^3P	$	& 487.4328  &	487.4328	&	487.4324	&	3.207[-4]	&	-1.804	&	1.855[11]	&	3.018[11]	&	0.3807	&	4.825[-8]	\\
$	2p3s~^1P	$	& 489.7694  &	489.7698	&	489.7691	&	3.259[-2]	&	-1.793	&	4.923[13]	&	2.782[11]	&	0.9944	&	1.732[-2]	\\
$	2p3d~^3P	$	& 490.7797  &	490.7797	&	490.7793	&	2.103[-4]	&	-1.659	&	1.096[9]	&	3.184[11]	&	0.0034	&	7.559[-10]	\\
$	2p3d~^1P	$	& 491.7520  &	491.7529	&	491.7526	&	8.539[-4]	&	-1.316	&	9.579[11]	&	3.395[11]	&	0.7383	&	8.018[-5]	\\
$	2s4p~^3P	$	& 499.5273  &	499.5273	&	499.5265	&	1.600[-3]	&	-1.665	&	2.228[12]	&	2.033[11]	&	0.9164	&	3.535[-7]	\\
$	2s4p~^1P	$	& -         &	499.1207	&	499.1199	&	2.693[-4]	&	-2.244	&	1.984[11]	&	2.107[11]	&	0.4849	&	5.979[-5]	\\
$	2p4s~^3P	$	& 499.8705  &	499.8705	&	499.8701	&	2.686[-4]	&	-1.767	&	1.075[11]	&	3.006[11]	&	0.2634	&	1.226[-7]	\\
$	2p4s~^1P	$	& -         &	500.6874	&	500.6869	&	1.201[-2]	&	-2.221	&	1.792[13]	&	3.288[11]	&	0.9820	&	8.260[-3]	\\
$	2p4d~^3P	$	& 501.1097  &	501.1097	&	501.1093	&	1.995[-4]	&	-1.708	&	2.818[9]	&	3.002[11]	&	0.0093	&	6.669[-9]	\\
$	2p4d~^1P	$	& -         &	501.5018	&	501.5014	&	5.234[-4]	&	-1.280	&	4.767[11]	&	3.185[11]	&	0.5994	&	2.674[-5]	\\
$	2s5p~^3P	$	& 505.1085  &	505.1085 	&	505.108 	&	8.503[-4]	&	-1.755	&	1.094[12]	&	1.976[11]	&	0.8470	&	7.308[-7]	\\
$	2s5p~^1P	$	& -         &	504.929 	&	504.928 	&	1.958[-4]	&	-2.226	&	1.242[11]	&	1.733[11]	&	0.4173	&	3.051[-5]	\\
$	2p5s~^3P	$	& 505.3219  &	505.3219 	&	505.322 	&	2.409[-4]	&	-1.752	&	6.376[10]	&	3.023[11]	&	0.1742	&	1.898[-7]	\\
$	2p5s/5d~^1P/^3D	$	& -     &	505.709	    &	505.709 	&	3.820[-3]	&	-3.148	&	5.457[12]	&	3.469[11]	&	0.9402	&	2.775[-3]	\\
$	2p5s/5d~^1P/^3D	$	& -     &	505.701	&	505.701 	&	3.821[-3]	&	-1.124	&	5.451[12]	&	3.550[11]	&	0.9388	&	3.945[-4]	\\
$	2p5d~^3P	$	& 505.9259  &	505.92599 	&	505.925 	&	1.974[-4]	&	-1.716	&	5.185[9]	&	2.948[11]	&	0.0173	&	2.437[-8]	\\
$	2p5d~^1P	$	& -         &	506.120 	&	506.119 	&	3.674[-4]	&	-1.251	&	2.479[11]	&	3.103[11]	&	0.4441	&	8.885[-6]	\\
$	2s6p~^3P	$	& 508.0743  &	508.0743 	&	508.073 	&	5.326[-4]	&	-1.740	&	6.182[11]	&	1.910[11]	&	0.7639	&	1.128[-6]	\\
$	2s6p~^1P	$	& -         &	507.979 	&	507.978 	&	2.133[-4]	&	-1.534	&	8.569[10]	&	2.383[11]	&	0.2645	&	4.443[-6]	\\
$	2p6s~^3P	$	& 508.2090  &	508.2090 	&	508.209 	&	2.245[-4]	&	-1.775	&	4.291[10]	&	2.982[11]	&	0.1258	&	2.796[-7]	\\
$	2p6s/6d~^1P/^3D	$	& -     &	508.426	    &	508.426 	&	2.090[-3]	&	-2.623	&	2.826[12]	&	3.491[11]	&	0.8900	&	1.272[-3]	\\
$	2p6s/6d~^1P/^3D	$	& -     &	508.417	&	508.417 	&	2.488[-3]	&	-1.493	&	3.429[12]	&	3.508[11]	&	0.9072	&	8.079[-4]	\\
$	2p6d~^3P	$	& 508.5514  &	508.5514 	&	508.551 	&	1.936[-4]	&	-1.853	&	8.317[9]	&	2.858[11]	&	0.0283	&	7.840[-8]	\\
$	2p6d~^1P	$	& -         &	508.661 	&	508.660 	&	2.966[-4]	&	-1.346	&	1.307[11]	&	3.199[11]	&	0.2900	&	4.409[-6]	\\
\end{tabular}
\end{ruledtabular}

\end{table*}

\subsection{\label{net_single} Net single photoionization of B$^{3+}$}
Two theoretical appoaches were used in the present study to obtain the cross sections for net single photoionization of B$^{3+}(1s^2~^1S)$ and B$^{3+}(1s2s~^3S)$.

\subsubsection{\label{Sec:sig34gs}  {\rm{B}}$^{3+}(1s^2~^1S)$}
\begin{figure}
\includegraphics[width=\columnwidth]{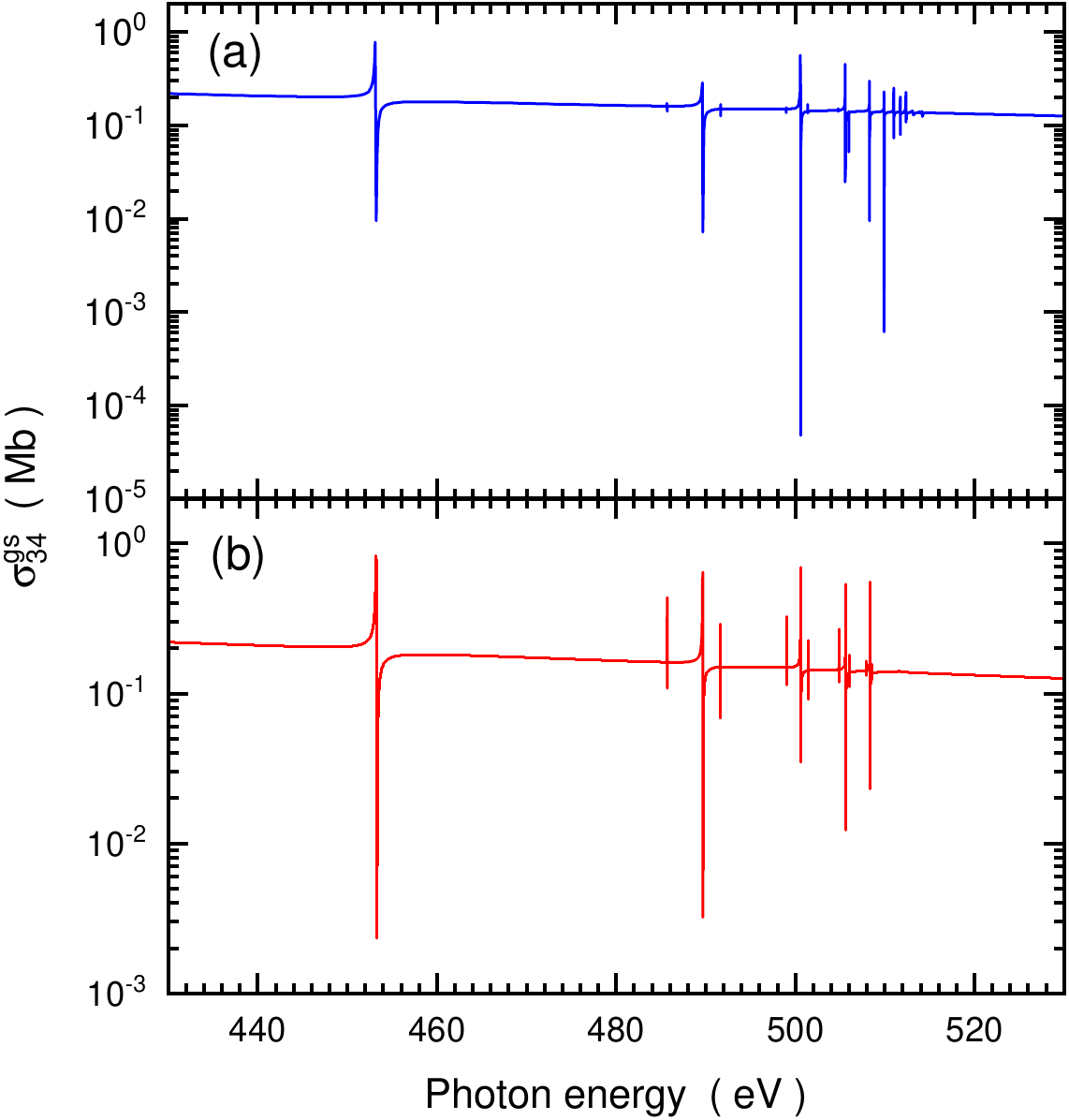}
\caption{\label{Fig:sig34gs} (color online) Theoretical cross sections $\sigma_{34}^{\mathrm{gs}}$ for net single photoionization of ground-level B$^{3+}(1s^2~^1S)$ ions in the region of double $K$-shell excitation resonances. Panel (a) presents the result of the present CCC calculations. Panel (b) shows the present MBPT results.
}
\end{figure}

\begin{figure*}[t]
\begin{center}
\includegraphics[width=17.5cm]{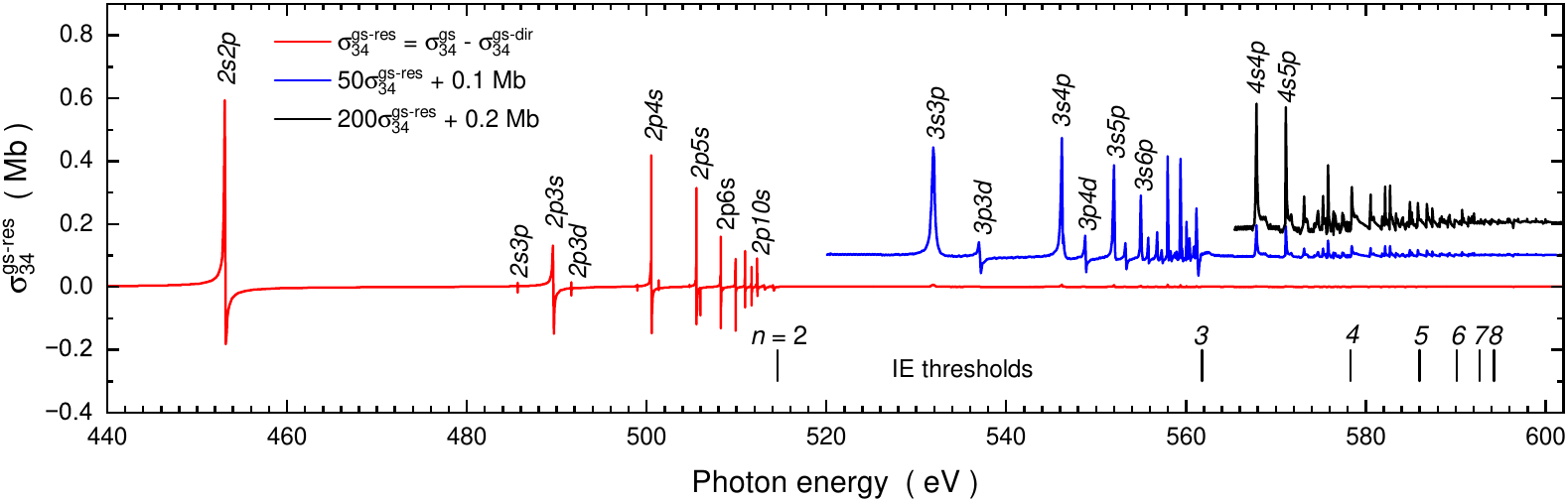}
\caption{\label{Fig:sig34gsresonancesonly} (color online) Resonance contributions $\sigma_{34}^{\mathrm{gs-res}}$ to the net single photoionization of heliumlike B$^{3+}(1s^2~^1S)$ ions obtained by subtracting $\sigma_{34}^{\mathrm{gs-dir}}$ for direct photoionization from CCC calculations of $\sigma_{34}^{\mathrm{gs}}$. The (red) solid is the overview spectrum of resonances. The next line up (blue) is the same curve multiplied by a factor 50 and offset by 0.1~Mb.  The spectrum of resonances was multiplied by a factor 200 and offset by 0.2~Mb to obtain the black line. The most important resonance features are identified by their leading $LS$ configuration. All resonances visible in the spectrum are associated with $^1P_1$ levels. The vertical bars labeled with numbers $n$ indicate the thresholds for IE processes leading to hydrogenlike B$^{4+}$ in levels with the electron in shells with principal quantum numbers $n$.}
\end{center}
\end{figure*}

The net single-ionization cross section $\sigma_{34}^{\mathrm{gs}}$ calculated using the present CCC theory is shown as the solid blue line in panel (a) of Fig.~\ref{Fig:sig34gs}. The same cross section calculated using the present MBPT method is represented by the red solid line in panel (b).
The CCC calculations comprise all photoionzation contributions that lead from the ground state of B$^{3+}$ to the final charge state B$^{4+}$. Within the constraints of the chosen Laguerre basis, the CCC theory includes resonances up to very highly excited DCH levels as evidenced by Fig.~\ref{Fig:sig34gsresonancesonly}. The unified CCC theory permits calculations of cross sections as a function of photon energy only on a point-by-point basis. This implies that narrow resonances can only be mapped out completely, if the grid of energies is sufficiently dense. Because of the existing constraints in computing resources, the number of energy steps had to be limited in the CCC calculations.

The MBPT method permits derivation of all relevant parameters associated with individual resonances. In order to keep the required calculatons tractable, the number of individual resonances was restricted to the 28 energetically lowest levels.  Because of the limitations of the CCC and MBPT calculations, the resonance contributions shown in Fig.~\ref{Fig:sig34gs} panels (a) and (b) differ from one another. For the 28 resonance levels investigated by the MBPT method, very accurate parameters could be obtained (see Table~\ref{Tab:gs-parameters}). Uncertainties of level energies are estimated to be less than 2~meV for the singlet states $2\ell n\ell'~^1P$ with $n \geq 4$ and below $\pm$1~meV for all other resonances investigated. As discussed in Sec.~\ref{MBPT}, the summation over higher partial waves generally converges more slowly for the singlet states and we estimate the $2s2p~^1P_1$ state to shift downwards by 1.2~meV when $\ell >10$ is considered, the $2p3s~^1P_1$ state to shift downwards by 0.4~meV, and the $2p3d~^1P_1$ state to shift downwards by 0.9~meV. For the $2s3p~^1P_1$ state as well as for the triplet states  the effect is  below $0.1$~meV. For levels with $n \geq 4$ we estimate that the contributions from higher $\ell$ values are less than 1~meV for the triplets and less than 2~meV for the singlets.

The role of the parameters in calculating the photoionization cross section has been described in detail previously~\cite{Mueller2018d}. The cross section $\sigma_{34}^{{\mathrm gs}}$ shown in Fig.~\ref{Fig:sig34gs} panel~(b) was evaluated by using the parameters provided in Table~\ref{Tab:gs-parameters}. In this table and throughout the present work, the  resonances are classified by their leading $LS$-term.  That this classification is meaningful can be deduced, e.g.,  from Table~\ref{Tab:gs-parameters} where it is evident that the strengths of the $^1P$ resonances (last column)  are generally larger than those of the triplet resonances (which will not be populated in the non-relativistic limit).  This is particularly true for the symmetrically excited resonances.   The reason is the size of the term splitting relative to the fine structure. For the lower resonances the fine-structure splitting of the inner electron ($\sim $0.03 eV) is much smaller than the term splitting: for principal quantum numbers  of the first and second electrons $n_1 = 2$ and $n_2 = 3$, respectively, the term splitting is around one eV, and for $(n_1 = 2, n_2 = 4)$,  it is  a few tenths of an eV.  This leads to a singlet-triplet mixing of below one percent in the former case and below ten percent in the latter. For $n_1 = 2$ and $n_2 = 5$ on the other hand, there are cases where the  term splitting is smaller than the fine-structure splitting. In this case we see  a breakdown of $LS$ coupling as indicated in Table~\ref{Tab:gs-parameters} for the resonances at $\sim 505.7$~eV.

In general, doubly excited states have to be described, even qualitatively,  by a combination of several strongly mixing configurations. One example can be the states dominated by
the nearly degenerate  $2s4p$, $2p4s$ and $2p4d$ configurations. The present calculation uses a model space consisting of all contributing $jj$ configurations. The result is then projected onto $LS$-coupled configurations and named after the leading one. This name should not be taken too seriously. For example, the state labeled  $2s4p_{J=1}$ in Table~\ref{Tab:msparameters}  is in fact  $51$\% $2s4p$, $44$\% $2p4s$, and $5$\% $2p4d$. The admixtures are, however, very similar for $J = 0$, $J = 1$, and J=2, and  the na\"{\i}ve expectation that the $2s4p$ triplet of states should be below the $2p4s$ triplet, which in turn is below the $2p4d$ triplet, indeed holds as seen in Table~\ref{Tab:msparameters}.

Alternative quantum labels have been developed for doubly excited states~\cite{Herrick1975,Lin1983}. In the well established notation $_N{\mathrm{(K,T)}}_n^{\mathrm{A}}$~\cite{Lin1984} the strongest resonances  seen in Table~\ref{Tab:gs-parameters} are classified by $_2{\mathrm{(0,1)}}_n^{\mathrm{+}}$,
while nearby states of the same overall symmetry ($^1P^o$),  which are significantly  more narrow and considerably weaker,  belong to the $_2{\mathrm{(1,0)}}_n^{\mathrm{-}}$  or $_2{\mathrm{(-1,0)}}_n^{\mathrm{0}}$ series.  The dominance of the population of the $_2{\mathrm{(0,1)}}_n^{\mathrm{+}}$ series
from the ground state can be explained by so-called propensity rules~\cite{Rost1997}.
In Table~\ref{Tab:msparameters}, where the doubly excited states are  populated from the $1s2s~^3S$ level,  the strongest resonances belong to the $_2{\mathrm{(1,0)}}_n^{\mathrm{+}}$ series, but also the $_2{\mathrm{(0,1)}}_n^{\mathrm{-}}$ resonances are prominent in the spectra, cf. Fig.~\ref{Fig:1s2sresonances}.
Here it is in fact the $\mathrm{A}=-1$ states that are favoured during the population, as predicted by the propensity rules,   but their  smaller branching ratio for Auger decay results still in the $\mathrm{A}=+1$ resonances being the strongest in the spectra.
The $_N{\mathrm{(K,T)}}_n^{\mathrm{A}}$ labels have some predictive power and are therefore helpful in characterizing the spectrum of doubly excited states in heliumlike systems. Reviews on the physics of two-electron systems have provided detailed explanations of the connection between new correlation quantum numbers and the new theoretical approaches behind them~\cite{Lin1995,Tanner2000a}. In the present paper, dealing with the lowest doubly excited levels of heliumlike B$^{3+}$, we stick to the simple $LS$ labels but also provide assignments with the KT$^\mathrm{A}$ classification scheme.
\begin{figure}
\includegraphics[width=\columnwidth]{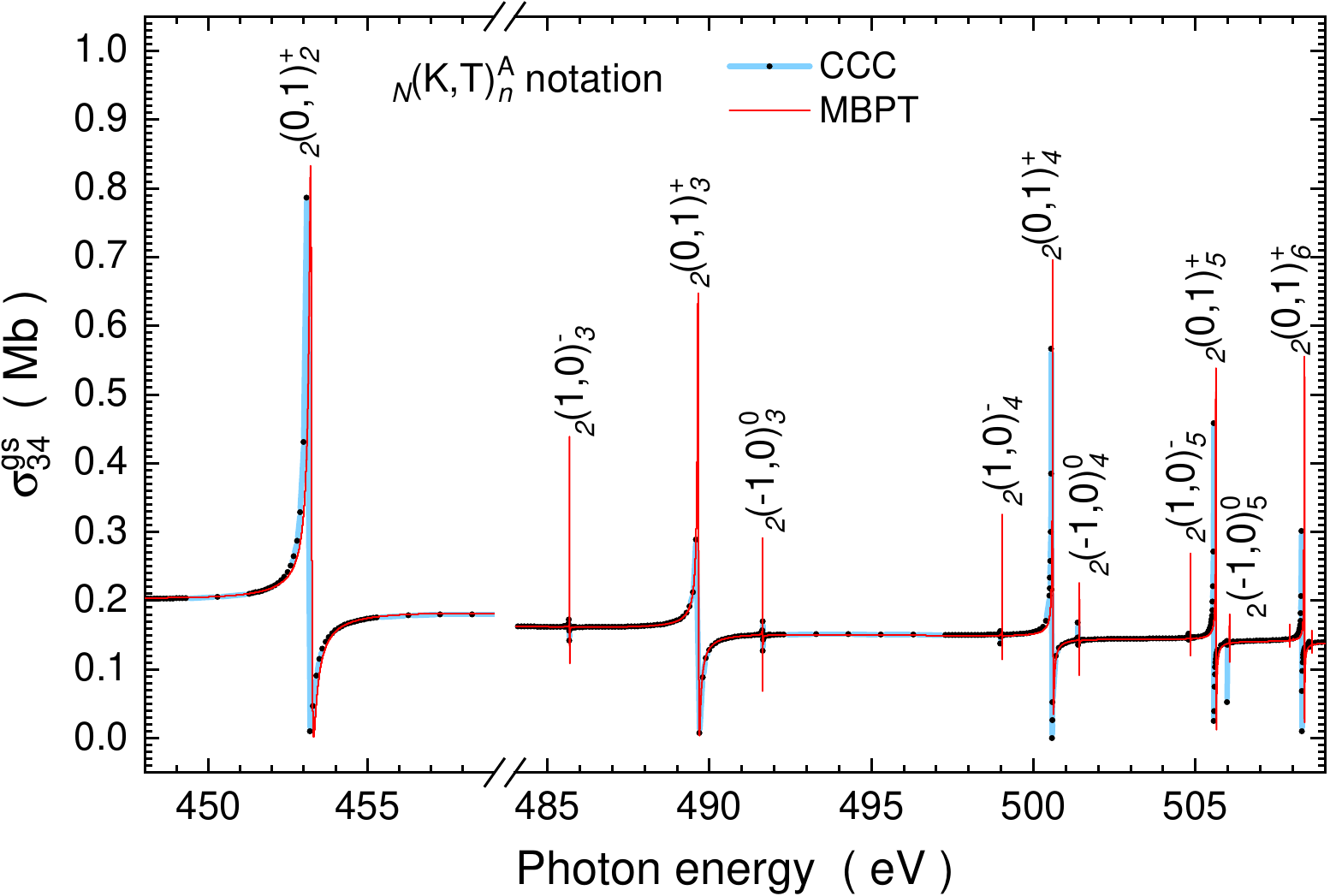}
\caption{\label{Fig:KTAcomparison} (color online) Theoretical cross sections $\sigma_{34}^{\mathrm{gs}}$ for net single photoionization of ground-level B$^{3+}(1s^2~^1S)$ ions in the energy range below the threshold for ionization with excitation of the remaining electron to the $L$ shell. The black dots connected by a light-blue solid line are the results of the present CCC calculations. The thin  solid red line shows the present MBPT results. The resonances are labeled using the $_N{\mathrm{(K,T)}}_n^{\mathrm{A}}$ notation.
}
\end{figure}

Figure~\ref{Fig:sig34gsresonancesonly} shows the partial cross section $\sigma_{34}^{\mathrm{gs-res}} = \sigma_{34}^{\mathrm{gs}}-\sigma_{34}^{\mathrm{gs-dir}}$. This is the contribution of resonances to net single photoionization associated with doubly excited levels in heliumlike B$^{3+}$. The spectrum for atomic number $Z = 5$ provided here, corresponds to the data for $Z = 2$ found in a series of previous publications on resonances in the photoionization of neutral He~\cite{Domke1991a,Domke1992a, Schulz1996a,Puettner2001}. In the present work, autoionizing two-electron levels are characterized by the leading configuration in the $LS$ coupling scheme as mentioned above. The IE thesholds labeled with the principal quantum numbers $n$ of the excited B$^{4+}$ product states clearly mark segments of the ionization resonance spectrum. The strongest resonance contributions are found below the $n = 2$ threshold. With increasing $n$ the resonance series contributing to net single ionization are associated with smaller cross sections.

For a more direct comparison of the results from the present CCC and MBPT calculations, Fig.~\ref{Fig:KTAcomparison} shows the cross sections $\sigma_{34}^{\mathrm{gs}}$ from both theoretical approaches. The energy range is limited to resonances with the ``inner'' electron in the $N=2$ shell and the ``outer'' electron in shell $n$ with $n\geq2$. The $_N{\mathrm{(K,T)}}_n^{\mathrm{A}}$ classification scheme is employed. There is a relatively good overall agreement between the two calculated cross sections. A closer inspection shows, however, that resonance energies do not quite match nor are the maxima and minima of resonance peaks identical. This is particularly true for the very narrow $_2(1,0)_n^{-}$ and $_2(-1,0)_n^{0}$ series of levels. A part of these discrepancies may be attributed to the fact that CCC calculations are only possible at discrete energies. With a chosen density of grid points, resonance strength may be missed that would otherwise be visible with smaller energy steps in the calculations.

\subsubsection{\label{Fig:sig34ms}  {\rm{B}}$^{3+}(1s2s~^3S)$}

\begin{figure}[t]
\includegraphics[width=\columnwidth]{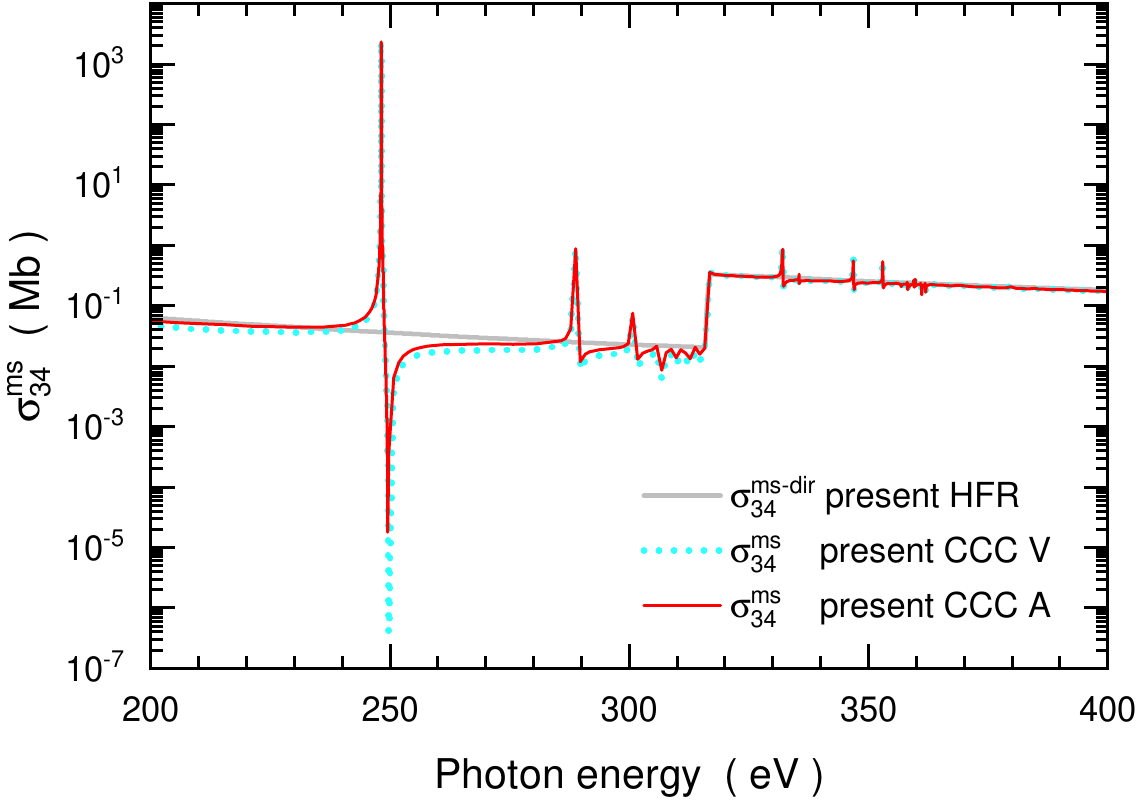}
\caption{\label{Fig:msCCC} (color online) Cross sections for net single photoionization of metastable B$^{3+}(1s2s~^3S)$ ions.
The solid red line and the dotted cyan line represent the cross section $\sigma_{34}^{\mathrm{ms}}$  obtained by CCC calculations in the acceleration and velocity gauges, respectively. The thick  solid light-gray line is the direct-ionization cross section resulting from the present HFR calculations.
}
\end{figure}

The net single-ionization cross section $\sigma_{34}^{\mathrm{ms}}$ calculated using the present CCC theory in the velocity gauge is shown as the dotted cyan line in Fig.~\ref{Fig:msCCC}. In the energy range below the $K$ edge, the calculations in the length, velocity, and acceleration gauges do not fully agree with each other indicating an incomplete representation of the initial state. For comparison, the result associated with the acceleration gauge is shown by the solid red line. The thick light-gray line is the direct-ionization cross section $\sigma_{34}^{\mathrm{ms-dir}}$ obtained by the present HFR calculations.

The step in the cross section at about 315~eV marks the onset of direct $K$-shell ionization of metastable B$^{3+}(1s2s~^3S)$. Below that step are the contributions of direct ionization of the 2s subshell and (resonant) single excitation of the $K$ shell. As mentioned in the context of Fig.~\ref{Fig:sig34gs} the CCC calculations did not fully map out individual resonance contributions which would have required a much higher density of energy-dependent cross-section points. Although ``incomplete'' in that sense, the CCC calculations do show, however, that the resonances at energies above the $K$ edge are very small reaching maxima in the range of 0.1~Mb. These resonances were not observed in the present experiments. They are associated with double excitations of both the $1s$ and the $2s$ electron in the initial B$^{3+}(1s2s~^3S)$ ion.  In contrast to these double excitations,  the strongest resonance below the $K$ edge at about 250~eV is associated with single $K$-shell excitation and reaches into the Gb range. This latter finding is confirmed by the MBPT approach.

\begin{figure}[t]
\includegraphics[width=\columnwidth]{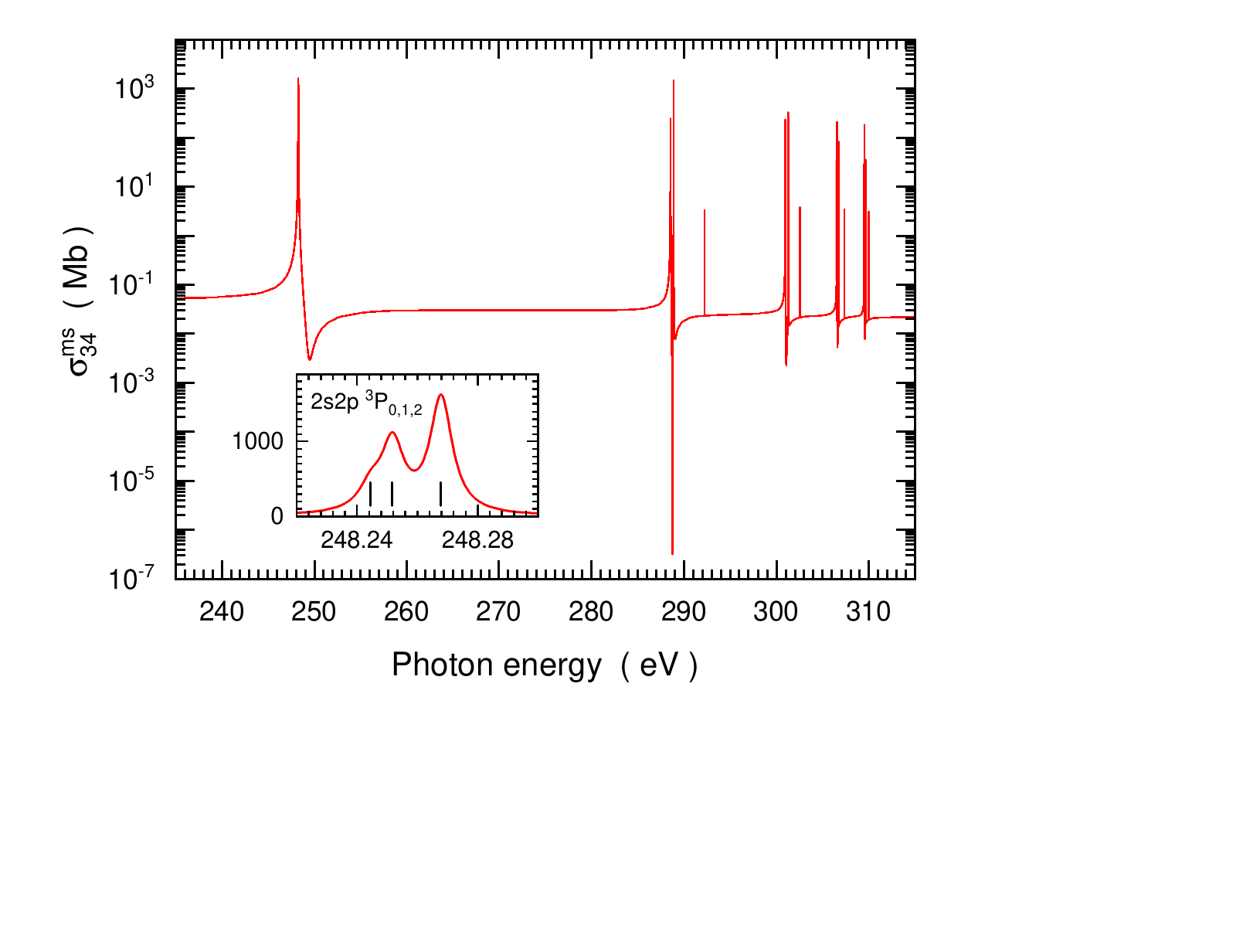}
\caption{\label{Fig:msMBPT} 
Cross section for single photoionization of metastable B$^{3+}(1s2s~^3S)$ ions obtained by MBPT calculations including the 40 energetically lowest $^3P$ resonance levels. The inset shows the first resonance feature in the spectrum on a linear scale. The vertical bars indicate the calculated resonance energies from Table~\ref{Tab:msparameters}.
}
\end{figure}

\begin{table*}
\caption{\label{Tab:msparameters} Calculated parameters (using MBPT) of the 40 lowest-energy resonance contributions to single photoionization of metastable B$^{3+}(1s2s~^3S)$. With the exception of the $2s2p~^1P_1$ level,  the excited states are associated with $^3P$ terms due to the selection rules for electric-dipole transitions. The first column provides the resonance level. Further entries are the resonance energies $E_\mathrm{res}$ (with converged partial-wave contributions) relative to the parent-ion initial state, the same energies without resonance QED corrections (and in the case of the $2s2p~^1P$ resonance also without the 1.2~meV extrapolation correction) $E_{\mathrm{res}^*}$, the natural (life-time) widths $\Gamma$, the Fano $q$ parameters~\cite{Fano1961},  the Auger decay rates $A_\mathrm{a}$, the total radiative rates $A_\mathrm{r}$, the branching ratios for Auger decay $B_\mathrm{a}$, and the ionization resonance strengths $S_\mathrm{ion}$. Numbers in square brackets are powers of 10.}
\begin{ruledtabular}
\begin{tabular}{cccccccccc}
level        &   $E_{\mathrm{res}}$ & $E_{\mathrm{res}^*}$ & $\Gamma$&   $q$         &$A_\mathrm{a}$ & $A_\mathrm{r}$& $B_\mathrm{a}$ &  $S_\mathrm{ion}$  \\
                    &      eV       &       eV      &    eV          &               &  s$^{-1}$     &   s$^{-1}$    &                &     Mb\,eV  \\
\hline
$	2s2p~^3P_0	$	&	248.2446	&	248.2430	&	9.272[-3]    &	-2.701[2]    &	1.376[13]    &	3.307[11]    &	9.7653[-1]    &	4.404[0]     \\
$	2s2p~^3P_1	$	&	248.2518	&	248.2501	&	9.151[-3]    &	-2.703[2]    &	1.357[13]    &	3.307[11]    &	9.7621[-1]    &	1.321[1]     \\
$	2s2p~^3P_2	$	&	248.2679	&	248.2662	&	9.045[-3]    &	-2.708[2]    &	1.341[13]    &	3.307[11]    &	9.7593[-1]    &	2.201[1]     \\
$	2s2p~^1P_1	$	&	254.5966	&	254.5962	&	8.442[-2]    &	-9.006[1]    &	1.279[14]    &	3.702[11]    &	9.9711[-1]    &	4.383[-5]    \\
$	2s3p~^3P_0	$	&	288.5296	&	288.5288	&	3.853[-3]    &	-1.082[2]    &	5.629[12]    &	2.253[11]    &	9.6151[-1]    &	2.639[-1]    \\
$	2s3p~^3P_1	$	&	288.5343	&	288.5334	&	3.810[-3]    &	-1.108[2]    &	5.566[12]    &	2.227[11]    &	9.6152[-1]    &	8.155[-1]    \\
$	2s3p~^3P_2	$	&	288.5441	&	288.5433	&	3.785[-3]    &	-1.161[2]    &	5.533[12]    &	2.176[11]    &	9.6216[-1]    &	1.440[0]     \\
$	2p3s~^3P_0	$	&	288.8600	&	288.8595	&	3.337[-4]    &	-2.742[3]    &	2.081[11]    &	2.990[11]    &	4.1037[-1]    &	1.899[-1]    \\
$	2p3s~^3P_1	$	&	288.8646	&	288.8641	&	3.207[-4]    &	-2.995[3]    &	1.855[11]    &	3.018[11]    &	3.8071[-1]    &	5.188[-1]    \\
$	2p3s~^3P_2	$	&	288.8739	&	288.8734	&	2.962[-4]    &	-3.630[3]    &	1.428[11]    &	3.073[11]    &	3.1720[-1]    &	6.930[-1]    \\
$	2p3d~^3P_0	$	&	292.2146	&	292.2143	&	2.098[-4]    &	1.981[4]     &	2.304[8]     &	3.185[11]    &	7.2293[-4]    &	7.616[-5]    \\
$	2p3d~^3P_1	$	&	292.2114	&	292.2111	&	2.103[-4]    &	1.988[4]     &	1.096[9]     &	3.184[11]    &	3.4318[-3]    &	1.089[-3]    \\
$	2p3d~^3P_2	$	&	292.2061	&	292.2058	&	2.096[-4]    &	1.981[4]     &	3.197[8]     &	3.181[11]    &	1.0041[-3]    &	5.360[-4]    \\
$	2s4p~^3P_0	$	&	300.9549	&	300.9541	&	1.617[-3]    &	-1.291[2]    &	2.251[12]    &	2.068[11]    &	9.1586[-1]    &	1.070[-1]    \\
$	2s4p~^3P_1	$	&	300.9591	&	300.9583	&	1.600[-3]    &	-1.314[2]    &	2.228[12]    &	2.033[11]    &	9.1637[-1]    &	3.306[-1]    \\
$	2s4p~^3P_2	$	&	300.9674	&	300.9665	&	1.590[-3]    &	-1.358[2]    &	2.220[12]    &	1.968[11]    &	9.1856[-1]    &	5.838[-1]    \\
$	2p4s~^3P_0	$	&	301.2983	&	301.2978	&	2.735[-4]    &	-9.044[2]    &	1.187[11]    &	2.967[11]    &	2.8576[-1]    &	3.932[-2]    \\
$	2p4s~^3P_1	$	&	301.3023	&	301.3018	&	2.686[-4]    &	-9.508[2]    &	1.075[11]    &	3.006[11]    &	2.6337[-1]    &	1.057[-1]    \\
$	2p4s~^3P_2	$	&	301.3104	&	301.3100	&	2.572[-4]    &	-1.065[3]    &	8.259[10]    &	3.082[11]    &	2.1133[-1]    &	1.334[-1]    \\
$	2p4d~^3P_0	$	&	302.5443	&	302.5439	&	1.980[-4]    &	-2.379[4]    &	2.421[8]     &	3.006[11]    &	8.0478[-4]    &	3.387[-5]    \\
$	2p4d~^3P_1	$	&	302.5415	&	302.5411	&	1.994[-4]    &	-2.186[4]    &	2.818[9]     &	3.002[11]    &	9.3000[-3]    &	1.185[-3]    \\
$	2p4d~^3P_2	$	&	302.5364	&	302.5360	&	1.972[-4]    &	-2.037[4]    &	2.417[8]     &	2.994[11]    &	8.0672[-4]    &	1.747[-4]    \\
$	2s5p~^3P_0	$	&	306.5362	&	306.5354	&	8.578[-4]    &	-1.338[2]    &	1.100[12]    &	2.037[11]    &	8.4367[-1]    &	4.818[-2]    \\
$	2s5p~^3P_1	$	&	306.5403	&	306.5394	&	8.502[-4]    &	-1.374[2]    &	1.094[12]    &	1.976[11]    &	8.4700[-1]    &	1.522[-1]    \\
$	2s5p~^3P_2	$	&	306.5477	&	306.5468	&	8.446[-4]    &	-1.441[2]    &	1.097[12]    &	1.864[11]    &	8.5477[-1]    &	2.794[-1]    \\
$	2p5s~^3P_0	$	&	306.7499	&	306.7495	&	2.418[-4]    &	-7.461[2]    &	7.213[10]    &	2.953[11]    &	1.9633[-1]    &	1.247[-2]    \\
$	2p5s~^3P_1	$	&	306.7537	&	306.7533	&	2.409[-4]    &	-8.002[2]    &	6.376[10]    &	3.023[11]    &	1.7417[-1]    &	3.153[-2]    \\
$	2p5s~^3P_2	$	&	306.7616	&	306.7612	&	2.342[-4]    &	-9.541[2]    &	3.981[10]    &	3.160[11]    &	1.1191[-1]    &	3.025[-2]    \\
$	2p5d~^3P_0	$	&	307.3603	&	307.3599	&	1.940[-4]    &	-2.225[4]    &	3.077[7]     &	2.948[11]    &	1.0437[-4]    &	2.121[-6]    \\
$	2p5d~^3P_1	$	&	307.3577	&	307.3572	&	1.974[-4]    &	-1.771[4]    &	5.185[9]     &	2.948[11]    &	1.7285[-2]    &	1.072[-3]    \\
$	2p5d~^3P_2	$	&	307.3527	&	307.3523	&	1.924[-4]    &	-1.538[4]    &	6.085[7]     &	2.923[11]    &	2.0810[-4]    &	2.231[-5]    \\
$	2s6p~^3P_0	$	&	309.5020	&	309.5012	&	5.380[-4]    &	-1.354[2]    &	6.152[11]    &	2.022[11]    &	7.5260[-1]    &	2.381[-2]    \\
$	2s6p~^3P_1	$	&	309.5061	&	309.5053	&	5.326[-4]    &	-1.412[2]    &	6.182[11]    &	1.910[11]    &	7.6391[-1]    &	7.830[-2]    \\
$	2s6p~^3P_2	$	&	309.5129	&	309.5120	&	5.248[-4]    &	-1.512[2]    &	6.234[11]    &	1.739[11]    &	7.8185[-1]    &	1.526[-1]    \\
$	2p6s~^3P_0	$	&	309.6371	&	309.6367	&	2.220[-4]    &	-6.758[2]    &	4.761[10]    &	2.898[11]    &	1.4111[-1]    &	5.051[-3]    \\
$	2p6s~^3P_1	$	&	309.6408	&	309.6404	&	2.245[-4]    &	-7.454[2]    &	4.291[10]    &	2.982[11]    &	1.2579[-1]    &	1.240[-2]    \\
$	2p6s~^3P_2	$	&	309.6488	&	309.6485	&	2.245[-4]    &	-9.972[2]    &	1.792[10]    &	3.232[11]    &	5.2523[-2]    &	7.126[-3]    \\
$	2p6d~^3P_0	$	&	309.9857	&	309.9853	&	1.935[-4]    &	6.530[4]     &	5.552[7]     &	2.940[11]    &	1.8881[-4]    &	2.056[-6]    \\
$	2p6d~^3P_1	$	&	309.9832	&	309.9827	&	1.936[-4]    &	-1.029[5]    &	8.317[9]     &	2.858[11]    &	2.8276[-2]    &	9.497[-4]    \\
$	2p6d~^3P_2	$	&	309.9784	&	309.9780	&	1.896[-4]    &	-4.056[4]    &	1.034[8]     &	2.879[11]    &	3.5904[-4]    &	2.135[-5]    \\
\end{tabular}
\end{ruledtabular}

\end{table*}

The MBPT calculations were restricted to the 40 lowest-energy resonances in the photoionization cross section $\sigma_{34}^{\mathrm{ms}}$. The results for the metastable B$^{3+}(1s2s~^3S)$ ion are shown in Fig.~\ref{Fig:msMBPT}. Different from Figs.~\ref{Fig:sig34gs}, \ref{Fig:msCCC}, and \ref{Fig:msMBPT} where data are shown on a logarithmic cross-section scale, the inset of Fig.~\ref{Fig:msMBPT} displays the first resonance group on a linear scale. The associated resonance energies obtained by the present MBPT calculations are indicated by the vertical bars. Due to the natural widths of the individual levels within the $^3P$ term, the triplet structure of the resonance feature cannot fully be resolved even with infinite resolving power.

The resonance energies were determined with very high accuracy with uncertainties of less than $\pm1$~meV for most of the investigated resonance levels. Other parameters were obtained with estimated uncertainties of less than 1\%. The calculated resonance parameters are provided in Table~\ref{Tab:msparameters}. The cross section shown in Fig.~\ref{Fig:msMBPT} was evaluated from these parameters with the resulting resonance profiles added to the calculated cross section for direct ionization of metastable B$^{3+}(1s2s~^3S)$.

As Table~\ref{Tab:msparameters} shows, the maximum life-time widths of the investigated $^3P_{0,1,2}$ resonances are only slightly above 9~meV. In contrast, life-time widths of the $^1P_1$ resonances populated by double-$K$-shell excitation reach up to about 84~meV. In principle, the triplet $P$ levels addressed by Table~\ref{Tab:msparameters} can also be populated from the ground state. However, in that case, they require inter-combination transitions and, hence, are characterized by small resonance strengths. The level-specific parameters, i.e., the natural widths $\Gamma$, the Auger and radiative decay rates $A_{\mathrm{a}}$ and $A_\mathrm{r}$, respectively, and the branching ratios of the $^3P_1$ levels for autoionization in the two tables are identical while the resonance energies, the resonance strengths $S_{\mathrm{ion}}$, and the Fano parameters $q$ are different since the transitions to the identical final levels start from different initial states.

\begin{table*}[htb]
\caption{\label{Tab:comparisons}
Comparison of parameters of $^{11}$B$^{3+}(2\ell 2\ell')$ resonances in single photoionization of metastable B$^{3+}(1s2s~^3S)$ and ground-state B$^{3+}(1s^2~^1S)$ originating from two different theoretical endeavors. Columns 2-5 provide  the  total binding energies for the doubly excited states resulting from the present MBPT approach, as well  as those obtained
by Zaytsev \textit{et al.}~\cite{Zaytsev2019}. In both cases the results  after inclusion of  the Coulomb and Breit interactions to all orders (DCB: Dirac-Coulomb-Breit) are given separately, and then the final results when recoil and QED effects are considered follow (see text for explanations).
Columns 6-9 give the transition energies (Trans. en.) relative to the  metastable B$^{3+}(1s2s~^3S)$ and the ground-state B$^{3+}(1s^2~^1S)$, using the results from Yerokhin \textit{et al.}~\cite{Yerokhin2015,Yerokhin2022}. Columns  10 and 11 present  Auger widths $\Gamma_\mathrm{A}$.
}
\begin{ruledtabular}
\begin{tabular}{l|cc|cc|cc|cc|cc}
Reso- & \multicolumn{2}{c|}{DCB energy}  & \multicolumn{2}{c|}{Total energy} & \multicolumn{2}{c|}{Trans.~en. from $1s2s^3S$}
& \multicolumn{2}{c|}{Trans.~en. from $1s^2~^1S$} & \multicolumn{2}{c}{$\Gamma_A$}
 \\
nance & present  & Ref.~\cite{Zaytsev2019} & present  & Ref.~\cite{Zaytsev2019}
& present  & Ref.~\cite{Zaytsev2019}& present  & Ref.~\cite{Zaytsev2019} & present  & Ref.~\cite{Zaytsev2019}\\
level & eV & eV & eV & eV & eV & eV & eV & eV & eV & eV\\
\hline
$2s2p~^3P_0$	& -152.79682	& -152.79682	& -152.7876	& -152.7877	& 248.2446	& 248.2445	& 			& & 9.05[-3]  &  9.02[-3]\\
$2s2p~^3P_1$	& -152.78968	& -152.78968	& -152.7804	& -152.7806	& 248.2518	& 248.2516	& 446.8200	& 446.8199	  & 8.93[-3]	& 8.92[-3]\\
$2s2p~^3P_2$	& -152.77364	& -152.77364	& -152.7644	& -152.7645	& 248.2679	& 248.2677	& 			& & 8.83[-3]  &  8.82[-3]\\
$2s2p~^1P_1$	& -146.44455(10)$^a$	& -146.44445	& -146.4356	& -146.4358	& 254.5966	& 254.5964	& 453.1648	& 453.1646	  & 8.42[-2]	& 8.41[-2]\\
\end{tabular}
$^a$ The result with $\ell_{\rm max}=10$ is $-146.44334$~eV, extrapolation from there gives the quoted result.
\end{ruledtabular}
\end{table*}

It is interesting to compare the present resonance energies with the results of other calculations and with experimental data. The first observation of doubly excited heliumlike B$^{3+}$ was reported by Kennedy and Carroll~\cite{Kennedy1978}. They found a wavelength of 49.934(5)~{\AA} for the transition $2s2p~^3P \to 1s2s~^3S$ without resolving the fine structure of the $^3P$ term.  The measured wavelength corresponds to a transition energy of 248.296(25)~eV which is much closer to the present energy than most of the existing theoretical values. There are quite many calculations on doubly excited levels of He-like ions in the literature. Comparisons of a number of previous results for resonance energies and resonance widths of He-like atoms and ions with atomic numbers between 2 and 10 have been provided by Gning \textit{et al.}~\cite{Gning2015}. The theoretical data for resonance energies show relatively large differences. For the $2\ell 2\ell'$ terms in the B$^{3+}$ ion, they scatter around their mean with a standard deviation of more than 500~meV reaching maximum deviations of more than 2~eV. For the same resonances the previously calculated total widths deviate from one another by up to 26\%.

A closer inspection shows that hardly any of the previous calculations match the high quality standards of the present theoretical work and, particularly, the present MBPT approach which is both relativistic and considers the most important quantum electrodynamic (QED) corrections. Only one theory publication on doubly excited states of B$^{3+}$ could be found that claims a level of accuracy sufficient to challenge the results of the present work. Zaytsev \textit{et al.}~\cite{Zaytsev2019} published high-accuracy energies and Auger widths for all the 10 levels possible in $2\ell 2\ell'$  doubly excited configurations of heliumlike ions with atomic numbers $Z\geq5$.

Like the present calculation, Ref.~\cite{Zaytsev2019} uses complex rotation to be able to treat autoionizing states, and includes correlation due to both the Coulomb and the Breit interactions. The methods differ, though, concerning the implementations.  Reference~\cite{Zaytsev2019} uses a B-spline  basis  and a configuration-interaction approach, while the present MBPT calculation employs  a  finite-difference basis combined with a many-body perturbation  expansion. Apart from numerical uncertainties these methods should, however, be equivalent for two-electron systems. Where possible, the results of Ref.~\cite{Zaytsev2019} for the $^{11}$B$^{3+}$ ion are compared with the present data in Table~\ref{Tab:comparisons}.

Zaytsev \textit{et al.} presented the total energies of the excited levels. In order to compare with the present results shown in Tables~\ref{Tab:gs-parameters} and \ref{Tab:msparameters}  it is necessary to determine the excitation energies relative to  the initial levels $^{11}$B$^{3+}(1s^2~^1S_0)$ and $^{11}$B$^{3+}(1s2s~^3S_1)$, respectively, which are relevant in the present context.
The total energies of those levels were taken or inferred from previous publications in which exceptionally small uncertainties are quoted. By extending calculations to include two-electron QED effects to all orders in the nuclear binding strength parameter $Z\alpha$, with the fine-structure constant $\alpha$, Yerokhin \textit{et al.}~\cite{Yerokhin2022} were able to obtain the most accurate theoretical predictions to date for the ionization energies of the lowest levels of heliumlike ions with $Z = 5$ to $Z = 30$. Numbers for B$^{3+}$ are provided in Table~\ref{Tab:thresholds}. The determination of the total energy of the doubly excited levels requires the additional knowledge of the binding energy of the electron in H-like $^{11}$B$^{4+}(1s~^2S_{1/2})$. Yerokhin and Shabaev~\cite{Yerokhin2015} provided ionization and excitation energies of $n = 1$ and $n = 2$ levels of hydrogenlike atoms and ions  on the basis of \textit{ab initio} QED calculations performed to all orders in the nuclear binding strength parameter $Z\alpha$ with  $Z$ ranging from 1 to 110. By combining the information from the two cited papers, it is possible to infer very accurate total energies as well as ionization and excitation energies (see Table~\ref{Tab:thresholds}) that can be employed for the comparison of the present findings with the data of Zaytsev \textit{et al.}

The results obtained with the present MBPT method and by the theoretical approach chosen by Zaytsev \textit{et al.}~\cite{Zaytsev2019} are in excellent agreement with one another. In fact the results on the Dirac-Coulomb-Breit (DCB) level [where the effects of the Coulomb and Breit interactions are fully accounted for but no effects due to the quantization of the electromagnetic field (QED effects for short) are considered yet] agree to all quoted digits for the triplet levels. Here, the summation over partial waves is converged and no extrapolation is needed. The partial-wave convergence is much slower for the $2s2p~^1P_1$
state and here we find that contributions from $\ell \geq 11$ change the results by about $1.2$~meV. With this contribution added, the present results agree to within a few tenths of a meV with (the also extrapolated) values from Ref.~\cite{Zaytsev2019} where the authors calculated QED and recoil effects from first principles. Here, we only add the reduced-mass effect and estimate the QED effects as explained in Sec.~\ref{MBPT}. Still our results agree to within $0.2$~meV with that from Ref.~\cite{Zaytsev2019} as detailed in Table~\ref{Tab:comparisons}.  This gives confidence in our values for the resonances $2\ell n\ell'$
with $n>2$, where no first-principle calculations of QED and recoil effects are available.

The differences of the calculated Auger widths $\Gamma_A$  obtained by the two theoretical approaches are remarkably small. They are only 0.01~meV with one exception, the width of the $2s2p~^3P_0$ resonance, where the result of the present calculation is 0.03~meV above that of Zaytsev \textit{et al.} The uncertainties of the latter results were quoted to be between 0.01 and 0.02~meV, thus essentially explaining the small differences. As mentioned above, calculated widths in previous publications differ from one another by up to 26\%. The differences between the present results and the calculations of Zaytsev \textit{et al.} are less than 0.3\%. Obviously, a milestone of accuracy has been reached in the present theoretical approach as well as in the St. Petersburg group's  methodology in calculating resonance parameters for doubly excited states of two-electron systems.

\subsection{\label{gs_ion-exci} Contributions of ionization-excitations to net single photoionization of B$^{3+}$}

Beside direct photoionization and indirect ionization processes involving the excitation of autoionizing levels in the parent ion and subsequent Auger decay, there is a third  mechanism, IE,  by which the parent ion finally ends up in the channel of net single ionization. In such processes, one of the electrons in the parent A$^{q+}$ ion is directly removed by a single photon while a second electron is lifted to an excited state of the A$^{(q+1)+}$ ion. In a He-like ion such as ground-state B$^{3+}$, the IE mechanism can be described as
\begin{equation}
\label{Eq:IE}
h\nu + \rm{B}^{3+}(1s^2) \to  \rm{B}^{4+}(n\ell) + e^- ,
\end{equation}
with the B$^{4+}$ ion in an excited state with principal quantum number $n \geq 2$ and orbital quantum number $\ell$. The CCC calculations reported here for single photoionization comprise all three net ionization channels discussed so far. In addition, CCC also addresses indirect processes that produce excited B$^{4+}(n\ell)$ levels. Such indirect processes may involve, for example, the population of intermediate doubly excited resonances such as B$^{3+}(3p nd)$ that can decay to B$^{4+}(2s)+e^-$ provided that $n \geq 3$.

\begin{figure}[t]
\includegraphics[width=\columnwidth]{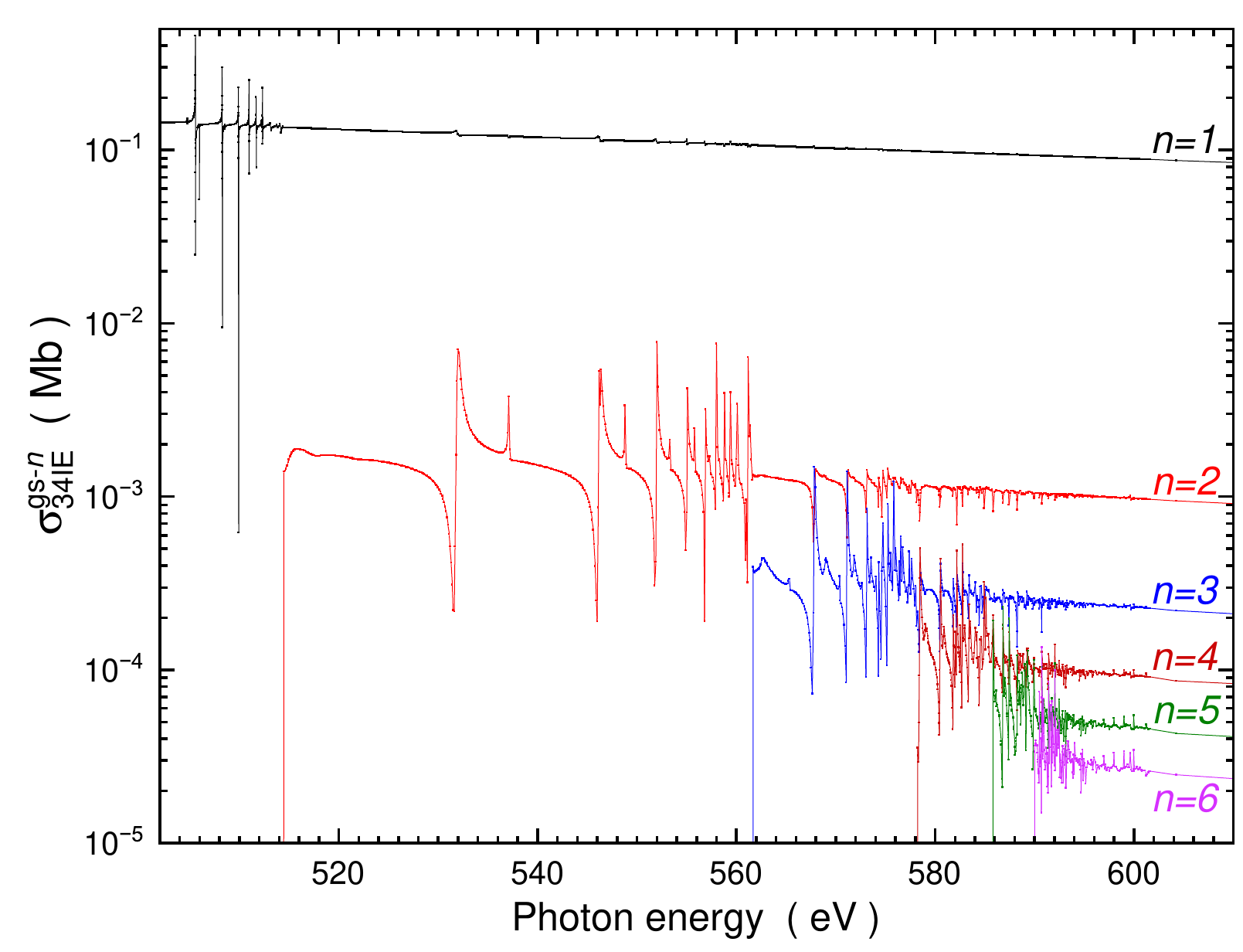}
\caption{\label{Fig:gs-IE} (color online) Cross sections $\sigma_{34{\mathrm{IE}}}^{{\mathrm{gs}}-n}$  for IE of B$^{3+}(1s^2~^1S)$ with production of B$^{4+}(n)$ where the remaining electron is found in shell $n$ with $n=2,3,...,6$. For comparison, the cross section for single ionization without excitation (the black curve labeled $n=1$) is also shown. The data were obtained by the present CCC calculations.
}
\end{figure}

By the present CCC calculations, cross sections $\sigma_{34{\mathrm{IE}}}^{{\mathrm{gs}}-n}$ were obtained for IE processes described by Eq.~\ref{Eq:IE} for the ground state of B$^{3+}$ without specifying the orbital quantum number $\ell$, that is, the cross sections are summed over all possible $\ell$ for each $n$. The principal quantum numbers $n$ range from $n=2$ to $n=6$. Similar calculations have been published previously for photoionization of the helium isoelectronic sequence~\cite{Kheifets1998}. In the present new calculations, the density of energy steps was increased relative to the calculations in Ref.~\cite{Kheifets1998} to see the effects of resonance contributions. Figure~\ref{Fig:gs-IE} shows the cross sections $\sigma_{34{\mathrm{IE}}}^{{\mathrm{gs}}-n}$ for different final principal quantum numbers $n$ as a function of the photon energy.
Clearly, the threshold for producing B$^{4+}$ increases with $n$. According to the calculations by  Yerokhin \textit{et al.}~\cite{Yerokhin2015,Yerokhin2022} the energy required to produce B$^{4+}(2s)$ is 
514.53533~eV (see Table~\ref{Tab:thresholds}). This is approximately the onset of the calculated cross section $\sigma_{34{\mathrm{IE}}}^{{\mathrm{gs}}-2}$. The onsets of the partial cross sections increase according to the excitation energies of hydrogenlike B$^{4+}(n\ell)$.

The first resonance in the cross section $\sigma_{34{\mathrm{IE}}}^{{\mathrm{gs}}-2}$ at an energy of approximately 532~eV with its pronounced Fano profile is associated with intermediate B$^{3+}(3s3p~^1P)$  autoionizing to B$^{4+}(n=2) + e^-$. Higher-lying resonances can be identified on the basis of the assignments provided in Fig.~\ref{Fig:sig34gsresonancesonly}. It is interesting to see the relatively large resonance contributions to the IE cross sections for excitations to higher principal quantum numbers $n$. When $n$ increases the continuum cross section for direct IE strongly decreases (approximately with $n^{-3}$ for $n \geq 2$ with the scaling getting better for higher $n$) so that the resonances riding on that continuum clearly stick out.

Especially important in the present context is the size of the IE cross sections. They are all well below $10^{-2}$~Mb and they rapidly drop with increasing $n$. The present CCC calculations also yielded $\sigma_{34{\mathrm{IE}}}^{{\mathrm{gs}}-1}$ which is the sum of the cross sections $\sigma_{34}^{\mathrm{gs-dir}}$ for direct single photoionization of the B$^{3+}$ ground-state ion and the associated resonance contributions or, in other words, the cross section  for net single ionization of B$^{3+}(1s^2~^1S)$ producing ground-level B$^{4+}(1s)$ without any excitation. Thus, obviously, $\sigma_{34{\mathrm{IE}}}^{{\mathrm{gs}}-1}$ does not contribute to the total IE cross section. The label ``IE'' just characterizes the partial, no-excitation cross section with $n=1$   from the series $\sigma_{34{\mathrm{IE}}}^{{\mathrm{gs}}-n}$.

\begin{figure}[htb]
\includegraphics[width=\columnwidth]{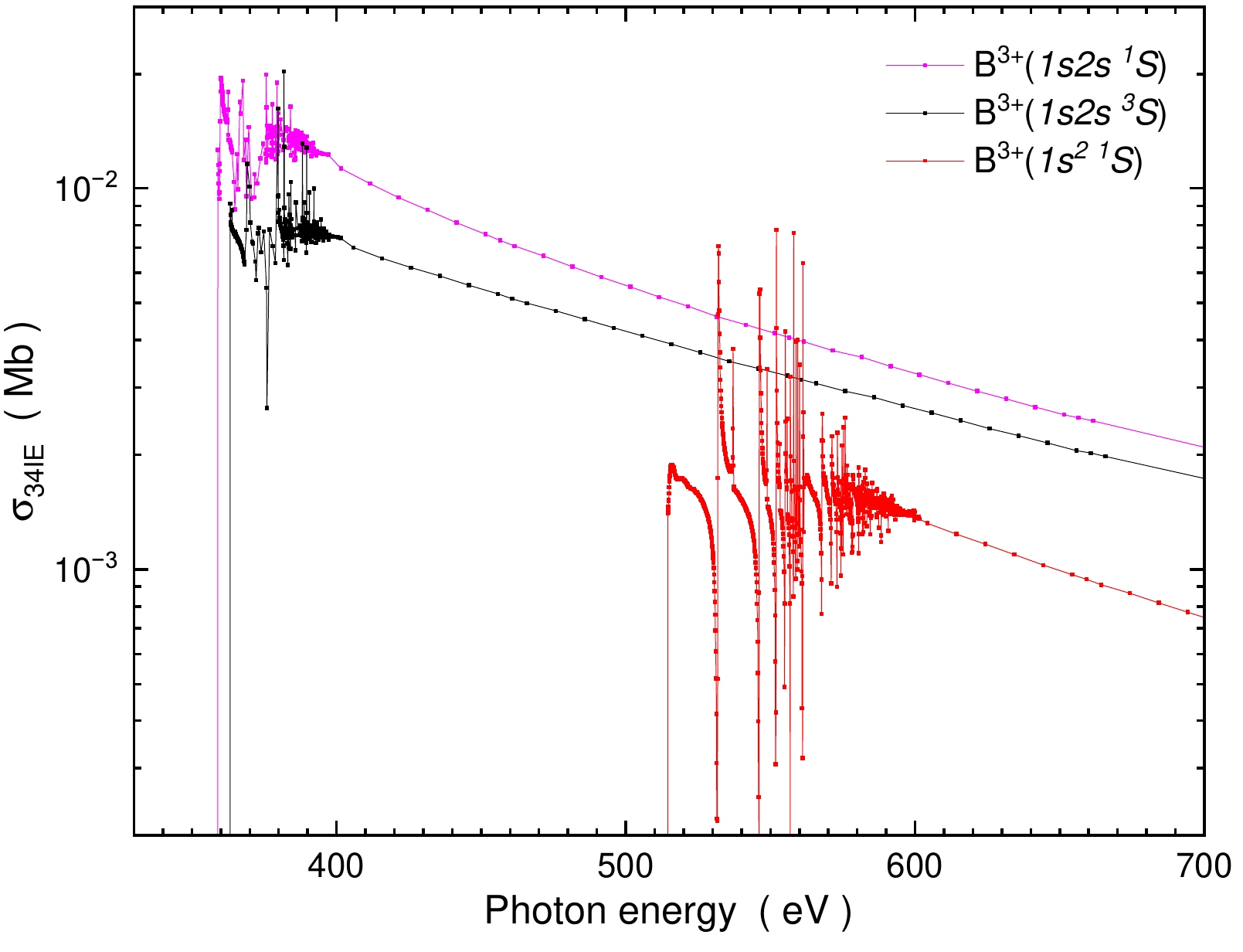}
\caption{\label{Fig:total-IE} (color online) Total cross sections $\sigma_{34{\mathrm{IE}}}$ for IE of B$^{3+}(1s^2~^1S)$, B$^{3+}(1s2s~^1S)$, and B$^{3+}(1s2s~^3S)$ ions. The lowest excitation function (red) belongs to the initial B$^{3+}$ ion in its ground state. The next above that (black)   belongs to the initial ion in the $1s2s~^3S$ level. The largest IE cross section is found for the $1s2s~^1S$ level. The small square dots are the cross sections obtained from CCC calculations. They are connected by straight colored lines to highlight the energy dependence of the cross sections.
}
\end{figure}
The CCC calculations show that $\sigma_{34{\mathrm{IE}}}^{{\mathrm{gs}}-1}$ is almost as large as $\sigma_{34}^{\mathrm{gs}}$, the cross section for net single ionization of B$^{3+}$ (shown in Fig.~\ref{Fig:sig34gs}). The difference of the latter two cross sections is the total cross section $\sigma_{34{\mathrm{IE}}}^{{\mathrm{gs}}}$ for IE of ground-state B$^{3+}$, i.e., for the process described by Eq.~\ref{Eq:IE} summed over all $n\geq2$
\begin{equation}
\label{Eq:netIE}
\sigma_{34{\mathrm{IE}}}^{{\mathrm{gs}}} = \sigma_{34}^{{\mathrm{gs}}} - \sigma_{34{\mathrm{IE}}}^{{\mathrm{gs}}-1}
\end{equation}
which is approximately only of the order of 1.5\% to 2\% of $\sigma_{34}^{{\mathrm{gs}}}$ and can, therefore, be neglected in the present investigation of single photoionization of ground-level B$^{3+}$.

IE calculations were also performed for the metastable levels $1s2s~^{1,3}S$ of the B$^{3+}$ ion. The results are similar to those for the ground-state ion in that the contributions of IE are also negligible relative to net single ionization of B$^{3+}$. The population of B$^{4+}(n)$ states with $n=1$ and $n=2$ is almost exclusively due to direct removal of the $2s$ electron or the $1s$ electron, respectively. At the threshold of $K$-shell ionization the contribution of direct removal of the $2s$  electron is only about 4\% of $\sigma_{34}^{\mathrm{ms}}$ for the initial $1s2s~^3S$ level and about 6.7\% for the initial $1s2s~^1S$ level.

Figure~\ref{Fig:total-IE} shows the initial-level-dependent total cross sections $\sigma_{34{\mathrm{IE}}}$ for IE of B$^{3+}(1s^2~^1S)$, B$^{3+}(1s2s~^1S)$, and B$^{3+}(1s2s~^3S)$ ions obtained from the CCC calculations. The total IE cross sections are determined by summing the partial IE cross sections over the principal quantum numbers $n$ of the final levels.  For the ground state the summation starts at $n=2$, for the metastable levels it starts at $n=3$. In reality, the infinite sums are obtained by following the recipe of Eq.~\ref{Eq:netIE}, i.e., by subtracting the contributions of single ionization without excitation from the net single ionization cross sections obtained by CCC calculations.

In the case of ground-state B$^{3+}$ the total IE cross section $\sigma_{34{\mathrm{IE}}}^{{\mathrm{gs}}}$ near its threshold oscillates around approximately 1.5 kb due to the effects of resonance contributions. The IE cross sections for the metastable levels are much larger at their considerably lower onsets, however, their relative contributions to the associated net single ionization are only of the order of a few percent. In conclusion, the contributions of ionization-excitation processes to the net single photoionization of B$^{3+}$ are all relatively small and can be neglected in the present context. Consequently, the net single ionization cross sections can be described as the sum of the well understood smooth direct-ionization continuum and the indirect, typically narrow, resonance contributions which reside on top of the smooth continuum cross section.

\section{\label{Sec:normalization} Normalization of the experimental single-ionization cross section}

\begin{figure}[t]
\begin{center}
\includegraphics[width=7.5cm]{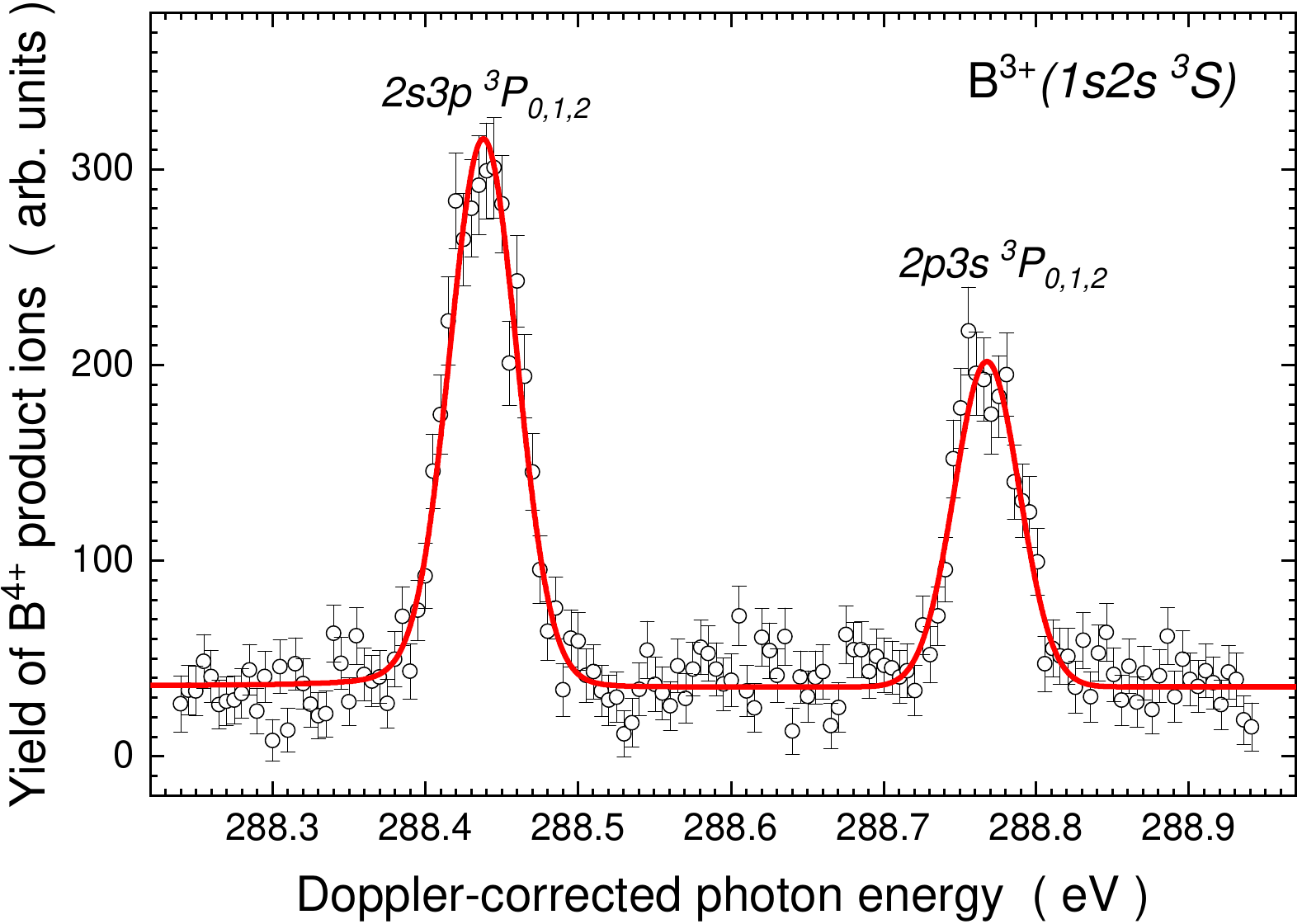}
\caption{\label{Fig:fraction_f} (color online) Measured B$^{4+}$ product-ion yield in the photon-energy range where $2s3p~^3P$ and $2p3s~^3P$ resonances are found that arise from single $K$-shell excitation of metastable B$^{3+}(1s2s~^3S)$. The data were taken with a 200~$\mu$m monochromator exit-slit size. The yields are provided in arbitrary units. The solid red line is a fit to the measured yield based on the results of MBPT theory. An energy resolution of 48(2)~meV was found by the fit.
}
\end{center}
\end{figure}

The findings of the previous section, and particularly the uniformity of the results shown in Fig.~\ref{Fig:gsdirect}, justify the normalization of the measured ion yields to the smooth total direct-ionization continuum of the net single-ionization cross section. Given the quality of the resonance parameters obtained from the MBPT calculations, it is also possible to determine the fraction $f$ of metastable $1s2s~^3S$ ions in the parent B$^{3+}$ beam. For this purpose, a scan of the photon-energy region 288.24 to 288.95~eV was analyzed. In this energy range, the contributions of three partial cross sections are responsible for the measured yield $Y$ of B$^{4+}$ photoionization products. The processes that can happen are (i) the direct ionization of the $2s$ subshell of metastable B$^{3+}$ ions in the parent-ion beam, (ii) the direct ionization of the $K$ shell of ground-state B$^{3+}$ parent ions, and (iii) the resonant $K$-shell excitation and subsequent autoionization of the metastable B$^{3+}$ parent ions. No $K$-shell direct ionization of metastable ions is possible in the chosen energy range as manifested by Table~\ref{Tab:thresholds}. Consequently, IE even only involving the minimum excitation to B$^{4+}(2p_{3/2})$ with its threshold at 315.9938~eV (see Table~\ref{Tab:thresholds}) is also energetically forbidden and could be neglected anyway. Hence, the apparent cross section behind the experimental B$^{4+}$ yield $Y$ in Fig.~\ref{Fig:fraction_f} can be represented as
\begin{equation}
\label{Eq:sig34at288eV}
\sigma_{34}^{\mathrm{app}} = (1-f) \sigma_{34}^{\mathrm{gs-dir}} + f \sigma_{34}^{\mathrm{ms-dir}} + f \sigma_{34}^{\mathrm{ms-res}}
\end{equation}
where the resonance contribution $\sigma_{34}^{\mathrm{ms-res}}$ resulting from $K$-shell excitation of metastable B$^{3+}$ is derived from the data provided in Table~\ref{Tab:msparameters}. From the measured resonance energies one may conclude that the two peak features are due to the $2s3p~^3P_{0,1,2}$ and $2p3s~^3P_{0,1,2}$ intermediate doubly excited states. Compared to the data in Table~\ref{Tab:msparameters} the measured resonances are slightly shifted by about 0.1~eV towards lower energy. The experimental energy scale has to be corrected accordingly. The smooth continuum is related to the cross sections $\sigma_{34}^{\mathrm{gs-dir}}$ and $\sigma_{34}^{\mathrm{ms-dir}}$ from Figs.~\ref{Fig:gsdirect} and \ref{Fig:msdirect}, respectively. A somewhat arbitrary representation of the ground-state direct ionization was chosen as the arithmetic mean of the results obtained by Verner \textit{et al.}~\cite{Verner1993a} and by Mikhailov \textit{et al.}~\cite{Mikhailov2007}. It falls right into the narrow span of theoretical cross-section data displayed in Fig.~\ref{Fig:gsdirect}. At 288.6~eV this cross section is $\sigma_{34}^{{\mathrm{gs}}-{\mathrm{dir}}}=0.558$~Mb while $\sigma_{34}^{{\mathrm{ms}}-{\mathrm{dir}}}$=0.025~Mb according to the present HFR calculations.

Based on Fig.~\ref{Fig:gsdirect},  an estimate of the relative uncertainty of the chosen model cross section for direct $K$-shell single ionization is $\pm 10\%$. At 288.6~eV this amounts to a possible error of 0.056~Mb. With the available data base being much smaller for the metastable ions, a more conservative estimate of $\pm 20\%$ is taken into account for the cross section $\sigma_{34}^{\mathrm{ms-dir}}$. From previous experiments a metastable fraction $f$ close to 0.1 has to be expected implying that the smooth continuum cross section behind the yield $Y$ shown in Fig.~\ref{Fig:fraction_f} is given approximately by $(0.9\times0.558 + 0.1\times0.025)~{\rm Mb} \approx 0.9\times 0.561~{\rm Mb}$. The contribution of direct ionization of the $2s$ subshell of metastable B$^{3+}(1s2s~^3S)$ is less than 0.6\% of the cross section $\sigma_{34}^{{\mathrm{gs}}-{\mathrm{dir}}}$ and can, therefore, be expected to have little effect in Eq.~\ref{Eq:sig34at288eV}.

The calculated strengths of the 6 resonance levels contributing to the yield curve in Fig.~\ref{Fig:fraction_f} can be taken from Table~\ref{Tab:msparameters}. In total, they amount to $S^{\mathrm{theo}}$ = 3.921~Mb\,eV. The experimental resonance strength $S^{\mathrm{exp}}$ can be derived from a fit to the experimental B$^{4+}$ yield $Y$ in Fig.~\ref{Fig:fraction_f}. The fit shown in the figure by a solid red line was performed by keeping the theoretical resonance energies $E_{\mathrm{res}}$, natural widths $\Gamma$, Fano parameters $q$, and ionization strengths $S_\mathrm{ion}$ constant while the photon-energy bandwidth $w_{\rm G}$, an energy shift $\Delta E_{\mathrm{ph}}$, a scale factor $A = S^{\mathrm{exp}}/S^{\mathrm{theo}}$, and a constant background $B$ were used as fit parameters. The fitted curve matches the experimental data very well as evidenced by Fig.~\ref{Fig:fraction_f}.

An energy shift $\Delta E_{\mathrm{ph}} = 0.1012(7)$~eV was found. It represents the photon-energy calibration error of the beamline in the vicinity of $E_{\mathrm{ph}}=290$~eV at the time of the specific measurement. This error was corrected for in the subsequent analysis. The energy resolution was found to be $w_{\rm G} = 48.2(1.7)$~meV for the measurement shown in Fig.~\ref{Fig:fraction_f} which was not sufficient to resolve the fine structure of the two resonance terms. The background $B = 35.59(1.40)$~y.u. (yield units) corresponds to the experimental yield $Y$ outside of the resonances, i.e., the contribution of non-resonant direct ionization. The scale factor came out to be $A = 6.35(19)$~y.u. Mb$^{-1}$. Introducing the factor $N$ for normalizing a measured yield to the absolute cross-section scale leads to a set of simple equations linking the fit parameters $A$ and $B$  to the experimental and theoretical data
\begin{eqnarray}
\label{Eq:eqnarray}
N B & = & (1-f) \sigma_{34}^{\mathrm{gs-dir}}+f \sigma_{34}^{\mathrm ms-dir},\\
N S^{\mathrm{exp}}& = & f S^{\mathrm{theo}},\\
A & = &S^{\mathrm{exp}}/S^{\mathrm{theo}} .
\end{eqnarray}
By dividing the two first equations by one another, the normalization factor drops out and the parameter $f$ can be determined from the numbers given in the text above. The resulting fraction of metastable ions is $f=0.091$ and the normalization factor $N$ = 0.0143~Mb (y.u.)$^{-1}$. The error propagation gives $\Delta f = 0.016$ which corresponds to a relative uncertainty $\Delta f/f \approx 0.18$. This result, $f = 0.091(16)$, is fully compatible with the findings of a previous experiment~\cite{Renwick2009a} that suggested $f = 0.105$ (see the discussion in Sec.~\ref{Sec:experiment}).

The normalization factor $N$ relating the measured yield and the absolute cross section depends on the photon energy $E_{\mathrm{ph}}$. This energy dependence is mainly caused by changes of the shape of the photon beam while $E_{\mathrm{ph}}$ is scanned over a wide energy range. As a result, the beam overlap, the so-called form factor~\cite{Phaneuf1999}, may change with $E_{\mathrm{ph}}$. Since that factor could not be measured during the present experimental campaign, the experimental yields were normalized to the theory for direct single ionization. In a long scan covering $E_{\mathrm{ph}}$ from about 250 to 1200~eV, the yield of B$^{4+}$ ions produced by photoionization of a mixed beam of B$^{3+}$ ions in the ground state and in the $^3S$ metastable state was measured in steps of 5~eV. The photon-energy resolution for that scan was chosen by setting the monochromator exit-slit width to 1000~$\mu$m. At $E_{\mathrm{ph}} \approx 450$~eV a resolution of 0.6~eV resulted from these settings. With the step width of 5~eV it is very unlikely to see evidence for resonance contributions to the measured yields and, indeed, the yield curve shows a smooth behavior.

The experimental yields are expected to follow the apparent cross section for a mixed B$^{3+}$ ion beam with a fraction $1-f = 0.909$, i.e., 90.9\%, of $1s^2~^1S$ ground-state ions and a fraction $f = 0.091$, i.e., 9.1\% of metastable $1s2s~^3S$ ions. Without considering the resonance contributions,
Eq.~\ref{Eq:sig34at288eV} takes the general form
\begin{equation}
\label{Eq:sig34direct}
\sigma_{34}^{\mathrm{app}} = 0.909 \sigma_{34}^{\mathrm{gs-dir}} + 0.091 (\sigma_{34}^{\mathrm{ms-dir}-1s} + \sigma_{34}^{\mathrm{ms-dir}-2s})
\end{equation}
where the partial cross sections for direct ionization of the metastable component by removal of the $1s$ or the $2s$ electron, respectively, are considered. By a smooth function $N(E_{\mathrm{ph}})$ the experimental yield curve is adjusted to $\sigma_{34}^{\mathrm{app}}$ given by Eq.~\ref{Eq:sig34direct}.

\begin{figure}[t]
\begin{center}
\includegraphics[width=\columnwidth]{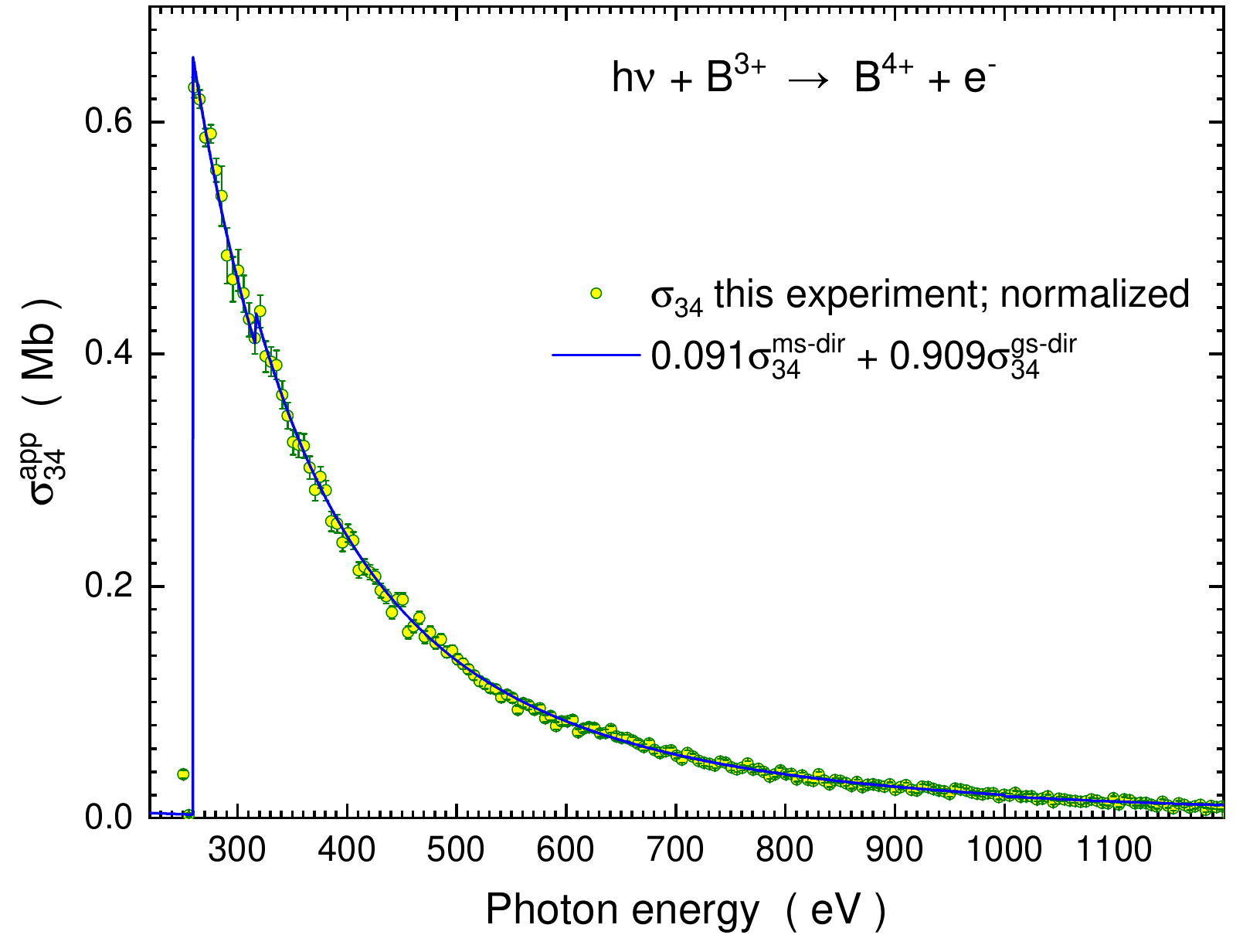}
\caption{\label{Fig:normalized_sig_34_apparent} (color online) Apparent experimental photoionization cross sections (yellow-shaded circles)
normalized to the theoretical cross section (solid blue line) described in Eq.~\ref{Eq:sig34direct}.
}
\end{center}
\end{figure}

The resulting experimental apparent cross section is shown in Fig.~\ref{Fig:normalized_sig_34_apparent}. Obviously, the smooth normalization function $N(E_{\mathrm{ph}})$ works well. The normalization developed for the cross-section overview  was then also employed to obtain the detailed cross-section results that deal with double-$K$-hole resonances.

\section{\label{Sec:comparison} comparison of theory and experiment}

The overview scan displayed in Fig.~\ref{Fig:normalized_sig_34_apparent} with photon-energy steps of 5~eV did not show any signs of resonance structures in the apparent cross section $\sigma_{34}^{\mathrm{app}}$. Such structures could only be revealed in measurements with step sizes smaller than 0.2~eV. After exploring the approximate positions of resonances with the monochromator exit slit wide open, dedicated measurements with increased resolution and smaller photon-energy steps were performed. The results of these measurements for metastable B$^{3+}(1s2s~^3S)$ and ground-state B$^{3+}(1s^2~^1S)$ are separately discussed in the following two subsections.

\subsection{\label{1s2scontributions} B$^{3+}(1s2s~^3S)$}

The metastable B$^{3+}(1s2s~^3S)$ ions with their high excitation energy of 198.568~eV~\cite{Yerokhin2022} already carry a $K$ vacancy. A single $K$-shell excitation produces a doubly excited autoionizing level with two $K$ holes, i.e., an empty $K$ shell. The cross sections for such single excitations are relatively large but drop with increasing excitation energies. Resonance contributions with single $K$-shell excitations of metastable B$^{3+}$ are to be expected in the energy range between the lowest-energy excitation to the $2s2p~^3P_{0,1,2}$ levels at about 248.245~eV and the $K$ edge at 315.97~eV. These numbers are obtained by combining results of previous theoretical work~\cite{Zaytsev2019,Yerokhin2015,Yerokhin2022}. See also Table~\ref{Tab:thresholds}. At energies beyond the $K$ edge, double excitations of both electrons in the $1s2s$ parent configuration can contribute to $\sigma_{34}^{\mathrm{app}}$. However, the sizes of the associated resonances are small as the present CCC calculations suggest (see Fig.~\ref{Fig:msCCC}). Moreover, the contributions of these resonances to the experimental cross section $\sigma_{34}^{\mathrm{app}}$ are weighted by the fraction $f$ of metastable ions in the parent-ion beam which strongly reduces their product ion yield while the continuum of direct ionization remains high. Therefore, the detection of these resonances was not attempted.
\begin{figure}[t]
\begin{center}
\includegraphics[width=\columnwidth]{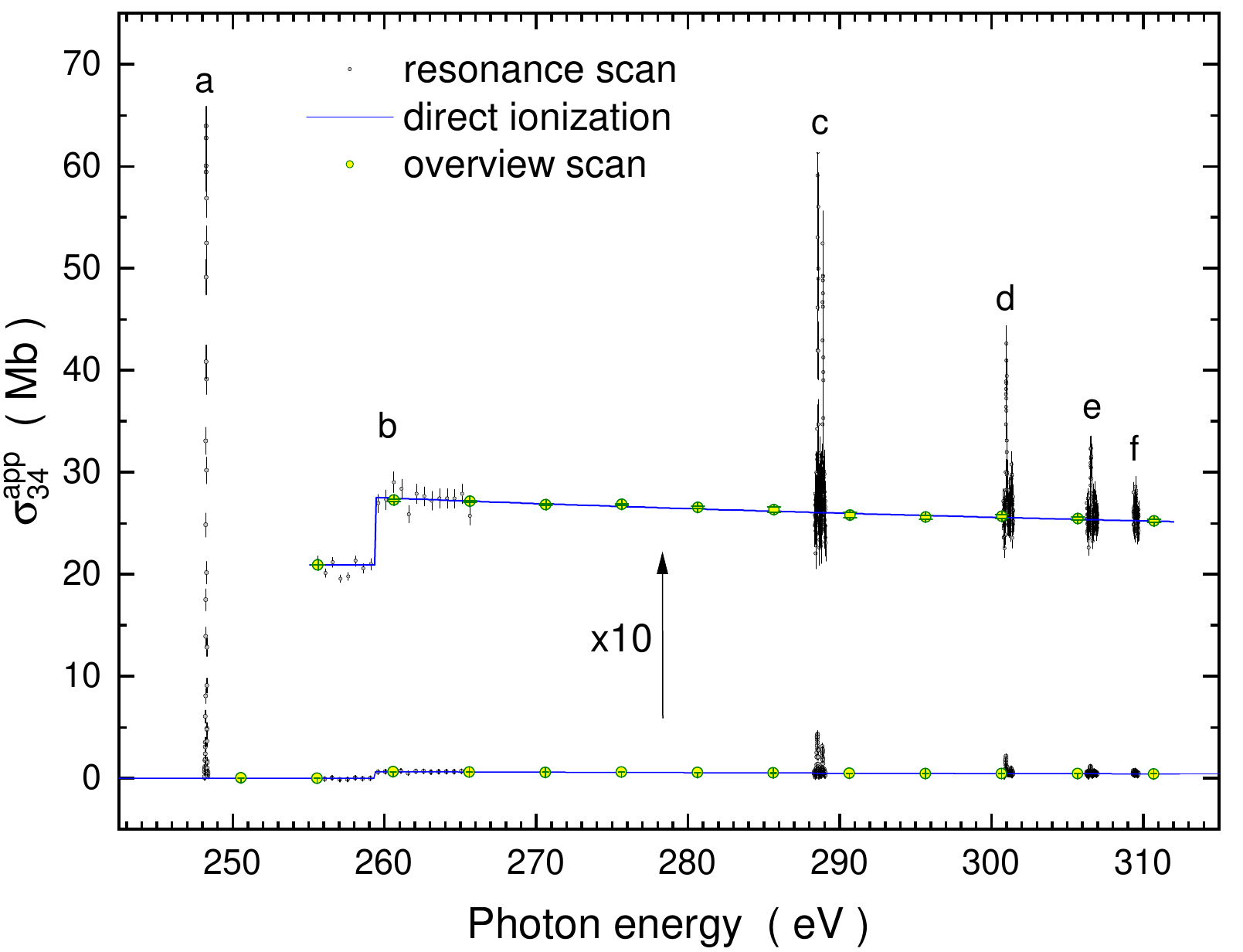}
\caption{\label{Fig:1s2sresonancesoverview} (color online) Apparent experimental B$^{3+}$ photoionization cross section normalized to the theoretical cross section (solid blue line) described in Eq.~\ref{Eq:sig34direct}. The parent-ion beam consisted of a fraction of 90.9\% of ground-state and 9.1\% of metastable ions. The part of the spectrum shown with an offset was multiplied by a factor 10 for better visibility of the details. The small letters stand for characteristic cross-section features. The step b is the $K$ edge of ground-state B$^{3+}$. The other features are highlighted in the subsequent two figures. The resonances are associated with $1s2s~^3S \to 2\ell n\ell'~^3P$ transitions with $\ell = s, p$, $n = 2,...,6$, and $\ell' = p, s$, respectively.
}
\end{center}
\end{figure}

\begin{figure*}[t]
\begin{center}
\includegraphics[width=17.6 cm]{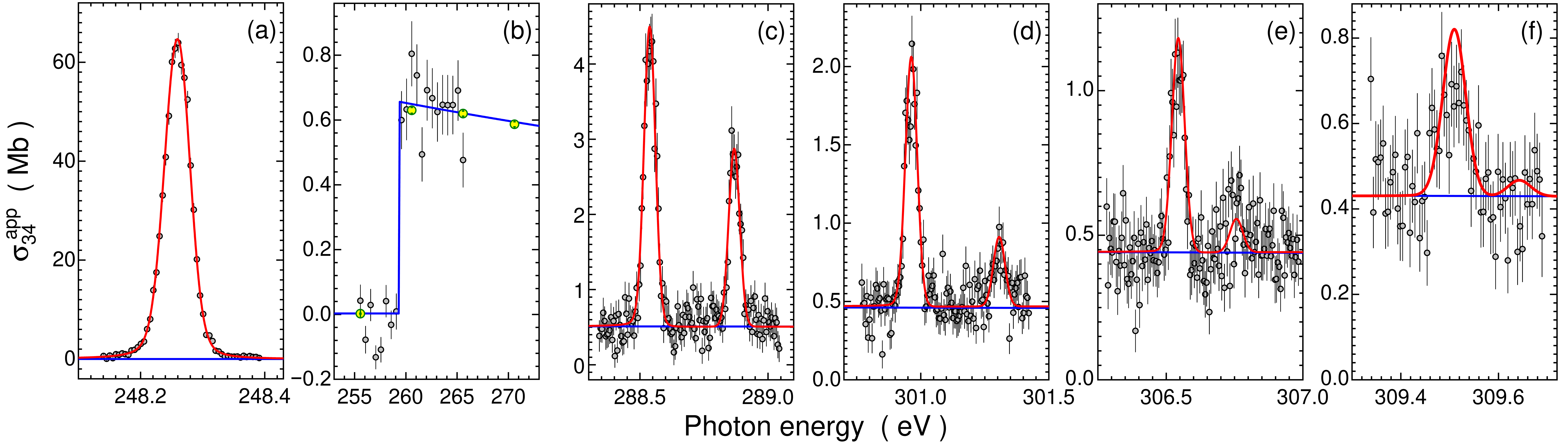}
\caption{\label{Fig:1s2sresonances} (color online) Characteristic features in the apparent experimental photoionization cross sections for a mixed beam of B$^{3+}$ ions. The labels of the individual panels correspond to the letters introduced in Fig.~\ref{Fig:1s2sresonancesoverview}. The resonances are associated with $K$-shell excitation of metastable $1s2s~^3S$ ions, the threshold in panel (b) is the $K$ edge of ground-state $1s^2~^1S$ ions. The solid blue line is the apparent direct ionization described by Eq.~\ref{Eq:sig34direct}. The red lines show the results of the MBPT calculations for the resonance contributions (see Table~\ref{Tab:msparameters}). The monochromator exit-slit width was 200~$\mu$m which resulted in a resolving power $E/\Delta E \approx 6000$.
}
\end{center}
\end{figure*}

Figure~\ref{Fig:1s2sresonancesoverview} shows the experimental results of a set of B$^{3+}$ resonance scans at moderate energy resolution obtained with a 200~$\mu$m exit-slit width of the monochromator. Photon-energy bandwidths were between approximately 37 and 55~meV, increasing with the resonance energy. Step sizes of 5~meV were chosen to scan the $1s2s~^3S \to 2\ell n\ell'~^3P$ resonances while the region of the $1s^2~^1S \to 1s~^2S + e^-$ ionization edge was covered with 0.5~eV steps. Along with the results of the fine resonance scans, the data of the overview scan of Fig.~\ref{Fig:normalized_sig_34_apparent} that fall into the present energy range are also displayed. Their statistical uncertainties are relatively small because the signal rates at 1000~$\mu$m exit slit were considerably higher than those at 200~$\mu$m. Figure~\ref{Fig:1s2sresonancesoverview} also shows the apparent direct-ionization cross section described in Eq.~\ref{Eq:sig34direct}. It is represented by the solid blue line. As mentioned before, the photoion yields observed in the experiments were normalized to that cross section and thereby put on an absolute scale.

The letters a-f in Fig.~\ref{Fig:1s2sresonancesoverview} are associated with certain characteristic features in the photoionization of the mix of B$^{3+}$ ions in the parent-ion beam. These features are shown separately in the six panels (a) - (f) of Fig.~\ref{Fig:1s2sresonances}. While (b) stands for the $K$-shell ionization threshold of the ground-state component of the parent B$^{3+}$ ion beam, the other letters represent resonances in the photoionization of metastable B$^{3+}(1s2s~^3S)$. They are associated with the 27 most strongly populated $^3P_{0,1,2}$ levels listed in Table~\ref{Tab:msparameters}. With a resolving power around 6000, the fine structure of the $^3P$ terms cannot be resolved. Thus, letter (a) represents the B$^{3+}(2s2p~^3P)$ term. Letter (c) comprises the $^3P$ terms of configurations $2s3p$ and $2p3s$. Letter (d) stands for $2s4p$ and $2p4s$, (e) represents $2s5p$ and $2p5s$, and (f) comprises $2s6p$ and $2p6s$. The population of $^1P$ levels is not strictly forbidden, however, intercombination transitions are strongly suppressed. The strongest such transition would have to be expected for the intermediate $2s2p~^1P$ level. The present MBPT calculations predict this level to have a resonance strength of $4.39 \times 10^{-5}$~Mb\,eV in the photoionization channel - approximately $10^{-6}$ of the $2s2p~^3P$ strength and, therefore,  too small to be seen in the present experiment at the expected resonance energy of 254.597~eV. The $2s2p~^1P$ resonance can, however, be populated from the ground state of B$^{3+}$ without spin flip at a resonance energy of 453.165~eV (see Table~\ref{Tab:gs-parameters}) and was clearly observed in the present experiments at higher photon energies (see below).

Table~\ref{Tab:msparameters} additionally comprises the resonance parameters for 12 $^3P$ states based on configurations $2pnd$ with $n=3,4,5,6$. Although the transitions from the metastable $^3S$ parent ion to these 12 levels are allowed in principle, their strengths are one to two orders of magnitude smaller than the strengths of all other transitions mentioned in Table~\ref{Tab:msparameters}. Accordingly, $2pnd$ resonances could not be observed experimentally. It should be mentioned that the configurations and spectroscopic level assigments used in the present paper only comprise the leading contributions in a multi-configuration representation.

The whole spectrum in Fig.~\ref{Fig:1s2sresonancesoverview} is dominated by the very strong $2s2p~^3P$ resonance term labeled a. The apparent cross section obtained with a mixed beam of ground-state and metastable B$^{3+}$ at an energy resolution of 37~meV reaches a maximum of 65~Mb. Considering the fact, that only 9.1\% of the parent-ion beam are responsible for the observed resonance a, the true cross section must have a maximum of more than 700~Mb. As Fig.~\ref{Fig:msMBPT} shows, the natural cross section (observable only at infinite resolving power) even reaches a maximum near 2~Gb.

Along with the experimental cross-section data, Fig.~\ref{Fig:1s2sresonances} shows the apparent direct-ionization cross section as a solid blue line that is already known from the preceding figures.  In addition, the results of the MBPT calculations are displayed as solid red lines in the different panels (a), (c), (d), (e), and (f). The theoretical cross sections were convoluted~\cite{Schippers2018a} with Gaussian functions of increasing full width at half maximum (FWHM) to model the experimental energy spreads. Since the experimental results were obtained with a metastable ion-beam fraction of only 9.1\% the theoretical cross sections were scaled down by a factor 0.091. The resulting red curves show a very good agreement with the experimental data.

The statistical quality of the experimental cross sections deteriorates with increasing excitation energy. The reason is in the decreasing cross sections and, hence, decreasing signal rates observed at increasing resonance energies. At the same time, the direct-ionization continuum becomes relatively more prominent as Fig.~\ref{Fig:1s2sresonances} clearly demonstrates and, therefore, isolating the resonant contribution to the apparent cross sections from that continuum becomes increasingly more difficult. We mention again that the experimental energy scale was calibrated to the theoretical resonance energies which are considered to be accurate within less than $\pm 1$~meV.

\begin{figure}[t]
\begin{center}
\includegraphics[width=7cm]{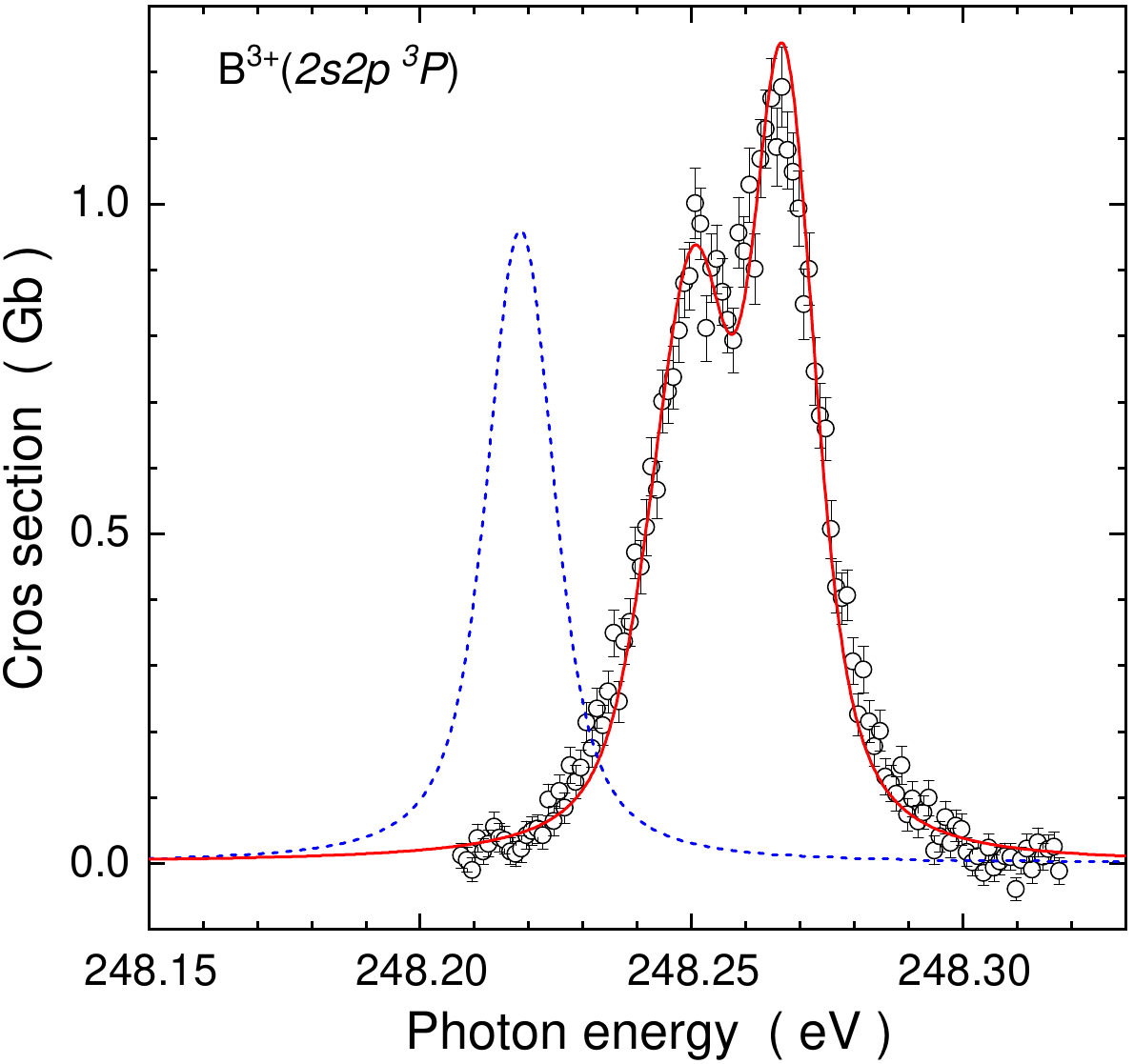}
\caption{\label{Fig:2s2phires} (color online) High-resolution measurement of the $1s2s~^3S_1 \to 2s2p~^3P_{0,1,2}$ resonance contributions to the B$^{3+}$ photoionization cross section normalized to represent the result for a pure ($f = 1$) metastable-ion ensemble. The data points were measured at an energy resolution of 8.5~meV. The solid red line is the cross section resulting from the MBPT calculations (see parameters in Table~\ref{Tab:msparameters}). It was convoluted with a 8.5~meV FWHM Gaussian function to simulate the experimental energy resolution ($E/\Delta E \approx 29000$). The dashed blue line represents the (non-relativistic) CCC calculation convoluted and normalized accordingly, but multiplied by 0.5 so that the cross section size fits the diagram.
}
\end{center}
\end{figure}

The size of the $2s2p~^3P$ resonance allowed us to increase the resolving power of the experiment and still get measureable signal rates. By reducing the monochromator exit-slit width to 8~$\mu$m and by optimizing the fixed-focus constant $c_{ff}$~\cite{Viefhaus2013,Reininger2005} of the variable-linespacing (VLS) grating in use at beamline P04, it was possible to carry out a measurement at a resolving power $E/{\Delta E} \approx 29000$. At the resulting photon-energy bandwidth of approximately 8.5~meV, the fine structure of the $^3P$ term could be partially resolved. The result of this measurement is shown in Fig.~\ref{Fig:2s2phires}. In this figure, the apparent experimental cross section was divided by a factor 0.091 in order to show the result for a parent-ion ensemble with a 100\% fraction of metastable ions. Now the cross-section maximum reaches 1.2~Gb.

The solid red line in Fig.~\ref{Fig:2s2phires} is the theoretical MBPT cross section convoluted with a 8.5~meV FWHM Gaussian function modeling the experimental resolution. The shape of the experimental cross-section function agrees very well with the theoretical model. The figure also shows the cross section obtained from the present CCC approach multiplied by a factor 0.5 and also convoluted with a 8.5~meV FWHM Gaussian function. It is represented by the dashed blue line. The CCC calculations are non-relativistic, i.e., the fine structure of the $^3P$ term is not described. Therefore, the whole calculated strength is agglomerated in one single resonance while in reality, it is distributed over three fine-structure levels. Therefore, the CCC cross-section maximum is about a factor of two above the maximum of the  MBPT result. In order to fit the CCC cross section on the same scale for display in Fig.~\ref{Fig:2s2phires}, it was multiplied by the factor of 0.5.

The CCC calculations do not provide direct access to the resonance parameters. For each photon energy a separate calculation is necessary to obtain the cross section at that energy. In the present study an energy grid with a 10~meV spacing was chosen for the region close to the resonance energy. The wings of the $2s2p~^3P$ resonance were covered with an energy grid of 100~meV. The spacing of the resulting data points is sufficient for a meaningful fit of a Fano profile to these points. The following resonance parameters were obtained from the Fano fit to the CCC calculations: $E_{\mathrm{res}} = 248.219$~eV, $\Gamma = 9.82$~meV, $q = -316$, and $S_{\mathrm{ion}} = 39.645$~Mb\,eV. The resonance strength is practically identical with that obtained by MBPT for the $^3P$ triplet (only 0.02~Mb\,eV less). The resonance energy is very slightly below  ($\approx 45$~meV) the center of gravity of the MBPT calculation and the natural width $\Gamma$  is a little greater (by less than 10\%) than that obtained by the MBPT calculations for each of the $^3P$ fine-structure components. The Fano parameter $q$ is  approximately 17\% lower than that resulting from the MBPT calculations. All in all, there is a remarkable agreement of the CCC and MBPT results for the $2s2p~^3P$ resonance.  However, at the level of a high experimental resolution such as the 8.5~meV accomplished in the present experiment, the relativistic MBPT calculations are clearly superior to the non-relativistic CCC calculations.

For the $^3P$ resonances with higher excitation energies the step size of the CCC calculations is too large to facilitate meaningful fits of Fano resonance profiles so that a detailed comparison with the experiments or the MBPT calculations is not possible.

\subsection{\label{1s2contributions} B$^{3+}(1s^2~^1S)$}

The resonant ionization of ground-state B$^{3+}(1s^2~^1S)$ ions requires double $K$-shell excitation. Cross sections for such transitions by absorption of a single photon are relatively small as evidenced by the results of theoretical calculations shown in panels (a) and (b) of Fig.~\ref{Fig:sig34gs}. The maxima of the double-excitation resonances are below 1~Mb which has to be compared with the $\approx 2$~Gb $2s2p~^3P$ resonance that can be excited from the metastable $1s2s~^3S$ level. Accordingly, the resonances arising from the ground level  are more difficult to detect in experiments. In the present measurements it was helpful that the primary ion beam consisted of about 10 times more ground-level compared to metastable B$^{3+}$ ions. Moreover, since the natural widths of the doubly excited autoionizing $^1P$ resonances are about a factor of 10 larger than the widths of the $^3P$ resonances, it is possible to reduce the experimental resolving power without losing too much information. With a monochromator exit-slit width of 400~$\mu$m and the use of a blazed grating with a lower resolution, the available photon flux could be greatly enhanced over the flux available for the measurement of the $^3P$ resonances excited from the metastable $^3S$ level of B$^{3+}$. For the $2s2p~^1P$ resonance a flux of $5.7 \times 10^{13}$~s$^{-1}$ (at about 453~eV) could be used under the chosen experimental conditions compared to a flux of $4 \times 10^{11}$~s$^{-1}$ (at about 248~eV) for the $2s2p~^3P$ resonance with an exit-slit width of 200~$\mu$m. Thus, it was possible to measure several double-excitation resonances in the photoionization of ground-state B$^{3+}$ ions.

\begin{figure}[bt]
\begin{center}
\includegraphics[width=\columnwidth]{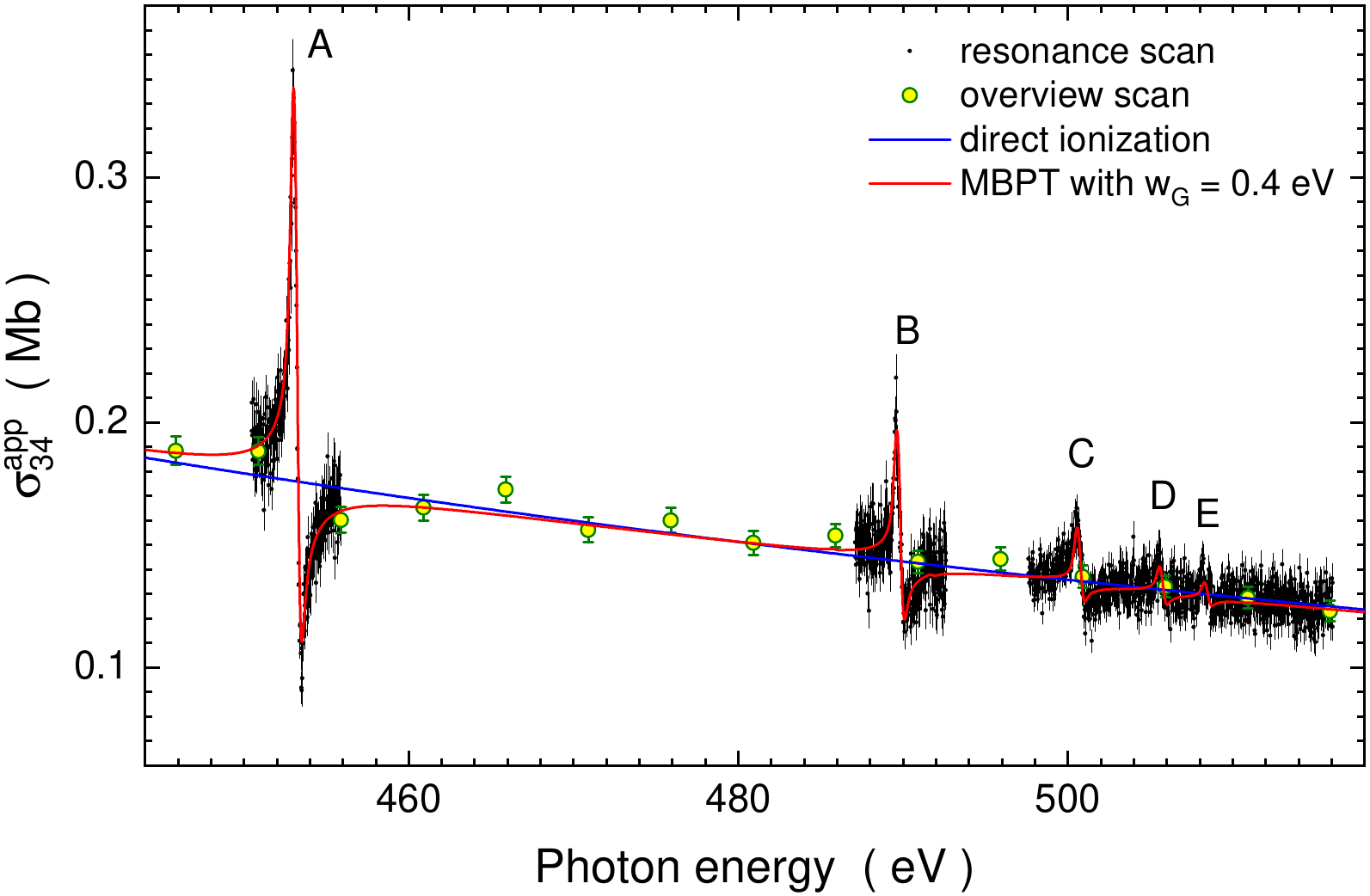}
\caption{\label{Fig:1s2resonancesoverview} (color online) Apparent experimental B$^{3+}$ photoionization cross section normalized to the theoretical cross section (solid blue line) described by Eq.~\ref{Eq:sig34direct}. The capital letters stand for resonance features that are highlighted in the subsequent figures. The resonances are mostly associated with $1s^2~^1S \to 2\ell n\ell'~^1P$ levels with $n$ = 2, 3, 4, 5, and 6. The solid red line represents the MBPT results for the resonances in the energy range of the calculations. The theoretical resonance spectrum (see parameters in Table~\ref{Tab:gs-parameters}) was convoluted with a 0.4~eV FWHM Gaussian function and multiplied with the fraction 0.909 of ground-state B$^{3+}$ ions in the parent-ion beam. The MBPT resonance spectrum was added to the direct-ionization contribution, i.e., the  solid blue line.
}
\end{center}
\end{figure}

An overview of the apparent cross section $\sigma_{34}^{\mathrm{app}}$ in the photon-energy range from 444 to 518~eV is provided by Fig.~\ref{Fig:1s2resonancesoverview}. This energy range contains the dominant $2\ell n\ell'$ resonances populated from the $1s^2$ ground level of B$^{3+}$. The figure shows the results of fine resonance scans measured with a step width of 20~meV. Resonance features are labeled with capital letters A -- E. Beside the fine-scan data the figure displays the results of the overview scan shown in Fig.~\ref{Fig:normalized_sig_34_apparent} that fall into the present energy range. The smooth direct-ionization continuum contribution to the apparent cross section was normalized to the theoretical model defined by Eq.~\ref{Eq:sig34direct} which is shown by the smooth solid blue line in Fig.~\ref{Fig:1s2resonancesoverview}. The normalization factor was then also applied to the resonance contributions.

The resonance contributions to $\sigma_{34}^{\mathrm{app}}$ obtained from the present MBPT calculations (the theoretical resonance parameters are listed in Table~\ref{Tab:gs-parameters}) were convoluted with a $w_{\mathrm{G}}$ = 0.4~eV FWHM Gaussian function to simulate the experimental resolution, multiplied by a factor 0.909 to account for the fraction of ground-level B$^{3+}$ ions in the parent-ion beam,  and then added to the direct-ionization continuum described by Eq.~\ref{Eq:sig34direct}. The resulting cross-section function is represented by the solid red curve in Fig.~\ref{Fig:1s2resonancesoverview}. It agrees very well with the experimental data. This is evidenced particularly by the following discussion and detailed comparisons.

\begin{figure}[tb]
\begin{center}
\includegraphics[width=\columnwidth]{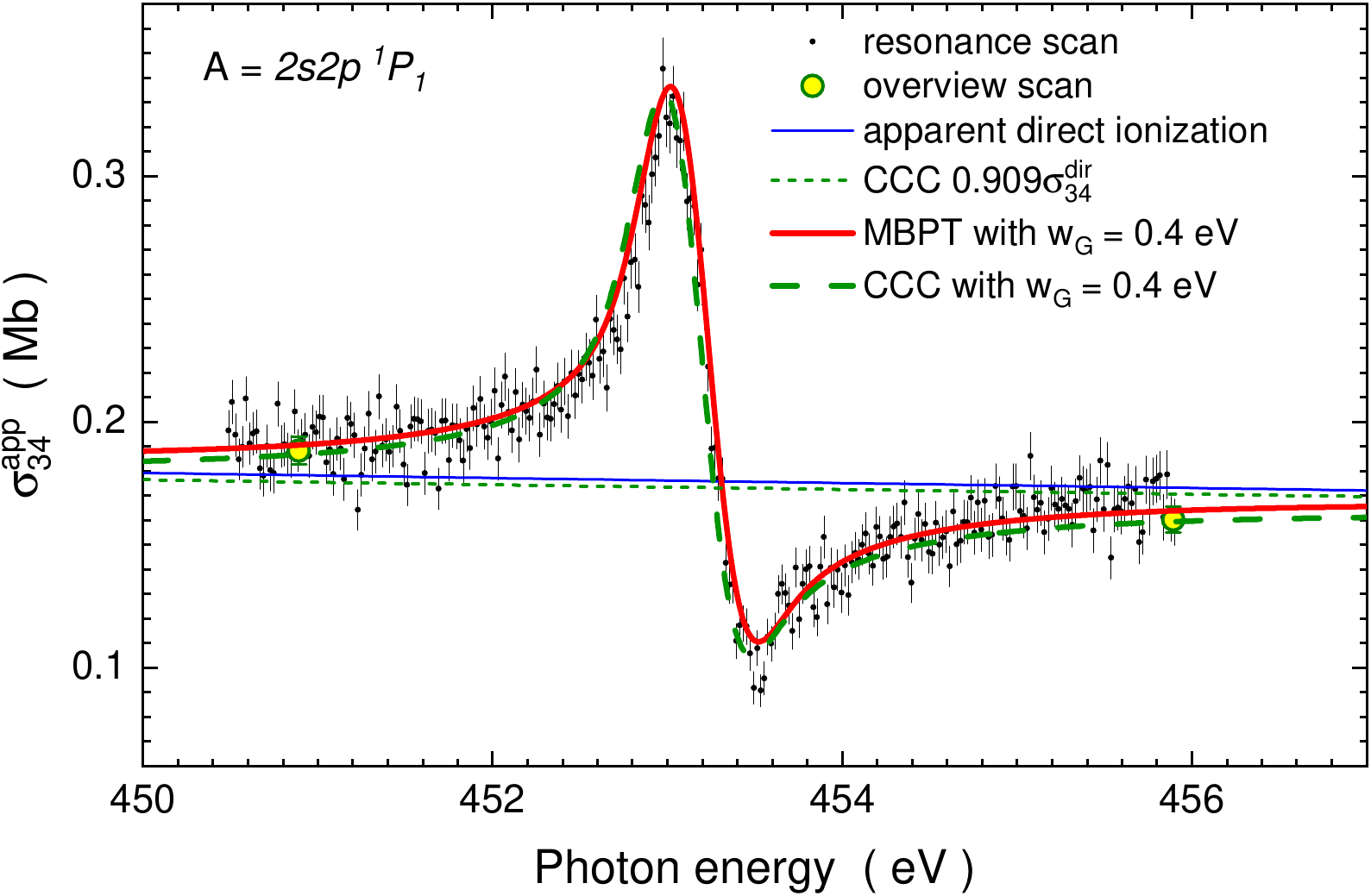}
\caption{\label{Fig:1s2to2s2presonance} (color online) The $1s^2~^1S \to 2s 2p~^1P_1$ resonance (labeled A in Fig.~\ref{Fig:1s2resonancesoverview}). The blue and red solid lines are the same as in Fig.~\ref{Fig:1s2resonancesoverview}. The green, almost horizontal short-dashed line is the direct-ionization contribution of the CCC calculation multiplied with a factor 0.909 to account for the fraction of ground-state ions in the parent B$^{3+}$ ion beam. The thick dashed green line is the result of a modified CCC calculation for the $2s 2p~^1P_1$ resonance added to the direct-ionization contribution.
}
\end{center}
\end{figure}
The resonance features observed in the overview scan are discussed below in more detail. They can be clearly identified on the basis of the MBPT calculations (see Table~\ref{Tab:gs-parameters}). The strongest resonance populated by $K$-shell double excitation of ground-level B$^{3+}$  is associated with the $2s2p~^1P$ level. It is labeled A in Fig.~\ref{Fig:1s2resonancesoverview}. A detailed comparison of the experimental and theoretical results for this resonance is provided in Fig.~\ref{Fig:1s2to2s2presonance}. The experimental data points together with their statistical uncertainties, the direct-ionization cross section represented by the smooth solid blue curve, and the MBPT result given by the solid red curve are the same as in Fig.~\ref{Fig:1s2resonancesoverview}, just zoomed in on the energy range from 450 to 457~eV. In addition, Fig.~\ref{Fig:1s2to2s2presonance} displays the results of the CCC calculations obtained from the calculated points after an additional treatment which is explained below. The thick long-dashed green curve shows the CCC result convoluted with a $w_{\mathrm{G}}$ = 0.4~eV FWHM Gaussian function and multiplied by a factor 0.909 (like the MBPT result). The short-dashed green line represents the smooth direct-ionization cross section obtained from the CCC calculation. At the present energy, this partial apparent cross section is by far dominated by single $K$-shell ionization of ground-level B$^{3+}$. Therefore, the CCC result was also multiplied by a factor 0.909 to account for the fraction of ground-level ions in the parent B$^{3+}$ ion beam. The short-dashed line agrees well with the present model for the direct-ionization contribution which is given by the smooth solid blue line. This supports the previous findings of very good agreement of theoretical direct-ionization cross sections (see Fig.~\ref{Fig:gsdirect}). Both the CCC and MBPT cross sections $\sigma_{34}^{\mathrm{app}}$ are in excellent agreement with the experimental data as well as with one another.

\begin{figure}[tb]
\begin{center}
\includegraphics[width=\columnwidth]{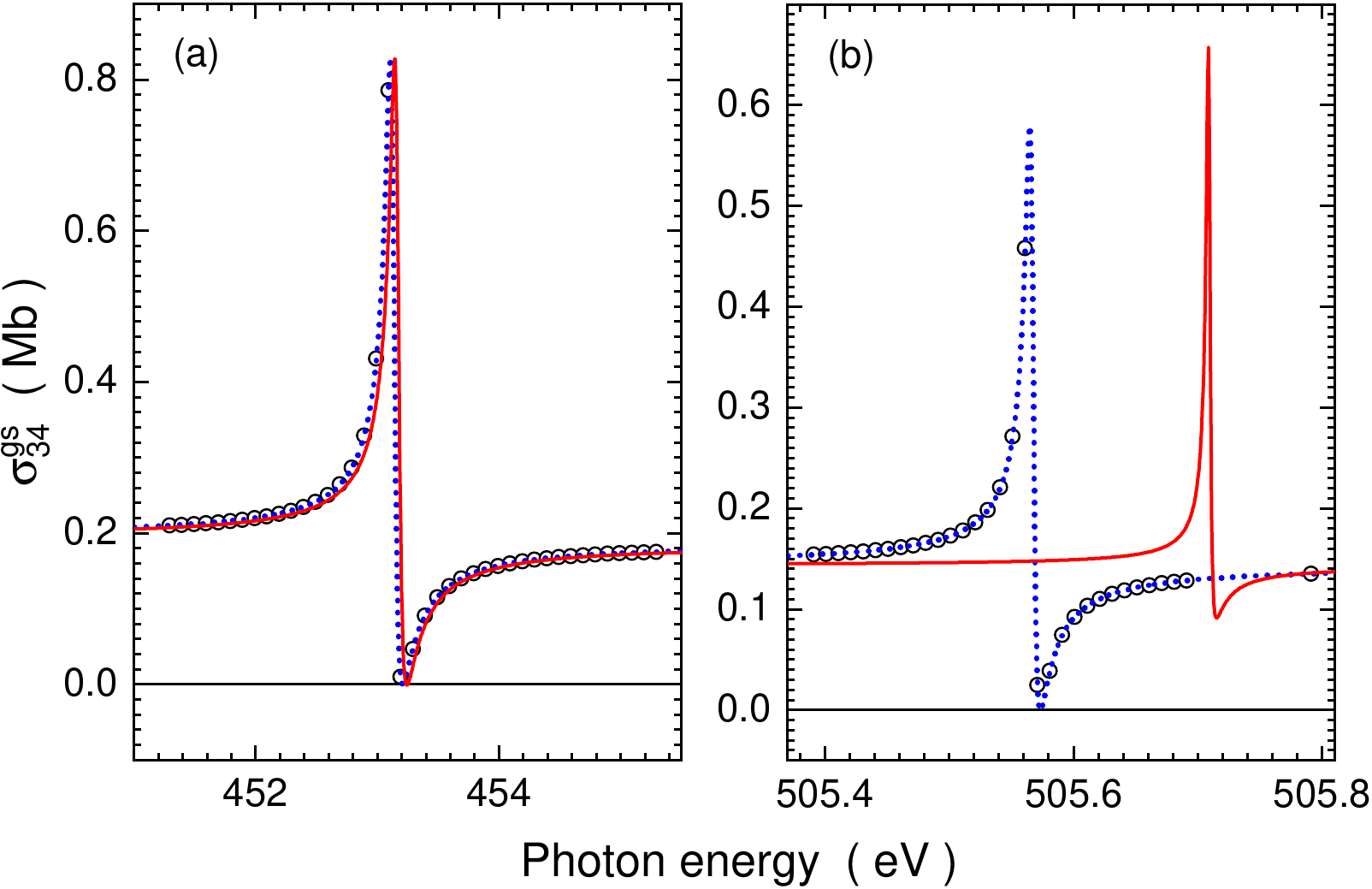}
\caption{\label{Fig:gsresCCCvsMBPT} (color online) Cross sections $\sigma_{34}^{\mathrm{gs}}$ for single photoionization of B$^{3+}(1s^2~^1S)$ obtained by the present CCC and MBPT calculations. Panel (a) shows $\sigma_{34}^{\mathrm{gs}}$ in the vicinity of the $2s2p~^1P_1$ level and panel (b) the same in the region of the $2p5s/5d~^1P/^3D$ resonance. The open circles are the results of CCC calculations at discrete photon energies. The  dashed blue lines are Fano profiles fitted to the CCC data points. The  solid red lines are the present MBPT results.
}
\end{center}
\end{figure}
The calculated CCC data points alone would not directly provide the level of agreement that is demonstrated in Fig.~\ref{Fig:1s2to2s2presonance}. As in the case of the $1s2s~^3S \to 2s2p~^3P$ transition, a Fano profile was first fitted to the calculated CCC data points. The individual data points and the fit function are shown in panel (a) of Fig.~\ref{Fig:gsresCCCvsMBPT}.   Clearly, the calculated points with their somewhat arbitrarily chosen energies do not map out the full Fano profile. However, the density and precision  of the calculated cross-section points facilitates the determination of the associated resonance parameters: the Fano fit curve delivers the resonance energy $E_{\mathrm{res}} = 453.129$~eV (compared to 453.1648~eV from MBPT), the natural width $\Gamma = 0.087$~eV (compared to 0.08441~eV from MBPT), the Fano parameter $q = -1.818$ (compared to {-1.833} from MBPT), and the ionization resonance strength $S_\mathrm{ion} = 0.0602$~Mb\,eV (compared to 0.05958~eV from MBPT). The agreement of the calculated resonance energies obtained from CCC and MBPT is remarkably good but is fortuitous to some extent as the further analysis shows. The other parameters are in fair agreement with one another but show deviations that cannot be attributed to the uncertainties in the fitting procedure.

Panel (b) of Fig.~\ref{Fig:gsresCCCvsMBPT} shows a second example of a direct comparison of CCC and MBPT calculations. In the particular case, the energy region around 505.5~eV is scrutinized where the $2p5s/5d~^1P/^3D$ resonance gives the dominant contribution to $\sigma_{34}^{\mathrm{gs}}$. As in panel (a), a Fano profile was fitted to the data points obtained by the CCC calculations at discrete photon energies. The resonance energy is about 0.15~eV different from that obtained by the MBPT calculations. The Fano $q$ parameter derived from the fit is about 55\% of the MBPT result and the lifetime width  is a factor of 2 above that of the MBPT calculation. Only the resonance strengths S$_{\mathrm{ion}}$ from the two calculations agree within 25\%.  Given such discrepancies, no further comparisons of the experimental data and the resonance features seen in the CCC calculations at photon energies higher than 455~eV are presented.

\begin{figure}[t]
\begin{center}
\includegraphics[width=\columnwidth]{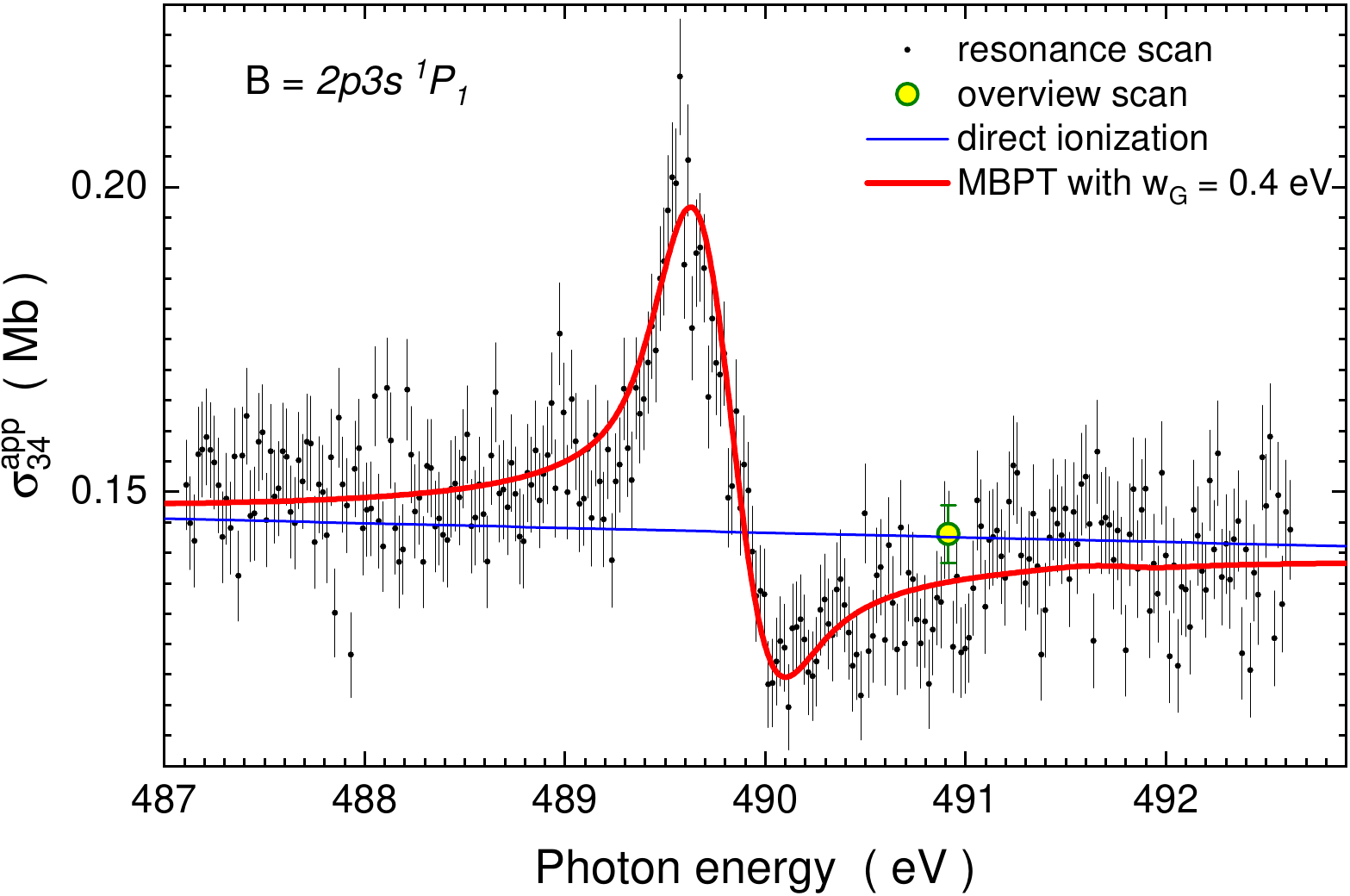}
\caption{\label{Fig:1s2to2p3sresonance} (color online) Detail of Fig.~\ref{Fig:1s2resonancesoverview} in the vicinity of peak feature B which corresponds to the $2p3s~^1P_1$ resonance.
}
\end{center}
\end{figure}

\begin{figure}[t]
\begin{center}
\includegraphics[width=\columnwidth]{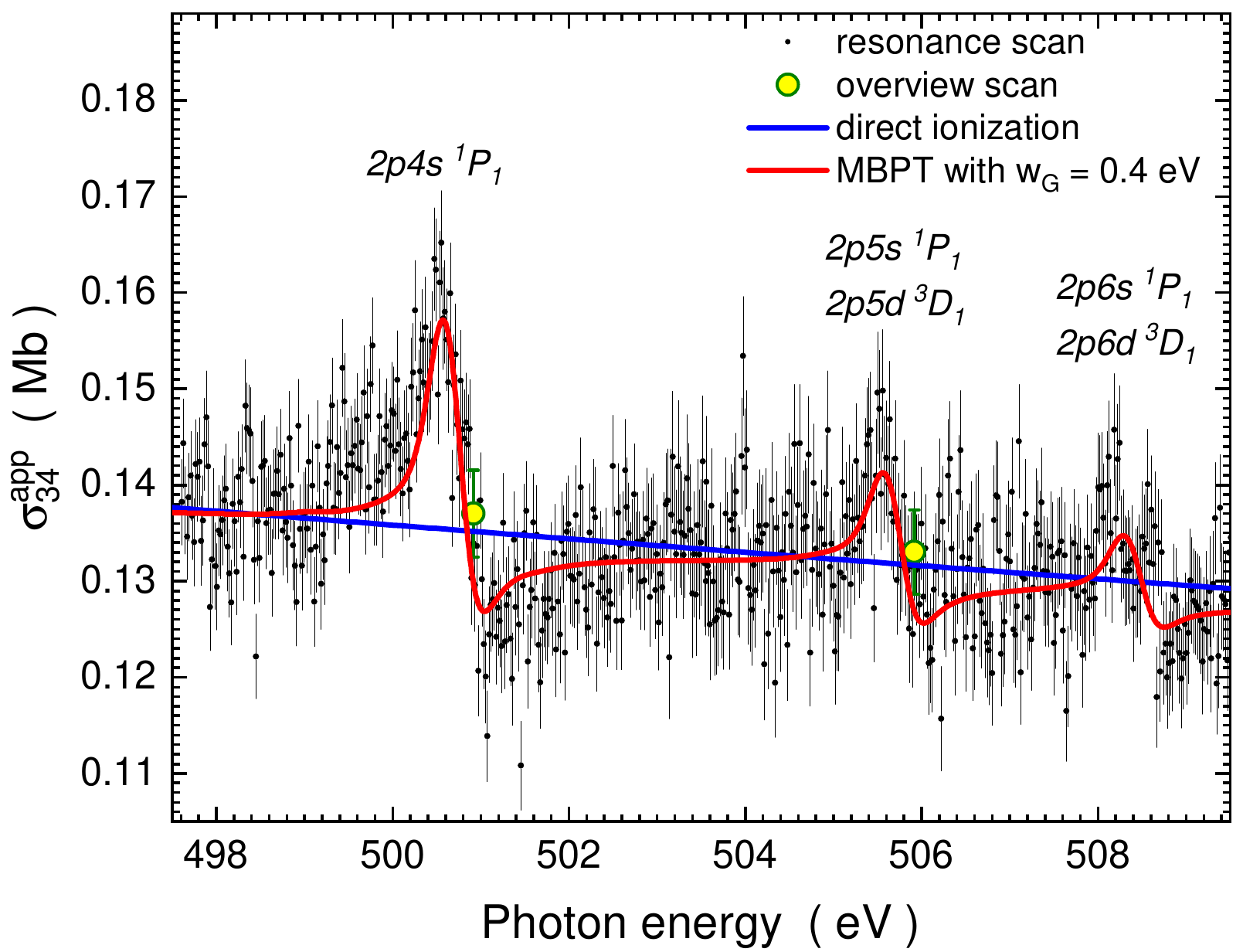}
\caption{\label{Fig:1s2to456resonances} (color online) Detail of Fig.~\ref{Fig:1s2resonancesoverview} in the vicinity of peak features C, D, and E: the leading levels are indicated in the figure. The red line comprises all resonances calculated in the displayed energy range obtained by applying MBPT.
}
\end{center}
\end{figure}

Figure~\ref{Fig:1s2to2p3sresonance} zooms in on the B$^{3+}(2p3s~^1P)$ resonance observed in the experiment. The data shown are the same as in Fig.~\ref{Fig:1s2resonancesoverview}. The statistical uncertainties of the measured apparent cross sections indicate the increasing difficulty to isolate the resonant contributions from the direct-ionization continuum as the excitation energy increases. This is all the more true for further increasing principal quantum numbers $n = 4, 5, 6$ as evidenced by Fig.~\ref{Fig:1s2to456resonances}. No signal could be found for $n \geq 7$. In all cases, the experimental results confirm the theoretical results of the present MBPT calculations. Theoretical energies, total widths, shapes, and strengths of the doubly excited resonances fit the present experimental observations.

It is interesting to note that 28 resonances were predicted to occur in the present energy range from 444 to 518~eV but only 5 of them are sufficiently strong to be observed in the experiments. The leading levels responsible for the labeled resonances in Fig.~\ref{Fig:1s2resonancesoverview} are A = $2s2p~^1P$ at 453.1648~eV (see Table~\ref{Tab:gs-parameters}), B = $2p3s~^1P$ at 489.7694~eV, C = $2p4s~^1P$ at 500.687~eV, D = $2p5s~^1P$ at 505.709~eV, and E = $2p6s~^1P$ at 508.426~eV. All the other resonances listed in Table~\ref{Tab:gs-parameters} are blended with stronger resonance contributions or are too small to significantly contribute to the observed spectrum of double-excitation resonances which were all measured with a resolution of approximately 0.4~eV. Accordingly, the MBPT results shown in Figs.~\ref{Fig:1s2resonancesoverview}, \ref{Fig:1s2to2s2presonance}, \ref{Fig:1s2to2p3sresonance}, and \ref{Fig:1s2to456resonances} were all convoluted with a 0.4~eV FWHM Gaussian function to simulate the experiment. They were also multiplied by a factor 0.909 to account for the 90.9\% fraction of $1s^2$ ground-level B$^{3+}$ ions that were present in the parent-ion beam.

\section{\label{Sec:summary} Summary and conclusions}

The present experimental and theoretical study investigates the role of double-$K$-hole resonances in the net single photoionization of heliumlike B$^{3+}$ ions  initially in their $1s^2~^1S_0$ ground level or initially in their $1s2s~^3S_1$ metastable excited level. The selection rules for electromagnetic transitions mediated by a single photon favor the production of $2\ell n\ell'~^1P_1$ doubly excited levels from ground-state ions and $2\ell n\ell'~^3P_{0,1,2}$ doubly excited levels from the metastable $^3S$ excited parent ions. In the experiments, resonances in both, the singlet and triplet manifolds were observed with $n = 2, 3, 4, 5, {\rm and} \, 6$. The experiments were accompanied by calculations employing the convergent close coupling (CCC) method and many-body perturbation theory (MBPT). Cross sections for direct single ionization of $1s$ and $2s$ electrons were calculated by employing relativistic Hartree-Fock (HFR) theory and the random-phase approximation with exchange (RPAE).

Direct single $K$-shell photoionization has been a topic of theoretical descriptions already since the early development of quantum mechanics. Comparison of altogether eight different approaches to direct single $K$-shell ionization of heliumlike B$^{3+}$ shows that the results are very consistent with one another. In fact, they are identical within maximum deviations of $\pm$10\%. This provides an accuracy for normalizing experimental yields of ionized ions that can hardly be matched by the conventional absolute measurement of cross sections in merged-beams experiments with photon and ion beams. By exploiting this observation, the present experimental B$^{4+}$ product-ion yields were normalized to the smooth direct-ionization continuum on which the resonant contributions reside. For a meaningful normalization, the fractions $1-f$ of ground-state and $f$ of  metastable B$^{3+}$ ions in the parent-ion beam had to be quantified. This was accomplished by comparing measured yields of B$^{4+}$ product ions with the results of the present MBPT calculations. A fraction $f = 0.091\pm 0.016 $ was found for the $1s2s~^3S$ component.

The MBPT results obtained in this study are of particularly high quality. Resonance energies are estimated to be correct within uncertainties of less than $\pm 1$~meV for nearly all of the investigated levels. For the $2\ell n\ell'~^1P$ levels with $n \geq 4$ the uncertainties are slightly higher than that but still estimated to be lower than 2~meV. The high accuracy of these resonance energies makes the resonances in the single photoionization of B$^{3+}$ ideal reference standards for calibrating the energies of photon beam lines as well as photon and (photo)electron spectrometers. The advantage of using He-like B$^{3+}$ for calibration can be illustrated by comparing with one of the most commonly used energy reference standards, the neutral N$_2$ molecule, where resonance features around 400~eV are known with a quoted uncertainty of $\pm 20$~meV~\cite{Sodhi1984a} which is roughly a factor of 20  larger than what has been accomplished in the present work. Common reference standards for the next higher energies around 530~eV employ $K$-shell excitations in oxygen-containing molecules such as neutral CO and have uncertainties even as large as 60 to 90 meV~\cite{Sodhi1984a}. Even the recent calibration work on Ne, CO$_2$, and SF$_6$ employing the theoretical sub-meV uncertainties of level energies in heliumlike ions has resulted in uncertainties of 100~meV, 40~meV, and 60~meV, respectively, for the calibration-reference gases~\cite{Stierhof2022}.

 The huge cross section of the $1s2s~^3S_1 \to 2s2p~^3P_{0,1,2}$ resonance term with small natural widths (approximately 9~meV) of each fine-structure component provides particularly comfortable signal rates in energy-calibration measurements at about 248~eV.                                                                                                                                                                                                                                                                                                                                                                                                                                                                                                                                                                                                                                                                                                                                                                                                                                                                                                                                                                                                                                                                                                                                                                                                                                                                                                                                                                                                                                                                                                                                                                                                                                                                                                                                                                                                                                                                                                                                                                                                                                                                                                                                                                                                                                                                                                                                                                                                                                                                                                                                                                                The calculated resonance parameters natural width $\Gamma$, Fano parameter $q$, and  ionization strength $S_{\mathrm{ion}}$  are estimated to be correct within $\pm 1$\%. A comparison of Auger widths of the $2s2p~^{1,3}P$ levels obtained by the present MBPT calculations with results~\cite{Zaytsev2019} obtained by the St. Petersburg theory group, that is renowned for highest-quality atomic-structure calculations, shows deviations of less than 0.3\%. This is particularly remarkable since previous calculations of these Auger widths show a scatter of about 30\%.

The present CCC calculations provided not only the cross sections for net single ionization of B$^{3+}(1s^2~^1S)$ and B$^{3+}(1s2s~^3S)$ but also yielded partial cross sections for single ionization accompanied by excitation of the remaining electron to different main shells with principal quantum numbers $n$ up to 6. The sum of ionization-excitation contributions to the net single photoionization cross section $\sigma_{34}^{{\mathrm{gs}}}$ of heliumlike ground-state B$^{3+}$ is only about 1.5\% to 2\%. The situation for metastable parent ions is similar.

The results obtained within the present investigation on photoionization of heliumlike B$^{3+}$ are much more compehensive than previous works on heliumlike Li$^+(1s^2~^1S)$~\cite{Scully2006a} and C$^{4+}(1s2s~^3S)$~\cite{Mueller2018d}.

Future work on photon interactions with heliumlike B$^{3+}$ ions will address the removal of both electrons by direct single-photon double ionization.

\section{\label{Sec:acknowledgement} Acknowledgement}

We acknowledge DESY (Hamburg, Germany), a member of the Helmholtz Association HGF, for the provision of experimental facilities. Parts of this research were carried out at beamline P04 of PETRA III. Beamtime was allocated for proposal I-20220764.  We gratefully acknowledge support from
Bundesministerium für Bildung und Forschung provided
within the “Verbundforschung” funding scheme (Contracts
No. 05K19GU3 and No. 05K19RG3). FT acknowledges funding by the Deutsche Forschungsgemeinschaft (DFG, German Research Foundation) - Project 509471550, Emmy Noether Programme. MM
acknowledges funding of SFB925/A3 by Deutsche Forschungsgemeinschaft
 via Grant No. 2012673. Support from the Swedish Research Council (Grant No. 2020-03315) is gratefully acknowledged.  ASK and IB acknowledge the support of
the Australian Research Council, the National Computer Infrastructure,
and the Pawsey Supercomputer Centre of Western Australia.


\begin{thebibliography}{93}%
\makeatletter
\providecommand \@ifxundefined [1]{%
 \@ifx{#1\undefined}
}%
\providecommand \@ifnum [1]{%
 \ifnum #1\expandafter \@firstoftwo
 \else \expandafter \@secondoftwo
 \fi
}%
\providecommand \@ifx [1]{%
 \ifx #1\expandafter \@firstoftwo
 \else \expandafter \@secondoftwo
 \fi
}%
\providecommand \natexlab [1]{#1}%
\providecommand \enquote  [1]{``#1''}%
\providecommand \bibnamefont  [1]{#1}%
\providecommand \bibfnamefont [1]{#1}%
\providecommand \citenamefont [1]{#1}%
\providecommand \href@noop [0]{\@secondoftwo}%
\providecommand \href [0]{\begingroup \@sanitize@url \@href}%
\providecommand \@href[1]{\@@startlink{#1}\@@href}%
\providecommand \@@href[1]{\endgroup#1\@@endlink}%
\providecommand \@sanitize@url [0]{\catcode `\\12\catcode `\$12\catcode
  `\&12\catcode `\#12\catcode `\^12\catcode `\_12\catcode `\%12\relax}%
\providecommand \@@startlink[1]{}%
\providecommand \@@endlink[0]{}%
\providecommand \url  [0]{\begingroup\@sanitize@url \@url }%
\providecommand \@url [1]{\endgroup\@href {#1}{\urlprefix }}%
\providecommand \urlprefix  [0]{URL }%
\providecommand \Eprint [0]{\href }%
\providecommand \doibase [0]{https://doi.org/}%
\providecommand \selectlanguage [0]{\@gobble}%
\providecommand \bibinfo  [0]{\@secondoftwo}%
\providecommand \bibfield  [0]{\@secondoftwo}%
\providecommand \translation [1]{[#1]}%
\providecommand \BibitemOpen [0]{}%
\providecommand \bibitemStop [0]{}%
\providecommand \bibitemNoStop [0]{.\EOS\space}%
\providecommand \EOS [0]{\spacefactor3000\relax}%
\providecommand \BibitemShut  [1]{\csname bibitem#1\endcsname}%
\let\auto@bib@innerbib\@empty
\bibitem [{\citenamefont {Cederbaum}\ \emph {et~al.}(1986)\citenamefont
  {Cederbaum}, \citenamefont {Tarantelli}, \citenamefont {Sgamellotti},\ and\
  \citenamefont {Schirmer}}]{Cederbaum1986}%
  \BibitemOpen
  \bibfield  {author} {\bibinfo {author} {\bibfnamefont {L.~S.}\ \bibnamefont
  {Cederbaum}}, \bibinfo {author} {\bibfnamefont {F.}~\bibnamefont
  {Tarantelli}}, \bibinfo {author} {\bibfnamefont {A.}~\bibnamefont
  {Sgamellotti}},\ and\ \bibinfo {author} {\bibfnamefont {J.}~\bibnamefont
  {Schirmer}},\ }\bibfield  {title} {\bibinfo {title} {On double vacancies in
  the core},\ }\href {https://doi.org/10.1063/1.451432} {\bibfield  {journal}
  {\bibinfo  {journal} {J. Chem. Phys.}\ }\textbf {\bibinfo {volume} {85}},\
  \bibinfo {pages} {6513 } (\bibinfo {year} {1986})}\BibitemShut {NoStop}%
\bibitem [{\citenamefont {Santra}\ \emph {et~al.}(2009)\citenamefont {Santra},
  \citenamefont {Kryzhevoi},\ and\ \citenamefont {Cederbaum}}]{Santra2009}%
  \BibitemOpen
  \bibfield  {author} {\bibinfo {author} {\bibfnamefont {R.}~\bibnamefont
  {Santra}}, \bibinfo {author} {\bibfnamefont {N.~V.}\ \bibnamefont
  {Kryzhevoi}},\ and\ \bibinfo {author} {\bibfnamefont {L.~S.}\ \bibnamefont
  {Cederbaum}},\ }\bibfield  {title} {\bibinfo {title} {X-ray two-photon
  photoelectron spectroscopy: {A} theoretical study of inner-shell spectra of
  the organic para-aminophenol molecule},\ }\href
  {https://doi.org/10.1103/PhysRevLett.103.013002} {\bibfield  {journal}
  {\bibinfo  {journal} {Phys. Rev. Lett.}\ }\textbf {\bibinfo {volume} {103}},\
  \bibinfo {pages} {013002} (\bibinfo {year} {2009})}\BibitemShut {NoStop}%
\bibitem [{\citenamefont {Berrah}\ \emph {et~al.}(2011)\citenamefont {Berrah},
  \citenamefont {Fang}, \citenamefont {Murphy}, \citenamefont {Osipov},
  \citenamefont {Ueda}, \citenamefont {Kukk}, \citenamefont {Feifel},
  \citenamefont {van~der Meulen}, \citenamefont {Salen}, \citenamefont
  {Schmidt}, \citenamefont {Thomas}, \citenamefont {Larsson}, \citenamefont
  {Richter}, \citenamefont {Prince}, \citenamefont {Bozek}, \citenamefont
  {Bostedt}, \citenamefont {Wada}, \citenamefont {Piancastelli}, \citenamefont
  {Tashiro},\ and\ \citenamefont {Ehara}}]{Berrah2011}%
  \BibitemOpen
  \bibfield  {author} {\bibinfo {author} {\bibfnamefont {N.}~\bibnamefont
  {Berrah}}, \bibinfo {author} {\bibfnamefont {L.}~\bibnamefont {Fang}},
  \bibinfo {author} {\bibfnamefont {B.}~\bibnamefont {Murphy}}, \bibinfo
  {author} {\bibfnamefont {T.}~\bibnamefont {Osipov}}, \bibinfo {author}
  {\bibfnamefont {K.}~\bibnamefont {Ueda}}, \bibinfo {author} {\bibfnamefont
  {E.}~\bibnamefont {Kukk}}, \bibinfo {author} {\bibfnamefont {R.}~\bibnamefont
  {Feifel}}, \bibinfo {author} {\bibfnamefont {P.}~\bibnamefont {van~der
  Meulen}}, \bibinfo {author} {\bibfnamefont {P.}~\bibnamefont {Salen}},
  \bibinfo {author} {\bibfnamefont {H.~T.}\ \bibnamefont {Schmidt}}, \bibinfo
  {author} {\bibfnamefont {R.~D.}\ \bibnamefont {Thomas}}, \bibinfo {author}
  {\bibfnamefont {M.}~\bibnamefont {Larsson}}, \bibinfo {author} {\bibfnamefont
  {R.}~\bibnamefont {Richter}}, \bibinfo {author} {\bibfnamefont {K.~C.}\
  \bibnamefont {Prince}}, \bibinfo {author} {\bibfnamefont {J.~D.}\
  \bibnamefont {Bozek}}, \bibinfo {author} {\bibfnamefont {C.}~\bibnamefont
  {Bostedt}}, \bibinfo {author} {\bibfnamefont {S.}~\bibnamefont {Wada}},
  \bibinfo {author} {\bibfnamefont {M.~N.}\ \bibnamefont {Piancastelli}},
  \bibinfo {author} {\bibfnamefont {M.}~\bibnamefont {Tashiro}},\ and\ \bibinfo
  {author} {\bibfnamefont {M.}~\bibnamefont {Ehara}},\ }\bibfield  {title}
  {\bibinfo {title} {Double-core-hole spectroscopy for chemical analysis with
  an intense x-ray femtosecond laser},\ }\href
  {https://doi.org/10.1073/pnas.1111380108} {\bibfield  {journal} {\bibinfo
  {journal} {Proc. Natl. Acad. Sci. U.S.A}\ }\textbf {\bibinfo {volume}
  {108}},\ \bibinfo {pages} {16912 } (\bibinfo {year} {2011})}\BibitemShut
  {NoStop}%
\bibitem [{\citenamefont {Nakano}\ \emph {et~al.}(2013)\citenamefont {Nakano},
  \citenamefont {Penent}, \citenamefont {Tashiro}, \citenamefont {Grozdanov},
  \citenamefont {$\check{\rm Z}$itnik}, \citenamefont {Carniato}, \citenamefont
  {Selles}, \citenamefont {Andric}, \citenamefont {Lablanquie}, \citenamefont
  {Palaudoux}, \citenamefont {Shigemasa}, \citenamefont {Iwayama},
  \citenamefont {Hikosaka}, \citenamefont {Soejima}, \citenamefont {Suzuki},
  \citenamefont {Kouchi},\ and\ \citenamefont {Ito}}]{Nakano2013}%
  \BibitemOpen
  \bibfield  {author} {\bibinfo {author} {\bibfnamefont {M.}~\bibnamefont
  {Nakano}}, \bibinfo {author} {\bibfnamefont {F.}~\bibnamefont {Penent}},
  \bibinfo {author} {\bibfnamefont {M.}~\bibnamefont {Tashiro}}, \bibinfo
  {author} {\bibfnamefont {T.~P.}\ \bibnamefont {Grozdanov}}, \bibinfo {author}
  {\bibfnamefont {M.}~\bibnamefont {$\check{\rm Z}$itnik}}, \bibinfo {author}
  {\bibfnamefont {S.}~\bibnamefont {Carniato}}, \bibinfo {author}
  {\bibfnamefont {P.}~\bibnamefont {Selles}}, \bibinfo {author} {\bibfnamefont
  {L.}~\bibnamefont {Andric}}, \bibinfo {author} {\bibfnamefont
  {P.}~\bibnamefont {Lablanquie}}, \bibinfo {author} {\bibfnamefont
  {J.}~\bibnamefont {Palaudoux}}, \bibinfo {author} {\bibfnamefont
  {E.}~\bibnamefont {Shigemasa}}, \bibinfo {author} {\bibfnamefont
  {H.}~\bibnamefont {Iwayama}}, \bibinfo {author} {\bibfnamefont
  {Y.}~\bibnamefont {Hikosaka}}, \bibinfo {author} {\bibfnamefont
  {K.}~\bibnamefont {Soejima}}, \bibinfo {author} {\bibfnamefont {I.~H.}\
  \bibnamefont {Suzuki}}, \bibinfo {author} {\bibfnamefont {N.}~\bibnamefont
  {Kouchi}},\ and\ \bibinfo {author} {\bibfnamefont {K.}~\bibnamefont {Ito}},\
  }\bibfield  {title} {\bibinfo {title} {Single photon ${K}^{-2}$ and ${K}^{-1}
  {K}^{-1}$ double core ionization in {C}$_2${H}$_{2n}$ ($n=1 - 3$), {CO}, and
  {N}$_2$ as a potential new tool for chemical analysis},\ }\href
  {https://doi.org/10.1103/PhysRevLett.110.163001} {\bibfield  {journal}
  {\bibinfo  {journal} {Phys. Rev. Lett.}\ }\textbf {\bibinfo {volume} {110}},\
  \bibinfo {pages} {163001} (\bibinfo {year} {2013})}\BibitemShut {NoStop}%
\bibitem [{\citenamefont {Ismail}\ \emph {et~al.}(2023)\citenamefont {Ismail},
  \citenamefont {Fert\'e}, \citenamefont {Penent}, \citenamefont {Guillemin},
  \citenamefont {Peng}, \citenamefont {Marchenko}, \citenamefont {Travnikova},
  \citenamefont {Inhester}, \citenamefont {Ta\"{\i}eb}, \citenamefont {Verma},
  \citenamefont {Velasquez}, \citenamefont {Kukk}, \citenamefont {Trinter},
  \citenamefont {Koulentianos}, \citenamefont {Mazza}, \citenamefont {Baumann},
  \citenamefont {Rivas}, \citenamefont {Ovcharenko}, \citenamefont {Boll},
  \citenamefont {Dold}, \citenamefont {De~Fanis}, \citenamefont {Ilchen},
  \citenamefont {Meyer}, \citenamefont {Goldsztejn}, \citenamefont {Li},
  \citenamefont {Doumy}, \citenamefont {Young}, \citenamefont {Sansone},
  \citenamefont {D\"orner}, \citenamefont {Piancastelli}, \citenamefont
  {Carniato}, \citenamefont {Bozek}, \citenamefont {P\"uttner},\ and\
  \citenamefont {Simon}}]{Ismail2023}%
  \BibitemOpen
  \bibfield  {author} {\bibinfo {author} {\bibfnamefont {I.}~\bibnamefont
  {Ismail}}, \bibinfo {author} {\bibfnamefont {A.}~\bibnamefont {Fert\'e}},
  \bibinfo {author} {\bibfnamefont {F.}~\bibnamefont {Penent}}, \bibinfo
  {author} {\bibfnamefont {R.}~\bibnamefont {Guillemin}}, \bibinfo {author}
  {\bibfnamefont {D.}~\bibnamefont {Peng}}, \bibinfo {author} {\bibfnamefont
  {T.}~\bibnamefont {Marchenko}}, \bibinfo {author} {\bibfnamefont
  {O.}~\bibnamefont {Travnikova}}, \bibinfo {author} {\bibfnamefont
  {L.}~\bibnamefont {Inhester}}, \bibinfo {author} {\bibfnamefont
  {R.}~\bibnamefont {Ta\"{\i}eb}}, \bibinfo {author} {\bibfnamefont
  {A.}~\bibnamefont {Verma}}, \bibinfo {author} {\bibfnamefont
  {N.}~\bibnamefont {Velasquez}}, \bibinfo {author} {\bibfnamefont
  {E.}~\bibnamefont {Kukk}}, \bibinfo {author} {\bibfnamefont {F.}~\bibnamefont
  {Trinter}}, \bibinfo {author} {\bibfnamefont {D.}~\bibnamefont
  {Koulentianos}}, \bibinfo {author} {\bibfnamefont {T.}~\bibnamefont {Mazza}},
  \bibinfo {author} {\bibfnamefont {T.~M.}\ \bibnamefont {Baumann}}, \bibinfo
  {author} {\bibfnamefont {D.~E.}\ \bibnamefont {Rivas}}, \bibinfo {author}
  {\bibfnamefont {Y.}~\bibnamefont {Ovcharenko}}, \bibinfo {author}
  {\bibfnamefont {R.}~\bibnamefont {Boll}}, \bibinfo {author} {\bibfnamefont
  {S.}~\bibnamefont {Dold}}, \bibinfo {author} {\bibfnamefont {A.}~\bibnamefont
  {De~Fanis}}, \bibinfo {author} {\bibfnamefont {M.}~\bibnamefont {Ilchen}},
  \bibinfo {author} {\bibfnamefont {M.}~\bibnamefont {Meyer}}, \bibinfo
  {author} {\bibfnamefont {G.}~\bibnamefont {Goldsztejn}}, \bibinfo {author}
  {\bibfnamefont {K.}~\bibnamefont {Li}}, \bibinfo {author} {\bibfnamefont
  {G.}~\bibnamefont {Doumy}}, \bibinfo {author} {\bibfnamefont
  {L.}~\bibnamefont {Young}}, \bibinfo {author} {\bibfnamefont
  {G.}~\bibnamefont {Sansone}}, \bibinfo {author} {\bibfnamefont
  {R.}~\bibnamefont {D\"orner}}, \bibinfo {author} {\bibfnamefont {M.~N.}\
  \bibnamefont {Piancastelli}}, \bibinfo {author} {\bibfnamefont
  {S.}~\bibnamefont {Carniato}}, \bibinfo {author} {\bibfnamefont {J.~D.}\
  \bibnamefont {Bozek}}, \bibinfo {author} {\bibfnamefont {R.}~\bibnamefont
  {P\"uttner}},\ and\ \bibinfo {author} {\bibfnamefont {M.}~\bibnamefont
  {Simon}},\ }\bibfield  {title} {\bibinfo {title} {Alternative pathway to
  double-core-hole states},\ }\href
  {https://doi.org/10.1103/PhysRevLett.131.253201} {\bibfield  {journal}
  {\bibinfo  {journal} {Phys. Rev. Lett.}\ }\textbf {\bibinfo {volume} {131}},\
  \bibinfo {pages} {253201} (\bibinfo {year} {2023})}\BibitemShut {NoStop}%
\bibitem [{\citenamefont {Trinter}\ \emph {et~al.}(2024)\citenamefont
  {Trinter}, \citenamefont {Inhester}, \citenamefont {P\"{u}ttner},
  \citenamefont {Malerz}, \citenamefont {Th\"{u}rmer}, \citenamefont
  {Marchenko}, \citenamefont {Piancastelli}, \citenamefont {Simon},
  \citenamefont {Winter},\ and\ \citenamefont {Hergenhahn}}]{Trinter2024}%
  \BibitemOpen
  \bibfield  {author} {\bibinfo {author} {\bibfnamefont {F.}~\bibnamefont
  {Trinter}}, \bibinfo {author} {\bibfnamefont {L.}~\bibnamefont {Inhester}},
  \bibinfo {author} {\bibfnamefont {R.}~\bibnamefont {P\"{u}ttner}}, \bibinfo
  {author} {\bibfnamefont {S.}~\bibnamefont {Malerz}}, \bibinfo {author}
  {\bibfnamefont {S.}~\bibnamefont {Th\"{u}rmer}}, \bibinfo {author}
  {\bibfnamefont {T.}~\bibnamefont {Marchenko}}, \bibinfo {author}
  {\bibfnamefont {M.~N.}\ \bibnamefont {Piancastelli}}, \bibinfo {author}
  {\bibfnamefont {M.}~\bibnamefont {Simon}}, \bibinfo {author} {\bibfnamefont
  {B.}~\bibnamefont {Winter}},\ and\ \bibinfo {author} {\bibfnamefont
  {U.}~\bibnamefont {Hergenhahn}},\ }\bibfield  {title} {\bibinfo {title}
  {Radiationless decay spectrum of {O} 1s double core holes in liquid water},\
  }\href {https://doi.org/10.1063/5.0205994} {\bibfield  {journal} {\bibinfo
  {journal} {J. Chem. Phys.}\ }\textbf {\bibinfo {volume} {160}},\ \bibinfo
  {pages} {194503} (\bibinfo {year} {2024})}\BibitemShut {NoStop}%
\bibitem [{\citenamefont {Drake}(1988)}]{Drake1988}%
  \BibitemOpen
  \bibfield  {author} {\bibinfo {author} {\bibfnamefont {G.~W.~F.}\
  \bibnamefont {Drake}},\ }\bibfield  {title} {\bibinfo {title} {Theoretical
  energies for the n=1 and n=2 states of the helium isoelectronic sequence up
  to {Z}=100},\ }\href {https://doi.org/10.1139/p88-100} {\bibfield  {journal}
  {\bibinfo  {journal} {Can. J. Phys.}\ }\textbf {\bibinfo {volume} {66}},\
  \bibinfo {pages} {586 } (\bibinfo {year} {1988})}\BibitemShut {NoStop}%
\bibitem [{\citenamefont {Madsen}(2003)}]{Madsen2003}%
  \BibitemOpen
  \bibfield  {author} {\bibinfo {author} {\bibfnamefont {L.~B.}\ \bibnamefont
  {Madsen}},\ }\bibfield  {title} {\bibinfo {title} {Triply excited states:
  electron-electron correlations in lithium},\ }\href
  {https://doi.org/10.1088/0953-4075/36/20/r01} {\bibfield  {journal} {\bibinfo
   {journal} {J. Phys. B: At. Mol. Opt. Phys.}\ }\textbf {\bibinfo {volume}
  {36}},\ \bibinfo {pages} {R223 } (\bibinfo {year} {2003})}\BibitemShut
  {NoStop}%
\bibitem [{\citenamefont {Indelicato}(2019)}]{Indelicato2019}%
  \BibitemOpen
  \bibfield  {author} {\bibinfo {author} {\bibfnamefont {P.}~\bibnamefont
  {Indelicato}},\ }\bibfield  {title} {\bibinfo {title} {{QED} tests with
  highly charged ions},\ }\href {https://doi.org/10.1088/1361-6455/ab42c9}
  {\bibfield  {journal} {\bibinfo  {journal} {J. Phys. B: At. Mol. Opt. Phys.}\
  }\textbf {\bibinfo {volume} {52}},\ \bibinfo {pages} {232001} (\bibinfo
  {year} {2019})}\BibitemShut {NoStop}%
\bibitem [{\citenamefont {Zaytsev}\ \emph {et~al.}(2023)\citenamefont
  {Zaytsev}, \citenamefont {Malyshev},\ and\ \citenamefont
  {Shabaev}}]{Zaytsev2023}%
  \BibitemOpen
  \bibfield  {author} {\bibinfo {author} {\bibfnamefont {V.~A.}\ \bibnamefont
  {Zaytsev}}, \bibinfo {author} {\bibfnamefont {A.~V.}\ \bibnamefont
  {Malyshev}},\ and\ \bibinfo {author} {\bibfnamefont {V.~M.}\ \bibnamefont
  {Shabaev}},\ }\bibfield  {title} {\bibinfo {title} {Complex-scaled ab initio
  {QED} approach to autoionizing states},\ }\href
  {https://doi.org/10.1103/PhysRevA.107.032801} {\bibfield  {journal} {\bibinfo
   {journal} {Phys. Rev. A}\ }\textbf {\bibinfo {volume} {107}},\ \bibinfo
  {pages} {032801} (\bibinfo {year} {2023})}\BibitemShut {NoStop}%
\bibitem [{\citenamefont {Madden}\ and\ \citenamefont
  {Codling}(1963)}]{Madden1963}%
  \BibitemOpen
  \bibfield  {author} {\bibinfo {author} {\bibfnamefont {R.~P.}\ \bibnamefont
  {Madden}}\ and\ \bibinfo {author} {\bibfnamefont {K.}~\bibnamefont
  {Codling}},\ }\bibfield  {title} {\bibinfo {title} {New autoionizing atomic
  energy levels in {H}e, {N}e, and {A}r},\ }\href
  {https://doi.org/10.1103/physrevlett.10.516} {\bibfield  {journal} {\bibinfo
  {journal} {Phys. Rev. Lett.}\ }\textbf {\bibinfo {volume} {10}},\ \bibinfo
  {pages} {516 } (\bibinfo {year} {1963})}\BibitemShut {NoStop}%
\bibitem [{\citenamefont {Rost}\ \emph {et~al.}(1997)\citenamefont {Rost},
  \citenamefont {Schulz}, \citenamefont {Domke},\ and\ \citenamefont
  {Kaindl}}]{Rost1997}%
  \BibitemOpen
  \bibfield  {author} {\bibinfo {author} {\bibfnamefont {J.~M.}\ \bibnamefont
  {Rost}}, \bibinfo {author} {\bibfnamefont {K.}~\bibnamefont {Schulz}},
  \bibinfo {author} {\bibfnamefont {M.}~\bibnamefont {Domke}},\ and\ \bibinfo
  {author} {\bibfnamefont {G.}~\bibnamefont {Kaindl}},\ }\bibfield  {title}
  {\bibinfo {title} {Resonance parameters of photo doubly excited helium},\
  }\href {https://doi.org/10.1088/0953-4075/30/21/010} {\bibfield  {journal}
  {\bibinfo  {journal} {J. Phys. B: At. Mol. Opt. Phys.}\ }\textbf {\bibinfo
  {volume} {30}},\ \bibinfo {pages} {4663 } (\bibinfo {year}
  {1997})}\BibitemShut {NoStop}%
\bibitem [{\citenamefont {Wuilleumier}(2000)}]{Wuilleumier2000}%
  \BibitemOpen
  \bibfield  {author} {\bibinfo {author} {\bibfnamefont {F.~J.}\ \bibnamefont
  {Wuilleumier}},\ }\bibfield  {title} {\bibinfo {title} {From doubly excited
  states of helium to triply excited states of lithium},\ }\href
  {https://physicsessays.org/browse-journal-2/product/1226-9-francois-j-wuilleumier-from-doubly-excited-states-of-helium-to-triply-excited-states-of-lithium.html}
  {\bibfield  {journal} {\bibinfo  {journal} {Physics Essays}\ }\textbf
  {\bibinfo {volume} {13}},\ \bibinfo {pages} {230 } (\bibinfo {year}
  {2000})}\BibitemShut {NoStop}%
\bibitem [{\citenamefont {Tanner}\ \emph {et~al.}(2000)\citenamefont {Tanner},
  \citenamefont {Richter},\ and\ \citenamefont {Rost}}]{Tanner2000a}%
  \BibitemOpen
  \bibfield  {author} {\bibinfo {author} {\bibfnamefont {G.}~\bibnamefont
  {Tanner}}, \bibinfo {author} {\bibfnamefont {K.}~\bibnamefont {Richter}},\
  and\ \bibinfo {author} {\bibfnamefont {J.-M.}\ \bibnamefont {Rost}},\
  }\bibfield  {title} {\bibinfo {title} {The theory of two-electron atoms:
  between ground state and complete fragmentation},\ }\href
  {https://doi.org/10.1103/revmodphys.72.497} {\bibfield  {journal} {\bibinfo
  {journal} {Rev. Mod. Phys.}\ }\textbf {\bibinfo {volume} {72}},\ \bibinfo
  {pages} {497 } (\bibinfo {year} {2000})}\BibitemShut {NoStop}%
\bibitem [{\citenamefont {Wehlitz}(2010)}]{Wehlitz2010}%
  \BibitemOpen
  \bibfield  {author} {\bibinfo {author} {\bibfnamefont {R.}~\bibnamefont
  {Wehlitz}},\ }\bibfield  {title} {\bibinfo {title} {Simultaneous emission of
  multiple electrons from atoms and molecules using synchrotron radiation},\
  }\href {https://doi.org/10.1016/S1049-250X(10)05806-4} {\bibfield  {journal}
  {\bibinfo  {journal} {Adv. At. Mol. Opt. Phys.}\ }\textbf {\bibinfo {volume}
  {58}},\ \bibinfo {pages} {1 } (\bibinfo {year} {2010})}\BibitemShut {NoStop}%
\bibitem [{\citenamefont {Hoszowska}\ \emph {et~al.}(2009)\citenamefont
  {Hoszowska}, \citenamefont {Kheifets}, \citenamefont {Dousse}, \citenamefont
  {Berset}, \citenamefont {Bray}, \citenamefont {Cao}, \citenamefont {Fennane},
  \citenamefont {Kayser}, \citenamefont {Kav\v{c}i\v{c}}, \citenamefont
  {Szlachetko},\ and\ \citenamefont {Szlachetko}}]{Hoszowska2009}%
  \BibitemOpen
  \bibfield  {author} {\bibinfo {author} {\bibfnamefont {J.}~\bibnamefont
  {Hoszowska}}, \bibinfo {author} {\bibfnamefont {A.~K.}\ \bibnamefont
  {Kheifets}}, \bibinfo {author} {\bibfnamefont {J.-C.}\ \bibnamefont
  {Dousse}}, \bibinfo {author} {\bibfnamefont {M.}~\bibnamefont {Berset}},
  \bibinfo {author} {\bibfnamefont {I.}~\bibnamefont {Bray}}, \bibinfo {author}
  {\bibfnamefont {W.}~\bibnamefont {Cao}}, \bibinfo {author} {\bibfnamefont
  {K.}~\bibnamefont {Fennane}}, \bibinfo {author} {\bibfnamefont
  {Y.}~\bibnamefont {Kayser}}, \bibinfo {author} {\bibfnamefont
  {M.}~\bibnamefont {Kav\v{c}i\v{c}}}, \bibinfo {author} {\bibfnamefont
  {J.}~\bibnamefont {Szlachetko}},\ and\ \bibinfo {author} {\bibfnamefont
  {M.}~\bibnamefont {Szlachetko}},\ }\bibfield  {title} {\bibinfo {title}
  {Physical mechanisms and scaling laws of ${K}$-shell double
  photoionization},\ }\href {https://doi.org/10.1103/physrevlett.102.073006}
  {\bibfield  {journal} {\bibinfo  {journal} {Phys. Rev. Lett.}\ }\textbf
  {\bibinfo {volume} {102}},\ \bibinfo {pages} {073006} (\bibinfo {year}
  {2009})}\BibitemShut {NoStop}%
\bibitem [{\citenamefont {Hoszowska}\ and\ \citenamefont
  {Dousse}(2013)}]{Hoszowska2013}%
  \BibitemOpen
  \bibfield  {author} {\bibinfo {author} {\bibfnamefont {J.}~\bibnamefont
  {Hoszowska}}\ and\ \bibinfo {author} {\bibfnamefont {J.-C.}\ \bibnamefont
  {Dousse}},\ }\bibfield  {title} {\bibinfo {title} {Photoinduced ${K}$-shell
  hollow atoms},\ }\href {https://doi.org/10.1016/j.elspec.2012.09.001}
  {\bibfield  {journal} {\bibinfo  {journal} {J. Electron. Spectrosc. Relat.
  Phenom.}\ }\textbf {\bibinfo {volume} {188}},\ \bibinfo {pages} {62 }
  (\bibinfo {year} {2013})}\BibitemShut {NoStop}%
\bibitem [{\citenamefont {Zaytsev}\ \emph {et~al.}(2019)\citenamefont
  {Zaytsev}, \citenamefont {Maltsev}, \citenamefont {Tupitsyn},\ and\
  \citenamefont {Shabaev}}]{Zaytsev2019}%
  \BibitemOpen
  \bibfield  {author} {\bibinfo {author} {\bibfnamefont {V.~A.}\ \bibnamefont
  {Zaytsev}}, \bibinfo {author} {\bibfnamefont {I.~A.}\ \bibnamefont
  {Maltsev}}, \bibinfo {author} {\bibfnamefont {I.~I.}\ \bibnamefont
  {Tupitsyn}},\ and\ \bibinfo {author} {\bibfnamefont {V.~M.}\ \bibnamefont
  {Shabaev}},\ }\bibfield  {title} {\bibinfo {title} {Complex-scaled
  relativistic configuration-interaction study of the ${LL}$ resonances in
  heliumlike ions: From boron to argon},\ }\href
  {https://doi.org/10.1103/PhysRevA.100.052504} {\bibfield  {journal} {\bibinfo
   {journal} {Phys. Rev. A}\ }\textbf {\bibinfo {volume} {100}},\ \bibinfo
  {pages} {052504} (\bibinfo {year} {2019})}\BibitemShut {NoStop}%
\bibitem [{\citenamefont {Lyashchenko}\ \emph {et~al.}(2021)\citenamefont
  {Lyashchenko}, \citenamefont {Andreev},\ and\ \citenamefont
  {Yu}}]{Lyashchenko2021}%
  \BibitemOpen
  \bibfield  {author} {\bibinfo {author} {\bibfnamefont {K.~N.}\ \bibnamefont
  {Lyashchenko}}, \bibinfo {author} {\bibfnamefont {O.~Y.}\ \bibnamefont
  {Andreev}},\ and\ \bibinfo {author} {\bibfnamefont {D.}~\bibnamefont {Yu}},\
  }\bibfield  {title} {\bibinfo {title} {{QED} calculation of two-electron
  one-photon transition probabilities in {H}e-like ions},\ }\href
  {https://doi.org/10.1103/PhysRevA.104.012818} {\bibfield  {journal} {\bibinfo
   {journal} {Phys. Rev. A}\ }\textbf {\bibinfo {volume} {104}},\ \bibinfo
  {pages} {012818} (\bibinfo {year} {2021})}\BibitemShut {NoStop}%
\bibitem [{\citenamefont {Wu}\ \emph {et~al.}(2023)\citenamefont {Wu},
  \citenamefont {Ding}, \citenamefont {Cao}, \citenamefont {Zhang},
  \citenamefont {Zhang}, \citenamefont {Xue}, \citenamefont {Yu},\ and\
  \citenamefont {Dong}}]{Wu2023}%
  \BibitemOpen
  \bibfield  {author} {\bibinfo {author} {\bibfnamefont {C.}~\bibnamefont
  {Wu}}, \bibinfo {author} {\bibfnamefont {X.}~\bibnamefont {Ding}}, \bibinfo
  {author} {\bibfnamefont {M.}~\bibnamefont {Cao}}, \bibinfo {author}
  {\bibfnamefont {D.}~\bibnamefont {Zhang}}, \bibinfo {author} {\bibfnamefont
  {M.}~\bibnamefont {Zhang}}, \bibinfo {author} {\bibfnamefont
  {Y.}~\bibnamefont {Xue}}, \bibinfo {author} {\bibfnamefont {D.}~\bibnamefont
  {Yu}},\ and\ \bibinfo {author} {\bibfnamefont {C.}~\bibnamefont {Dong}},\
  }\bibfield  {title} {\bibinfo {title} {Energy levels and radiative transition
  properties of the $2s2p$ double ${K}$-shell vacancy state in {H}e-like ions
  $(4\leq{Z}\leq54)$},\ }\href {https://doi.org/10.1016/j.adt.2023.101602}
  {\bibfield  {journal} {\bibinfo  {journal} {At. Data Nucl. Data}\ }\textbf
  {\bibinfo {volume} {154}},\ \bibinfo {pages} {101602} (\bibinfo {year}
  {2023})}\BibitemShut {NoStop}%
\bibitem [{\citenamefont {Becker}\ and\ \citenamefont
  {Dahler}(1964)}]{Becker1964}%
  \BibitemOpen
  \bibfield  {author} {\bibinfo {author} {\bibfnamefont {P.~M.}\ \bibnamefont
  {Becker}}\ and\ \bibinfo {author} {\bibfnamefont {J.~S.}\ \bibnamefont
  {Dahler}},\ }\bibfield  {title} {\bibinfo {title} {Double excitation of
  helium by electron impact},\ }\href {https://doi.org/10.1103/physrev.136.a73}
  {\bibfield  {journal} {\bibinfo  {journal} {Phys. Rev.}\ }\textbf {\bibinfo
  {volume} {136}},\ \bibinfo {pages} {A73 } (\bibinfo {year}
  {1964})}\BibitemShut {NoStop}%
\bibitem [{\citenamefont {Westerveld}\ \emph {et~al.}(1979)\citenamefont
  {Westerveld}, \citenamefont {Kets}, \citenamefont {Heideman},\ and\
  \citenamefont {van Eck}}]{Westerveld1979}%
  \BibitemOpen
  \bibfield  {author} {\bibinfo {author} {\bibfnamefont {W.~B.}\ \bibnamefont
  {Westerveld}}, \bibinfo {author} {\bibfnamefont {F.~B.}\ \bibnamefont
  {Kets}}, \bibinfo {author} {\bibfnamefont {H.~G.~M.}\ \bibnamefont
  {Heideman}},\ and\ \bibinfo {author} {\bibfnamefont {J.}~\bibnamefont {van
  Eck}},\ }\bibfield  {title} {\bibinfo {title} {Electron impact excitation of
  the $(2p^2)~^3{P}$ doubly excited state of helium},\ }\href
  {https://doi.org/10.1088/0022-3700/12/15/018} {\bibfield  {journal} {\bibinfo
   {journal} {J. Phys. B: Atom. Mol. Phys.}\ }\textbf {\bibinfo {volume}
  {12}},\ \bibinfo {pages} {2575} (\bibinfo {year} {1979})}\BibitemShut
  {NoStop}%
\bibitem [{\citenamefont {Padhy}\ and\ \citenamefont {Rai}(1990)}]{Padhy1990}%
  \BibitemOpen
  \bibfield  {author} {\bibinfo {author} {\bibfnamefont {B.}~\bibnamefont
  {Padhy}}\ and\ \bibinfo {author} {\bibfnamefont {D.~K.}\ \bibnamefont
  {Rai}},\ }\bibfield  {title} {\bibinfo {title} {Two-electron excitation in
  helium-like ions by electron impact},\ }\href
  {https://doi.org/10.1007/BF02846597} {\bibfield  {journal} {\bibinfo
  {journal} {Pramana - J. Phys.}\ }\textbf {\bibinfo {volume} {35}},\ \bibinfo
  {pages} {341 } (\bibinfo {year} {1990})}\BibitemShut {NoStop}%
\bibitem [{\citenamefont {Mikhailov}\ \emph {et~al.}(2013)\citenamefont
  {Mikhailov}, \citenamefont {Mikhailov}, \citenamefont {Nefiodov},\ and\
  \citenamefont {Plunien}}]{Mikhailov2013}%
  \BibitemOpen
  \bibfield  {author} {\bibinfo {author} {\bibfnamefont {A.}~\bibnamefont
  {Mikhailov}}, \bibinfo {author} {\bibfnamefont {I.}~\bibnamefont
  {Mikhailov}}, \bibinfo {author} {\bibfnamefont {A.}~\bibnamefont
  {Nefiodov}},\ and\ \bibinfo {author} {\bibfnamefont {G.}~\bibnamefont
  {Plunien}},\ }\bibfield  {title} {\bibinfo {title} {Excitation of
  autoionizing states of helium-like ions by scattering of high-energy
  particles},\ }\href {https://doi.org/10.1134/S1063776113020155} {\bibfield
  {journal} {\bibinfo  {journal} {J. Exp. Theor. Phys.}\ }\textbf {\bibinfo
  {volume} {116}},\ \bibinfo {pages} {363 } (\bibinfo {year}
  {2013})}\BibitemShut {NoStop}%
\bibitem [{\citenamefont {Woods}\ \emph {et~al.}(1975)\citenamefont {Woods},
  \citenamefont {Kauffman}, \citenamefont {Jamison}, \citenamefont
  {Stolterfoht},\ and\ \citenamefont {Richard}}]{Woods1975a}%
  \BibitemOpen
  \bibfield  {author} {\bibinfo {author} {\bibfnamefont {C.~W.}\ \bibnamefont
  {Woods}}, \bibinfo {author} {\bibfnamefont {R.~L.}\ \bibnamefont {Kauffman}},
  \bibinfo {author} {\bibfnamefont {K.~A.}\ \bibnamefont {Jamison}}, \bibinfo
  {author} {\bibfnamefont {N.}~\bibnamefont {Stolterfoht}},\ and\ \bibinfo
  {author} {\bibfnamefont {P.}~\bibnamefont {Richard}},\ }\bibfield  {title}
  {\bibinfo {title} {${K}$-shell {A}uger-electron hypersatellites of {N}e},\
  }\href {https://doi.org/10.1103/PhysRevA.12.1393} {\bibfield  {journal}
  {\bibinfo  {journal} {Phys. Rev. A}\ }\textbf {\bibinfo {volume} {12}},\
  \bibinfo {pages} {1393} (\bibinfo {year} {1975})}\BibitemShut {NoStop}%
\bibitem [{\citenamefont {R{\o}dbro}\ \emph {et~al.}(1979)\citenamefont
  {R{\o}dbro}, \citenamefont {Bruch},\ and\ \citenamefont
  {Bisgaard}}]{Rodbro1979}%
  \BibitemOpen
  \bibfield  {author} {\bibinfo {author} {\bibfnamefont {M.}~\bibnamefont
  {R{\o}dbro}}, \bibinfo {author} {\bibfnamefont {R.}~\bibnamefont {Bruch}},\
  and\ \bibinfo {author} {\bibfnamefont {P.}~\bibnamefont {Bisgaard}},\
  }\bibfield  {title} {\bibinfo {title} {High-resolution projectile {A}uger
  spectroscopy for {Li, Be, B and C} excited in single gas collisions {I}. line
  energies for prompt decay},\ }\href
  {https://doi.org/10.1088/0022-3700/12/15/009} {\bibfield  {journal} {\bibinfo
   {journal} {J. Phys. B: At. Mol. Opt. Phys.}\ }\textbf {\bibinfo {volume}
  {12}},\ \bibinfo {pages} {2413} (\bibinfo {year} {1979})}\BibitemShut
  {NoStop}%
\bibitem [{\citenamefont {Mack}\ \emph {et~al.}(1989)\citenamefont {Mack},
  \citenamefont {Nijland}, \citenamefont {{van der Straten}}, \citenamefont
  {Niehaus},\ and\ \citenamefont {Morgenstern}}]{Mack1989}%
  \BibitemOpen
  \bibfield  {author} {\bibinfo {author} {\bibfnamefont {M.}~\bibnamefont
  {Mack}}, \bibinfo {author} {\bibfnamefont {J.~H.}\ \bibnamefont {Nijland}},
  \bibinfo {author} {\bibfnamefont {P.}~\bibnamefont {{van der Straten}}},
  \bibinfo {author} {\bibfnamefont {A.}~\bibnamefont {Niehaus}},\ and\ \bibinfo
  {author} {\bibfnamefont {R.}~\bibnamefont {Morgenstern}},\ }\bibfield
  {title} {\bibinfo {title} {Correlation in double electron capture in
  collisions of fully stripped ions on {H}e and {H}$_2$},\ }\href
  {https://doi.org/10.1103/physreva.39.3846} {\bibfield  {journal} {\bibinfo
  {journal} {Phys. Rev. A}\ }\textbf {\bibinfo {volume} {39}},\ \bibinfo
  {pages} {3846 } (\bibinfo {year} {1989})}\BibitemShut {NoStop}%
\bibitem [{\citenamefont {Barat}\ and\ \citenamefont
  {Roncin}(1992)}]{Barat1992}%
  \BibitemOpen
  \bibfield  {author} {\bibinfo {author} {\bibfnamefont {M.}~\bibnamefont
  {Barat}}\ and\ \bibinfo {author} {\bibfnamefont {P.}~\bibnamefont {Roncin}},\
  }\bibfield  {title} {\bibinfo {title} {Multiple electron capture by highly
  charged ions at ke{V} energies},\ }\href
  {https://doi.org/10.1088/0953-4075/25/10/006} {\bibfield  {journal} {\bibinfo
   {journal} {J. Phys. B: At. Mol. Opt. Phys.}\ }\textbf {\bibinfo {volume}
  {25}},\ \bibinfo {pages} {2205} (\bibinfo {year} {1992})}\BibitemShut
  {NoStop}%
\bibitem [{\citenamefont {Benis}\ \emph {et~al.}(2004)\citenamefont {Benis},
  \citenamefont {Zouros}, \citenamefont {Gorczyca}, \citenamefont
  {Gonz{\'a}lez},\ and\ \citenamefont {Richard}}]{Benis2004a}%
  \BibitemOpen
  \bibfield  {author} {\bibinfo {author} {\bibfnamefont {E.~P.}\ \bibnamefont
  {Benis}}, \bibinfo {author} {\bibfnamefont {T.~J.~M.}\ \bibnamefont
  {Zouros}}, \bibinfo {author} {\bibfnamefont {T.~W.}\ \bibnamefont
  {Gorczyca}}, \bibinfo {author} {\bibfnamefont {A.~D.}\ \bibnamefont
  {Gonz{\'a}lez}},\ and\ \bibinfo {author} {\bibfnamefont {P.}~\bibnamefont
  {Richard}},\ }\bibfield  {title} {\bibinfo {title} {Elastic resonant and
  nonresonant differential scattering of quasifree electrons from
  {B}$^{4+}(1s)$ and {B}$^{3+}(1s^2)$ ions},\ }\href
  {https://doi.org/10.1103/PhysRevA.69.052718} {\bibfield  {journal} {\bibinfo
  {journal} {Phys. Rev. A}\ }\textbf {\bibinfo {volume} {69}},\ \bibinfo
  {pages} {052718} (\bibinfo {year} {2004})}\BibitemShut {NoStop}%
\bibitem [{\citenamefont {Bruch}\ \emph {et~al.}(1975)\citenamefont {Bruch},
  \citenamefont {Paul}, \citenamefont {Andr\"a},\ and\ \citenamefont
  {Lipsky}}]{Bruch1975}%
  \BibitemOpen
  \bibfield  {author} {\bibinfo {author} {\bibfnamefont {R.}~\bibnamefont
  {Bruch}}, \bibinfo {author} {\bibfnamefont {G.}~\bibnamefont {Paul}},
  \bibinfo {author} {\bibfnamefont {J.}~\bibnamefont {Andr\"a}},\ and\ \bibinfo
  {author} {\bibfnamefont {L.}~\bibnamefont {Lipsky}},\ }\bibfield  {title}
  {\bibinfo {title} {Autoionization of foil-excited states in {Li I and Li
  II}},\ }\href {https://doi.org/10.1103/physreva.12.1808} {\bibfield
  {journal} {\bibinfo  {journal} {Phys.Rev. A}\ }\textbf {\bibinfo {volume}
  {12}},\ \bibinfo {pages} {1808} (\bibinfo {year} {1975})}\BibitemShut
  {NoStop}%
\bibitem [{\citenamefont {Kasthurirangan}\ \emph {et~al.}(2013)\citenamefont
  {Kasthurirangan}, \citenamefont {Saha}, \citenamefont {Agnihotri},
  \citenamefont {Bhattacharyya}, \citenamefont {Misra}, \citenamefont {Kumar},
  \citenamefont {Mukherjee}, \citenamefont {Santos}, \citenamefont {Costa},
  \citenamefont {Indelicato}, \citenamefont {Mukherjee},\ and\ \citenamefont
  {Tribedi}}]{Kasthurirangan2013}%
  \BibitemOpen
  \bibfield  {author} {\bibinfo {author} {\bibfnamefont {S.}~\bibnamefont
  {Kasthurirangan}}, \bibinfo {author} {\bibfnamefont {J.~K.}\ \bibnamefont
  {Saha}}, \bibinfo {author} {\bibfnamefont {A.~N.}\ \bibnamefont {Agnihotri}},
  \bibinfo {author} {\bibfnamefont {S.}~\bibnamefont {Bhattacharyya}}, \bibinfo
  {author} {\bibfnamefont {D.}~\bibnamefont {Misra}}, \bibinfo {author}
  {\bibfnamefont {A.}~\bibnamefont {Kumar}}, \bibinfo {author} {\bibfnamefont
  {P.~K.}\ \bibnamefont {Mukherjee}}, \bibinfo {author} {\bibfnamefont {J.~P.}\
  \bibnamefont {Santos}}, \bibinfo {author} {\bibfnamefont {A.~M.}\
  \bibnamefont {Costa}}, \bibinfo {author} {\bibfnamefont {P.}~\bibnamefont
  {Indelicato}}, \bibinfo {author} {\bibfnamefont {T.~K.}\ \bibnamefont
  {Mukherjee}},\ and\ \bibinfo {author} {\bibfnamefont {L.~C.}\ \bibnamefont
  {Tribedi}},\ }\bibfield  {title} {\bibinfo {title} {Observation of
  $2p3d~^1{P}^o \to 1s3d~^1{D}^e$ radiative transition in {He-like Si, S, and
  Cl} ions},\ }\href {https://doi.org/10.1103/PhysRevLett.111.243201}
  {\bibfield  {journal} {\bibinfo  {journal} {Phys. Rev. Lett.}\ }\textbf
  {\bibinfo {volume} {111}},\ \bibinfo {pages} {243201} (\bibinfo {year}
  {2013})}\BibitemShut {NoStop}%
\bibitem [{\citenamefont {Kennedy}\ \emph {et~al.}(2004)\citenamefont
  {Kennedy}, \citenamefont {Costello}, \citenamefont {Mosnier},\ and\
  \citenamefont {{van Kampen}}}]{Kennedy2004a}%
  \BibitemOpen
  \bibfield  {author} {\bibinfo {author} {\bibfnamefont {E.~T.}\ \bibnamefont
  {Kennedy}}, \bibinfo {author} {\bibfnamefont {J.~T.}\ \bibnamefont
  {Costello}}, \bibinfo {author} {\bibfnamefont {J.~P.}\ \bibnamefont
  {Mosnier}},\ and\ \bibinfo {author} {\bibfnamefont {P.}~\bibnamefont {{van
  Kampen}}},\ }\bibfield  {title} {\bibinfo {title} {{VUV/EUV} ionising
  radiation and atoms and ions: dual laser plasma investigations},\ }\href
  {https://doi.org/10.1016/j.radphyschem.2003.12.018} {\bibfield  {journal}
  {\bibinfo  {journal} {Rad. Phys. Chem.}\ }\textbf {\bibinfo {volume} {70}},\
  \bibinfo {pages} {291} (\bibinfo {year} {2004})}\BibitemShut {NoStop}%
\bibitem [{\citenamefont {Kilgus}\ \emph {et~al.}(1990)\citenamefont {Kilgus},
  \citenamefont {Berger}, \citenamefont {Blatt}, \citenamefont {Grieser},
  \citenamefont {Habs}, \citenamefont {Hochadel}, \citenamefont {Jaeschke},
  \citenamefont {Kr{\"a}mer}, \citenamefont {Neumann}, \citenamefont
  {Neureither}, \citenamefont {Ott}, \citenamefont {Schwalm}, \citenamefont
  {Steck}, \citenamefont {Stokstad}, \citenamefont {Szmola}, \citenamefont
  {Wolf}, \citenamefont {Schuch}, \citenamefont {M{\"u}ller},\ and\
  \citenamefont {Wagner}}]{Kilgus1990}%
  \BibitemOpen
  \bibfield  {author} {\bibinfo {author} {\bibfnamefont {G.}~\bibnamefont
  {Kilgus}}, \bibinfo {author} {\bibfnamefont {J.}~\bibnamefont {Berger}},
  \bibinfo {author} {\bibfnamefont {P.}~\bibnamefont {Blatt}}, \bibinfo
  {author} {\bibfnamefont {M.}~\bibnamefont {Grieser}}, \bibinfo {author}
  {\bibfnamefont {D.}~\bibnamefont {Habs}}, \bibinfo {author} {\bibfnamefont
  {B.}~\bibnamefont {Hochadel}}, \bibinfo {author} {\bibfnamefont
  {E.}~\bibnamefont {Jaeschke}}, \bibinfo {author} {\bibfnamefont
  {D.}~\bibnamefont {Kr{\"a}mer}}, \bibinfo {author} {\bibfnamefont
  {R.}~\bibnamefont {Neumann}}, \bibinfo {author} {\bibfnamefont
  {G.}~\bibnamefont {Neureither}}, \bibinfo {author} {\bibfnamefont
  {W.}~\bibnamefont {Ott}}, \bibinfo {author} {\bibfnamefont {D.}~\bibnamefont
  {Schwalm}}, \bibinfo {author} {\bibfnamefont {M.}~\bibnamefont {Steck}},
  \bibinfo {author} {\bibfnamefont {R.}~\bibnamefont {Stokstad}}, \bibinfo
  {author} {\bibfnamefont {E.}~\bibnamefont {Szmola}}, \bibinfo {author}
  {\bibfnamefont {A.}~\bibnamefont {Wolf}}, \bibinfo {author} {\bibfnamefont
  {R.}~\bibnamefont {Schuch}}, \bibinfo {author} {\bibfnamefont
  {A.}~\bibnamefont {M{\"u}ller}},\ and\ \bibinfo {author} {\bibfnamefont
  {M.}~\bibnamefont {Wagner}},\ }\bibfield  {title} {\bibinfo {title}
  {Dielectronic recombination of hydrogenlike oxygen in a heavy-ion storage
  ring},\ }\href {https://doi.org/10.1103/physrevlett.64.737} {\bibfield
  {journal} {\bibinfo  {journal} {Phys. Rev. Lett.}\ }\textbf {\bibinfo
  {volume} {64}},\ \bibinfo {pages} {737} (\bibinfo {year} {1990})}\BibitemShut
  {NoStop}%
\bibitem [{\citenamefont {DeWitt}\ \emph {et~al.}(1995)\citenamefont {DeWitt},
  \citenamefont {Lindroth}, \citenamefont {Schuch}, \citenamefont {Gao},
  \citenamefont {Quinteros},\ and\ \citenamefont {Zong}}]{Dewitt1995}%
  \BibitemOpen
  \bibfield  {author} {\bibinfo {author} {\bibfnamefont {D.~R.}\ \bibnamefont
  {DeWitt}}, \bibinfo {author} {\bibfnamefont {E.}~\bibnamefont {Lindroth}},
  \bibinfo {author} {\bibfnamefont {R.}~\bibnamefont {Schuch}}, \bibinfo
  {author} {\bibfnamefont {H.}~\bibnamefont {Gao}}, \bibinfo {author}
  {\bibfnamefont {T.}~\bibnamefont {Quinteros}},\ and\ \bibinfo {author}
  {\bibfnamefont {W.}~\bibnamefont {Zong}},\ }\bibfield  {title} {\bibinfo
  {title} {Spectroscopy of highly doubly-excited states of helium through
  dielectronic recombination},\ }\href
  {https://doi.org/10.1088/0953-4075/28/5/007} {\bibfield  {journal} {\bibinfo
  {journal} {J. Phys. B: At. Mol. Opt. Phys.}\ }\textbf {\bibinfo {volume}
  {28}},\ \bibinfo {pages} {L147} (\bibinfo {year} {1995})}\BibitemShut
  {NoStop}%
\bibitem [{\citenamefont {Bernhardt}\ \emph {et~al.}(2011)\citenamefont
  {Bernhardt}, \citenamefont {Brandau}, \citenamefont {Harman}, \citenamefont
  {Kozhuharov}, \citenamefont {M{\"u}ller}, \citenamefont {Scheid},
  \citenamefont {Schippers}, \citenamefont {Schmidt}, \citenamefont {Yu},
  \citenamefont {Artemyev}, \citenamefont {Tupitsyn}, \citenamefont {B{\"o}hm},
  \citenamefont {Bosch}, \citenamefont {Currell}, \citenamefont {Franzke},
  \citenamefont {Gumberidze}, \citenamefont {Jacobi}, \citenamefont {Mokler},
  \citenamefont {Nolden}, \citenamefont {Spillman}, \citenamefont {Stachura},
  \citenamefont {Steck},\ and\ \citenamefont {St{\"o}hlker}}]{Bernhardt2011}%
  \BibitemOpen
  \bibfield  {author} {\bibinfo {author} {\bibfnamefont {D.}~\bibnamefont
  {Bernhardt}}, \bibinfo {author} {\bibfnamefont {C.}~\bibnamefont {Brandau}},
  \bibinfo {author} {\bibfnamefont {Z.}~\bibnamefont {Harman}}, \bibinfo
  {author} {\bibfnamefont {C.}~\bibnamefont {Kozhuharov}}, \bibinfo {author}
  {\bibfnamefont {A.}~\bibnamefont {M{\"u}ller}}, \bibinfo {author}
  {\bibfnamefont {W.}~\bibnamefont {Scheid}}, \bibinfo {author} {\bibfnamefont
  {S.}~\bibnamefont {Schippers}}, \bibinfo {author} {\bibfnamefont {E.~W.}\
  \bibnamefont {Schmidt}}, \bibinfo {author} {\bibfnamefont {D.}~\bibnamefont
  {Yu}}, \bibinfo {author} {\bibfnamefont {A.~N.}\ \bibnamefont {Artemyev}},
  \bibinfo {author} {\bibfnamefont {I.~I.}\ \bibnamefont {Tupitsyn}}, \bibinfo
  {author} {\bibfnamefont {S.}~\bibnamefont {B{\"o}hm}}, \bibinfo {author}
  {\bibfnamefont {F.}~\bibnamefont {Bosch}}, \bibinfo {author} {\bibfnamefont
  {F.~J.}\ \bibnamefont {Currell}}, \bibinfo {author} {\bibfnamefont
  {B.}~\bibnamefont {Franzke}}, \bibinfo {author} {\bibfnamefont
  {A.}~\bibnamefont {Gumberidze}}, \bibinfo {author} {\bibfnamefont
  {J.}~\bibnamefont {Jacobi}}, \bibinfo {author} {\bibfnamefont {P.~H.}\
  \bibnamefont {Mokler}}, \bibinfo {author} {\bibfnamefont {F.}~\bibnamefont
  {Nolden}}, \bibinfo {author} {\bibfnamefont {U.}~\bibnamefont {Spillman}},
  \bibinfo {author} {\bibfnamefont {Z.}~\bibnamefont {Stachura}}, \bibinfo
  {author} {\bibfnamefont {M.}~\bibnamefont {Steck}},\ and\ \bibinfo {author}
  {\bibfnamefont {T.}~\bibnamefont {St{\"o}hlker}},\ }\bibfield  {title}
  {\bibinfo {title} {Breit interaction in dielectronic recombination of
  hydrogenlike uranium},\ }\href {https://doi.org/10.1103/physreva.83.020701}
  {\bibfield  {journal} {\bibinfo  {journal} {Phys. Rev. A}\ }\textbf {\bibinfo
  {volume} {83}},\ \bibinfo {pages} {020701(R)} (\bibinfo {year}
  {2011})}\BibitemShut {NoStop}%
\bibitem [{\citenamefont {Wang}\ \emph {et~al.}(2024)\citenamefont {Wang},
  \citenamefont {Brandau}, \citenamefont {Fritzsche}, \citenamefont {Fuchs},
  \citenamefont {Harman}, \citenamefont {Kozhuharov}, \citenamefont
  {M\"{u}ller}, \citenamefont {Steck},\ and\ \citenamefont
  {Schippers}}]{Wang2024}%
  \BibitemOpen
  \bibfield  {author} {\bibinfo {author} {\bibfnamefont {S.-X.}\ \bibnamefont
  {Wang}}, \bibinfo {author} {\bibfnamefont {C.}~\bibnamefont {Brandau}},
  \bibinfo {author} {\bibfnamefont {S.}~\bibnamefont {Fritzsche}}, \bibinfo
  {author} {\bibfnamefont {S.}~\bibnamefont {Fuchs}}, \bibinfo {author}
  {\bibfnamefont {Z.}~\bibnamefont {Harman}}, \bibinfo {author} {\bibfnamefont
  {C.}~\bibnamefont {Kozhuharov}}, \bibinfo {author} {\bibfnamefont
  {A.}~\bibnamefont {M\"{u}ller}}, \bibinfo {author} {\bibfnamefont
  {M.}~\bibnamefont {Steck}},\ and\ \bibinfo {author} {\bibfnamefont
  {S.}~\bibnamefont {Schippers}},\ }\bibfield  {title} {\bibinfo {title} {Breit
  interaction in dielectronic recombination of hydrogenlike xenon ions:
  {S}torage-ring experiment and theory},\ }\href
  {https://doi.org/10.1140/epjd/s10053-024-00914-7} {\bibfield  {journal}
  {\bibinfo  {journal} {Eur. Phys. J. D}\ }\textbf {\bibinfo {volume} {78}},\
  \bibinfo {pages} {122} (\bibinfo {year} {2024})}\BibitemShut {NoStop}%
\bibitem [{\citenamefont {M\"{u}ller}\ \emph {et~al.}(2014)\citenamefont
  {M\"{u}ller}, \citenamefont {{Borovik Jr.}}, \citenamefont {Huber},
  \citenamefont {Schippers}, \citenamefont {Fursa},\ and\ \citenamefont
  {Bray}}]{Mueller2014g}%
  \BibitemOpen
  \bibfield  {author} {\bibinfo {author} {\bibfnamefont {A.}~\bibnamefont
  {M\"{u}ller}}, \bibinfo {author} {\bibfnamefont {A.}~\bibnamefont {{Borovik
  Jr.}}}, \bibinfo {author} {\bibfnamefont {K.}~\bibnamefont {Huber}}, \bibinfo
  {author} {\bibfnamefont {S.}~\bibnamefont {Schippers}}, \bibinfo {author}
  {\bibfnamefont {D.~V.}\ \bibnamefont {Fursa}},\ and\ \bibinfo {author}
  {\bibfnamefont {I.}~\bibnamefont {Bray}},\ }\bibfield  {title} {\bibinfo
  {title} {Double-${K}$-vacancy states in electron-impact single ionization of
  metastable two-electron {N}$^{5+}(1s2s~^3{S}_1$) ions},\ }\href
  {https://doi.org/10.1103/PhysRevA.90.010701} {\bibfield  {journal} {\bibinfo
  {journal} {Phys. Rev. A}\ }\textbf {\bibinfo {volume} {90}},\ \bibinfo
  {pages} {010701(R)} (\bibinfo {year} {2014})}\BibitemShut {NoStop}%
\bibitem [{\citenamefont {M\"{u}ller}\ \emph
  {et~al.}(2018{\natexlab{a}})\citenamefont {M\"{u}ller}, \citenamefont
  {{Borovik Jr.}}, \citenamefont {Huber}, \citenamefont {Schippers},
  \citenamefont {Fursa},\ and\ \citenamefont {Bray}}]{Mueller2018b}%
  \BibitemOpen
  \bibfield  {author} {\bibinfo {author} {\bibfnamefont {A.}~\bibnamefont
  {M\"{u}ller}}, \bibinfo {author} {\bibfnamefont {A.}~\bibnamefont {{Borovik
  Jr.}}}, \bibinfo {author} {\bibfnamefont {K.}~\bibnamefont {Huber}}, \bibinfo
  {author} {\bibfnamefont {S.}~\bibnamefont {Schippers}}, \bibinfo {author}
  {\bibfnamefont {D.~V.}\ \bibnamefont {Fursa}},\ and\ \bibinfo {author}
  {\bibfnamefont {I.}~\bibnamefont {Bray}},\ }\bibfield  {title} {\bibinfo
  {title} {Indirect contributions to electron-impact ionization of
  {L}i$^+(1s2s~^3{S}_1$) ions: {R}ole of intermediate double-${K}$-vacancy
  states},\ }\href {https://doi.org/10.1103/PhysRevA.97.022709} {\bibfield
  {journal} {\bibinfo  {journal} {Phys. Rev. A}\ }\textbf {\bibinfo {volume}
  {97}},\ \bibinfo {pages} {022709} (\bibinfo {year}
  {2018}{\natexlab{a}})}\BibitemShut {NoStop}%
\bibitem [{\citenamefont {Diehl}\ \emph {et~al.}(1999)\citenamefont {Diehl},
  \citenamefont {Cubaynes}, \citenamefont {Bizau}, \citenamefont {Wuilleumier},
  \citenamefont {Kennedy}, \citenamefont {Mosnier},\ and\ \citenamefont
  {Morgan}}]{Diehl1999a}%
  \BibitemOpen
  \bibfield  {author} {\bibinfo {author} {\bibfnamefont {S.}~\bibnamefont
  {Diehl}}, \bibinfo {author} {\bibfnamefont {D.}~\bibnamefont {Cubaynes}},
  \bibinfo {author} {\bibfnamefont {J.-M.}\ \bibnamefont {Bizau}}, \bibinfo
  {author} {\bibfnamefont {F.~J.}\ \bibnamefont {Wuilleumier}}, \bibinfo
  {author} {\bibfnamefont {E.~T.}\ \bibnamefont {Kennedy}}, \bibinfo {author}
  {\bibfnamefont {J.-P.}\ \bibnamefont {Mosnier}},\ and\ \bibinfo {author}
  {\bibfnamefont {T.~J.}\ \bibnamefont {Morgan}},\ }\bibfield  {title}
  {\bibinfo {title} {New high-resolution measurements of doubly excited states
  of {L}i$^{+}$},\ }\href {https://doi.org/10.1088/0953-4075/32/17/305}
  {\bibfield  {journal} {\bibinfo  {journal} {J. Phys. B: At. Mol. Opt. Phys.}\
  }\textbf {\bibinfo {volume} {32}},\ \bibinfo {pages} {4193} (\bibinfo {year}
  {1999})}\BibitemShut {NoStop}%
\bibitem [{\citenamefont {Carroll}\ and\ \citenamefont
  {Kennedy}(1977)}]{Carroll1977}%
  \BibitemOpen
  \bibfield  {author} {\bibinfo {author} {\bibfnamefont {P.~K.}\ \bibnamefont
  {Carroll}}\ and\ \bibinfo {author} {\bibfnamefont {E.~T.}\ \bibnamefont
  {Kennedy}},\ }\bibfield  {title} {\bibinfo {title} {Doubly excited
  autoionization resonances in the absorption spectrum of {L}i$^+$ formed in a
  laser-produced plasma},\ }\href {https://doi.org/10.1103/physrevlett.38.1068}
  {\bibfield  {journal} {\bibinfo  {journal} {Phys. Rev. Lett.}\ }\textbf
  {\bibinfo {volume} {38}},\ \bibinfo {pages} {1068 } (\bibinfo {year}
  {1977})}\BibitemShut {NoStop}%
\bibitem [{\citenamefont {Kiernan}\ \emph {et~al.}(1994)\citenamefont
  {Kiernan}, \citenamefont {Kennedy}, \citenamefont {Mosnier}, \citenamefont
  {Costello},\ and\ \citenamefont {Sonntag}}]{Kiernan1994}%
  \BibitemOpen
  \bibfield  {author} {\bibinfo {author} {\bibfnamefont {L. M.}~\bibnamefont
  {Kiernan}}, \bibinfo {author} {\bibfnamefont {E.~T.}\ \bibnamefont
  {Kennedy}}, \bibinfo {author} {\bibfnamefont {J.-P.}\ \bibnamefont
  {Mosnier}}, \bibinfo {author} {\bibfnamefont {J.~T.}\ \bibnamefont
  {Costello}},\ and\ \bibinfo {author} {\bibfnamefont {B.~F.}\ \bibnamefont
  {Sonntag}},\ }\bibfield  {title} {\bibinfo {title} {First observation of a
  photon induced triply excited state in atomic lithium},\ }\href
  {https://doi.org/10.1103/physrevlett.72.2359} {\bibfield  {journal} {\bibinfo
   {journal} {Phys. Rev. Lett.}\ }\textbf {\bibinfo {volume} {72}},\ \bibinfo
  {pages} {2359 } (\bibinfo {year} {1994})}\BibitemShut {NoStop}%
\bibitem [{\citenamefont {Jannitti}\ \emph {et~al.}(1984)\citenamefont
  {Jannitti}, \citenamefont {Nicolosi},\ and\ \citenamefont
  {Tondello}}]{Jannitti1984}%
  \BibitemOpen
  \bibfield  {author} {\bibinfo {author} {\bibfnamefont {E.}~\bibnamefont
  {Jannitti}}, \bibinfo {author} {\bibfnamefont {P.}~\bibnamefont {Nicolosi}},\
  and\ \bibinfo {author} {\bibfnamefont {G.}~\bibnamefont {Tondello}},\
  }\bibfield  {title} {\bibinfo {title} {Photoionization and double excitation
  spectrum of {B}e$^{2+}$},\ }\href
  {https://doi.org/10.1016/0030-4018(84)90322-5} {\bibfield  {journal}
  {\bibinfo  {journal} {Opt. Commun.}\ }\textbf {\bibinfo {volume} {50}},\
  \bibinfo {pages} {225 } (\bibinfo {year} {1984})}\BibitemShut {NoStop}%
\bibitem [{\citenamefont {Scully}\ \emph {et~al.}(2006)\citenamefont {Scully},
  \citenamefont {{\' A}lvarez}, \citenamefont {Cisneros}, \citenamefont
  {Emmons}, \citenamefont {Gharaibeh}, \citenamefont {Leitner}, \citenamefont
  {Lubell}, \citenamefont {M{\"u}ller}, \citenamefont {Phaneuf}, \citenamefont
  {P{\"u}ttner}, \citenamefont {Schlachter}, \citenamefont {Schippers},\ and\
  \citenamefont {McLaughlin}}]{Scully2006a}%
  \BibitemOpen
  \bibfield  {author} {\bibinfo {author} {\bibfnamefont {S.~W.~J.}\
  \bibnamefont {Scully}}, \bibinfo {author} {\bibfnamefont {I.}~\bibnamefont
  {{\' A}lvarez}}, \bibinfo {author} {\bibfnamefont {C.}~\bibnamefont
  {Cisneros}}, \bibinfo {author} {\bibfnamefont {E.~D.}\ \bibnamefont
  {Emmons}}, \bibinfo {author} {\bibfnamefont {M.~F.}\ \bibnamefont
  {Gharaibeh}}, \bibinfo {author} {\bibfnamefont {D.}~\bibnamefont {Leitner}},
  \bibinfo {author} {\bibfnamefont {M.~S.}\ \bibnamefont {Lubell}}, \bibinfo
  {author} {\bibfnamefont {A.}~\bibnamefont {M{\"u}ller}}, \bibinfo {author}
  {\bibfnamefont {R.~A.}\ \bibnamefont {Phaneuf}}, \bibinfo {author}
  {\bibfnamefont {R.}~\bibnamefont {P{\"u}ttner}}, \bibinfo {author}
  {\bibfnamefont {A.~S.}\ \bibnamefont {Schlachter}}, \bibinfo {author}
  {\bibfnamefont {S.}~\bibnamefont {Schippers}},\ and\ \bibinfo {author}
  {\bibfnamefont {B.~M.}\ \bibnamefont {McLaughlin}},\ }\bibfield  {title}
  {\bibinfo {title} {Doubly excited resonances in the photoionization spectrum
  of {L}i$^+$: Experiment and theory},\ }\href
  {https://doi.org/10.1088/0953-4075/39/18/024} {\bibfield  {journal} {\bibinfo
   {journal} {J. Phys. B: At. Mol. Opt. Phys.}\ }\textbf {\bibinfo {volume}
  {39}},\ \bibinfo {pages} {3957} (\bibinfo {year} {2006})}\BibitemShut
  {NoStop}%
\bibitem [{\citenamefont {M\"{u}ller}\ \emph
  {et~al.}(2018{\natexlab{b}})\citenamefont {M\"{u}ller}, \citenamefont
  {Lindroth}, \citenamefont {Bari}, \citenamefont {{Borovik, Jr.}},
  \citenamefont {Hillenbrand}, \citenamefont {Holste}, \citenamefont
  {Indelicato}, \citenamefont {Kilcoyne}, \citenamefont {Klumpp}, \citenamefont
  {Martins}, \citenamefont {Viefhaus}, \citenamefont {Wilhelm},\ and\
  \citenamefont {Schippers}}]{Mueller2018d}%
  \BibitemOpen
  \bibfield  {author} {\bibinfo {author} {\bibfnamefont {A.}~\bibnamefont
  {M\"{u}ller}}, \bibinfo {author} {\bibfnamefont {E.}~\bibnamefont
  {Lindroth}}, \bibinfo {author} {\bibfnamefont {S.}~\bibnamefont {Bari}},
  \bibinfo {author} {\bibfnamefont {A.}~\bibnamefont {{Borovik, Jr.}}},
  \bibinfo {author} {\bibfnamefont {P.-M.}\ \bibnamefont {Hillenbrand}},
  \bibinfo {author} {\bibfnamefont {K.}~\bibnamefont {Holste}}, \bibinfo
  {author} {\bibfnamefont {P.}~\bibnamefont {Indelicato}}, \bibinfo {author}
  {\bibfnamefont {A.~L.~D.}\ \bibnamefont {Kilcoyne}}, \bibinfo {author}
  {\bibfnamefont {S.}~\bibnamefont {Klumpp}}, \bibinfo {author} {\bibfnamefont
  {M.}~\bibnamefont {Martins}}, \bibinfo {author} {\bibfnamefont
  {J.}~\bibnamefont {Viefhaus}}, \bibinfo {author} {\bibfnamefont
  {P.}~\bibnamefont {Wilhelm}},\ and\ \bibinfo {author} {\bibfnamefont
  {S.}~\bibnamefont {Schippers}},\ }\bibfield  {title} {\bibinfo {title}
  {Photoionization of metastable heliumlike {C}$^{4+}(1s2s\, ^3{S}_1)$ ions:
  {P}recision study of intermediate doubly excited states},\ }\href
  {https://doi.org/10.1103/PhysRevA.98.033416} {\bibfield  {journal} {\bibinfo
  {journal} {Phys. Rev. A}\ }\textbf {\bibinfo {volume} {98}},\ \bibinfo
  {pages} {033416} (\bibinfo {year} {2018}{\natexlab{b}})}\BibitemShut
  {NoStop}%
\bibitem [{\citenamefont {Viefhaus}\ \emph {et~al.}(2013)\citenamefont
  {Viefhaus}, \citenamefont {Scholz}, \citenamefont {Deinert}, \citenamefont
  {Glaser}, \citenamefont {Ilchen}, \citenamefont {Seltmann}, \citenamefont
  {Walter},\ and\ \citenamefont {Siewert}}]{Viefhaus2013}%
  \BibitemOpen
  \bibfield  {author} {\bibinfo {author} {\bibfnamefont {J.}~\bibnamefont
  {Viefhaus}}, \bibinfo {author} {\bibfnamefont {F.}~\bibnamefont {Scholz}},
  \bibinfo {author} {\bibfnamefont {S.}~\bibnamefont {Deinert}}, \bibinfo
  {author} {\bibfnamefont {L.}~\bibnamefont {Glaser}}, \bibinfo {author}
  {\bibfnamefont {M.}~\bibnamefont {Ilchen}}, \bibinfo {author} {\bibfnamefont
  {J.}~\bibnamefont {Seltmann}}, \bibinfo {author} {\bibfnamefont
  {P.}~\bibnamefont {Walter}},\ and\ \bibinfo {author} {\bibfnamefont
  {F.}~\bibnamefont {Siewert}},\ }\bibfield  {title} {\bibinfo {title} {The
  variable polarization {XUV} beamline {P04} at {PETRA III}: optics, mechanics
  and their performance},\ }\href {https://doi.org/10.1016/j.nima.2012.10.110}
  {\bibfield  {journal} {\bibinfo  {journal} {Nucl. Instrum. Methods Phys.
  Res.}\ }\textbf {\bibinfo {volume} {A710}},\ \bibinfo {pages} {151 }
  (\bibinfo {year} {2013})}\BibitemShut {NoStop}%
\bibitem [{\citenamefont {Schroer}\ \emph {et~al.}(2022)\citenamefont
  {Schroer}, \citenamefont {Wille}, \citenamefont {Seeck}, \citenamefont
  {Bagschik}, \citenamefont {Schulte-Schrepping}, \citenamefont {Tischer},
  \citenamefont {Graafsma}, \citenamefont {Laasch}, \citenamefont {Baev},
  \citenamefont {Klumpp}, \citenamefont {Bartolini}, \citenamefont {Reichert},
  \citenamefont {Leemans},\ and\ \citenamefont {Weckert}}]{Schroer2022}%
  \BibitemOpen
  \bibfield  {author} {\bibinfo {author} {\bibfnamefont {C.~G.}\ \bibnamefont
  {Schroer}}, \bibinfo {author} {\bibfnamefont {H.-C.}\ \bibnamefont {Wille}},
  \bibinfo {author} {\bibfnamefont {O.~H.}\ \bibnamefont {Seeck}}, \bibinfo
  {author} {\bibfnamefont {K.}~\bibnamefont {Bagschik}}, \bibinfo {author}
  {\bibfnamefont {H.}~\bibnamefont {Schulte-Schrepping}}, \bibinfo {author}
  {\bibfnamefont {M.}~\bibnamefont {Tischer}}, \bibinfo {author} {\bibfnamefont
  {H.}~\bibnamefont {Graafsma}}, \bibinfo {author} {\bibfnamefont
  {W.}~\bibnamefont {Laasch}}, \bibinfo {author} {\bibfnamefont
  {K.}~\bibnamefont {Baev}}, \bibinfo {author} {\bibfnamefont {S.}~\bibnamefont
  {Klumpp}}, \bibinfo {author} {\bibfnamefont {R.}~\bibnamefont {Bartolini}},
  \bibinfo {author} {\bibfnamefont {H.}~\bibnamefont {Reichert}}, \bibinfo
  {author} {\bibfnamefont {W.}~\bibnamefont {Leemans}},\ and\ \bibinfo {author}
  {\bibfnamefont {E.}~\bibnamefont {Weckert}},\ }\bibfield  {title} {\bibinfo
  {title} {The synchrotron radiation source {PETRA III} and its future
  ultra-low-emittance upgrade {PETRA IV}},\ }\href
  {https://doi.org/10.1140/epjp/s13360-022-03517-6} {\bibfield  {journal}
  {\bibinfo  {journal} {Eur. Phys. J. Plus}\ }\textbf {\bibinfo {volume}
  {137}},\ \bibinfo {pages} {1312} (\bibinfo {year} {2022})}\BibitemShut
  {NoStop}%
\bibitem [{\citenamefont {Schippers}\ \emph {et~al.}(2020)\citenamefont
  {Schippers}, \citenamefont {Buhr}, \citenamefont {{Borovik Jr.}},
  \citenamefont {Holste}, \citenamefont {Perry-Sassmannshausen}, \citenamefont
  {Mertens}, \citenamefont {Reinwardt}, \citenamefont {Martins}, \citenamefont
  {Klumpp}, \citenamefont {Schubert}, \citenamefont {Bari}, \citenamefont
  {Beerwerth}, \citenamefont {Fritzsche}, \citenamefont {Ricz}, \citenamefont
  {Hellhund},\ and\ \citenamefont {M\"{u}ller}}]{Schippers2020a}%
  \BibitemOpen
  \bibfield  {author} {\bibinfo {author} {\bibfnamefont {S.}~\bibnamefont
  {Schippers}}, \bibinfo {author} {\bibfnamefont {T.}~\bibnamefont {Buhr}},
  \bibinfo {author} {\bibfnamefont {A.}~\bibnamefont {{Borovik Jr.}}}, \bibinfo
  {author} {\bibfnamefont {K.}~\bibnamefont {Holste}}, \bibinfo {author}
  {\bibfnamefont {A.}~\bibnamefont {Perry-Sassmannshausen}}, \bibinfo {author}
  {\bibfnamefont {K.}~\bibnamefont {Mertens}}, \bibinfo {author} {\bibfnamefont
  {S.}~\bibnamefont {Reinwardt}}, \bibinfo {author} {\bibfnamefont
  {M.}~\bibnamefont {Martins}}, \bibinfo {author} {\bibfnamefont
  {S.}~\bibnamefont {Klumpp}}, \bibinfo {author} {\bibfnamefont
  {K.}~\bibnamefont {Schubert}}, \bibinfo {author} {\bibfnamefont
  {S.}~\bibnamefont {Bari}}, \bibinfo {author} {\bibfnamefont {R.}~\bibnamefont
  {Beerwerth}}, \bibinfo {author} {\bibfnamefont {S.}~\bibnamefont
  {Fritzsche}}, \bibinfo {author} {\bibfnamefont {S.}~\bibnamefont {Ricz}},
  \bibinfo {author} {\bibfnamefont {J.}~\bibnamefont {Hellhund}},\ and\
  \bibinfo {author} {\bibfnamefont {A.}~\bibnamefont {M\"{u}ller}},\ }\bibfield
   {title} {\bibinfo {title} {The photon-ion merged-beams experiment {PIPE} at
  {PETRA III} - {T}he first five years},\ }\href
  {https://doi.org/10.1002/xrs.3035} {\bibfield  {journal} {\bibinfo  {journal}
  {X-Ray Spectrom.}\ }\textbf {\bibinfo {volume} {49}},\ \bibinfo {pages} {11 }
  (\bibinfo {year} {2020})}\BibitemShut {NoStop}%
\bibitem [{\citenamefont {Phaneuf}\ \emph {et~al.}(1999)\citenamefont
  {Phaneuf}, \citenamefont {Havener}, \citenamefont {Dunn},\ and\ \citenamefont
  {M{\"u}ller}}]{Phaneuf1999}%
  \BibitemOpen
  \bibfield  {author} {\bibinfo {author} {\bibfnamefont {R.~A.}\ \bibnamefont
  {Phaneuf}}, \bibinfo {author} {\bibfnamefont {C.~C.}\ \bibnamefont
  {Havener}}, \bibinfo {author} {\bibfnamefont {G.~H.}\ \bibnamefont {Dunn}},\
  and\ \bibinfo {author} {\bibfnamefont {A.}~\bibnamefont {M{\"u}ller}},\
  }\bibfield  {title} {\bibinfo {title} {Merged-beams experiments in atomic and
  molecular physics},\ }\href {https://doi.org/10.1088/0034-4885/62/7/202}
  {\bibfield  {journal} {\bibinfo  {journal} {Rep. Prog. Phys.}\ }\textbf
  {\bibinfo {volume} {62}},\ \bibinfo {pages} {1143} (\bibinfo {year}
  {1999})}\BibitemShut {NoStop}%
\bibitem [{\citenamefont {Schippers}\ \emph {et~al.}(2014)\citenamefont
  {Schippers}, \citenamefont {Ricz}, \citenamefont {Buhr}, \citenamefont
  {{Borovik Jr.}}, \citenamefont {Hellhund}, \citenamefont {Holste},
  \citenamefont {Huber}, \citenamefont {Sch\"{a}fer}, \citenamefont {Schury},
  \citenamefont {Klumpp}, \citenamefont {Mertens}, \citenamefont {Martins},
  \citenamefont {Flesch}, \citenamefont {Ulrich}, \citenamefont {R\"{u}hl},
  \citenamefont {Jahnke}, \citenamefont {Lower}, \citenamefont {Metz},
  \citenamefont {Schmidt}, \citenamefont {Sch\"{o}ffler}, \citenamefont
  {Williams}, \citenamefont {Glaser}, \citenamefont {Scholz}, \citenamefont
  {Seltmann}, \citenamefont {Viefhaus}, \citenamefont {Dorn}, \citenamefont
  {Wolf}, \citenamefont {Ullrich},\ and\ \citenamefont
  {M\"{u}ller}}]{Schippers2014}%
  \BibitemOpen
  \bibfield  {author} {\bibinfo {author} {\bibfnamefont {S.}~\bibnamefont
  {Schippers}}, \bibinfo {author} {\bibfnamefont {S.}~\bibnamefont {Ricz}},
  \bibinfo {author} {\bibfnamefont {T.}~\bibnamefont {Buhr}}, \bibinfo {author}
  {\bibfnamefont {A.}~\bibnamefont {{Borovik Jr.}}}, \bibinfo {author}
  {\bibfnamefont {J.}~\bibnamefont {Hellhund}}, \bibinfo {author}
  {\bibfnamefont {K.}~\bibnamefont {Holste}}, \bibinfo {author} {\bibfnamefont
  {K.}~\bibnamefont {Huber}}, \bibinfo {author} {\bibfnamefont {H.-J.}\
  \bibnamefont {Sch\"{a}fer}}, \bibinfo {author} {\bibfnamefont
  {D.}~\bibnamefont {Schury}}, \bibinfo {author} {\bibfnamefont
  {S.}~\bibnamefont {Klumpp}}, \bibinfo {author} {\bibfnamefont
  {K.}~\bibnamefont {Mertens}}, \bibinfo {author} {\bibfnamefont
  {M.}~\bibnamefont {Martins}}, \bibinfo {author} {\bibfnamefont
  {R.}~\bibnamefont {Flesch}}, \bibinfo {author} {\bibfnamefont
  {G.}~\bibnamefont {Ulrich}}, \bibinfo {author} {\bibfnamefont
  {E.}~\bibnamefont {R\"{u}hl}}, \bibinfo {author} {\bibfnamefont
  {T.}~\bibnamefont {Jahnke}}, \bibinfo {author} {\bibfnamefont
  {J.}~\bibnamefont {Lower}}, \bibinfo {author} {\bibfnamefont
  {D.}~\bibnamefont {Metz}}, \bibinfo {author} {\bibfnamefont {L.~P.~H.}\
  \bibnamefont {Schmidt}}, \bibinfo {author} {\bibfnamefont {M.}~\bibnamefont
  {Sch\"{o}ffler}}, \bibinfo {author} {\bibfnamefont {J.~B.}\ \bibnamefont
  {Williams}}, \bibinfo {author} {\bibfnamefont {L.}~\bibnamefont {Glaser}},
  \bibinfo {author} {\bibfnamefont {F.}~\bibnamefont {Scholz}}, \bibinfo
  {author} {\bibfnamefont {J.}~\bibnamefont {Seltmann}}, \bibinfo {author}
  {\bibfnamefont {J.}~\bibnamefont {Viefhaus}}, \bibinfo {author}
  {\bibfnamefont {A.}~\bibnamefont {Dorn}}, \bibinfo {author} {\bibfnamefont
  {A.}~\bibnamefont {Wolf}}, \bibinfo {author} {\bibfnamefont {J.}~\bibnamefont
  {Ullrich}},\ and\ \bibinfo {author} {\bibfnamefont {A.}~\bibnamefont
  {M\"{u}ller}},\ }\bibfield  {title} {\bibinfo {title} {Absolute cross
  sections for photoionization of {X}e$^{q+}$ ions (1 $\leq q \leq$ 5) at the
  $3d$ ionization threshold},\ }\href
  {https://doi.org/10.1088/0953-4075/47/11/115602} {\bibfield  {journal}
  {\bibinfo  {journal} {J. Phys. B: At. Mol. Opt. Phys.}\ }\textbf {\bibinfo
  {volume} {47}},\ \bibinfo {pages} {115602} (\bibinfo {year}
  {2014})}\BibitemShut {NoStop}%
\bibitem [{\citenamefont {M\"{u}ller}\ \emph {et~al.}(2017)\citenamefont
  {M\"{u}ller}, \citenamefont {Bernhardt}, \citenamefont {{Borovik Jr.}},
  \citenamefont {Buhr}, \citenamefont {Hellhund}, \citenamefont {Holste},
  \citenamefont {Kilcoyne}, \citenamefont {Klumpp}, \citenamefont {Martins},
  \citenamefont {Ricz}, \citenamefont {Seltmann}, \citenamefont {Viefhaus},\
  and\ \citenamefont {Schippers}}]{Mueller2017}%
  \BibitemOpen
  \bibfield  {author} {\bibinfo {author} {\bibfnamefont {A.}~\bibnamefont
  {M\"{u}ller}}, \bibinfo {author} {\bibfnamefont {D.}~\bibnamefont
  {Bernhardt}}, \bibinfo {author} {\bibfnamefont {A.}~\bibnamefont {{Borovik
  Jr.}}}, \bibinfo {author} {\bibfnamefont {T.}~\bibnamefont {Buhr}}, \bibinfo
  {author} {\bibfnamefont {J.}~\bibnamefont {Hellhund}}, \bibinfo {author}
  {\bibfnamefont {K.}~\bibnamefont {Holste}}, \bibinfo {author} {\bibfnamefont
  {A.~L.~D.}\ \bibnamefont {Kilcoyne}}, \bibinfo {author} {\bibfnamefont
  {S.}~\bibnamefont {Klumpp}}, \bibinfo {author} {\bibfnamefont
  {M.}~\bibnamefont {Martins}}, \bibinfo {author} {\bibfnamefont
  {S.}~\bibnamefont {Ricz}}, \bibinfo {author} {\bibfnamefont {J.}~\bibnamefont
  {Seltmann}}, \bibinfo {author} {\bibfnamefont {J.}~\bibnamefont {Viefhaus}},\
  and\ \bibinfo {author} {\bibfnamefont {S.}~\bibnamefont {Schippers}},\
  }\bibfield  {title} {\bibinfo {title} {Photoionization of {N}e atoms and
  {N}e$^+$ ions near the ${K}$ edge: {P}recision spectroscopy and absolute
  cross-sections},\ }\href {https://doi.org/10.3847/1538-4357/836/2/166}
  {\bibfield  {journal} {\bibinfo  {journal} {Astrophys. J.}\ }\textbf
  {\bibinfo {volume} {836}},\ \bibinfo {pages} {166} (\bibinfo {year}
  {2017})}\BibitemShut {NoStop}%
\bibitem [{\citenamefont {Br{\"o}tz}\ \emph {et~al.}(2001)\citenamefont
  {Br{\"o}tz}, \citenamefont {Trassl}, \citenamefont {McCullough},
  \citenamefont {Arnold},\ and\ \citenamefont {Salzborn}}]{Broetz2001}%
  \BibitemOpen
  \bibfield  {author} {\bibinfo {author} {\bibfnamefont {F.}~\bibnamefont
  {Br{\"o}tz}}, \bibinfo {author} {\bibfnamefont {R.}~\bibnamefont {Trassl}},
  \bibinfo {author} {\bibfnamefont {R.~W.}\ \bibnamefont {McCullough}},
  \bibinfo {author} {\bibfnamefont {W.}~\bibnamefont {Arnold}},\ and\ \bibinfo
  {author} {\bibfnamefont {E.}~\bibnamefont {Salzborn}},\ }\bibfield  {title}
  {\bibinfo {title} {Design of compact all-permanent magnet electron cyclotron
  resonance {(ECR)} ion sources for atomic physics experiments},\ }\href
  {https://doi.org/10.1238/physica.topical.092a00278} {\bibfield  {journal}
  {\bibinfo  {journal} {Phys. Scr.}\ }\textbf {\bibinfo {volume} {T92}},\
  \bibinfo {pages} {278} (\bibinfo {year} {2001})}\BibitemShut {NoStop}%
\bibitem [{\citenamefont {Drake}(1971)}]{Drake1971}%
  \BibitemOpen
  \bibfield  {author} {\bibinfo {author} {\bibfnamefont {G.~W.~F.}\
  \bibnamefont {Drake}},\ }\bibfield  {title} {\bibinfo {title} {Theory of
  relativistic magnetic dipole transitions: Lifetime of the metastable $2^3{S}$
  state of the heliumlike ions},\ }\href
  {https://doi.org/10.1103/PhysRevA.3.908} {\bibfield  {journal} {\bibinfo
  {journal} {Phys. Rev. A}\ }\textbf {\bibinfo {volume} {3}},\ \bibinfo {pages}
  {908 } (\bibinfo {year} {1971})}\BibitemShut {NoStop}%
\bibitem [{\citenamefont {Renwick}\ \emph {et~al.}(2009)\citenamefont
  {Renwick}, \citenamefont {Bray}, \citenamefont {Fursa}, \citenamefont
  {Jacobi}, \citenamefont {Knopp}, \citenamefont {Schippers},\ and\
  \citenamefont {M\"{u}ller}}]{Renwick2009a}%
  \BibitemOpen
  \bibfield  {author} {\bibinfo {author} {\bibfnamefont {A.~C.}\ \bibnamefont
  {Renwick}}, \bibinfo {author} {\bibfnamefont {I.}~\bibnamefont {Bray}},
  \bibinfo {author} {\bibfnamefont {D.~V.}\ \bibnamefont {Fursa}}, \bibinfo
  {author} {\bibfnamefont {J.}~\bibnamefont {Jacobi}}, \bibinfo {author}
  {\bibfnamefont {H.}~\bibnamefont {Knopp}}, \bibinfo {author} {\bibfnamefont
  {S.}~\bibnamefont {Schippers}},\ and\ \bibinfo {author} {\bibfnamefont
  {A.}~\bibnamefont {M\"{u}ller}},\ }\bibfield  {title} {\bibinfo {title}
  {Electron-impact ionization of {B}$^{3+}$ ions},\ }\href
  {https://doi.org/10.1088/0953-4075/42/17/175203} {\bibfield  {journal}
  {\bibinfo  {journal} {J. Phys. B: At. Mol. Opt. Phys.}\ }\textbf {\bibinfo
  {volume} {42}},\ \bibinfo {pages} {175203} (\bibinfo {year}
  {2009})}\BibitemShut {NoStop}%
\bibitem [{\citenamefont {Derevianko}\ and\ \citenamefont
  {Johnson}(1997)}]{Derevianko1997}%
  \BibitemOpen
  \bibfield  {author} {\bibinfo {author} {\bibfnamefont {A.}~\bibnamefont
  {Derevianko}}\ and\ \bibinfo {author} {\bibfnamefont {W.~R.}\ \bibnamefont
  {Johnson}},\ }\bibfield  {title} {\bibinfo {title} {Two-photon decay of 2
  $^{1}{S}_{0}$ and 2 $^{3}{S}_{1}$ states of heliumlike ions},\ }\href
  {https://doi.org/10.1103/PhysRevA.56.1288} {\bibfield  {journal} {\bibinfo
  {journal} {Phys. Rev. A}\ }\textbf {\bibinfo {volume} {56}},\ \bibinfo
  {pages} {1288} (\bibinfo {year} {1997})}\BibitemShut {NoStop}%
\bibitem [{\citenamefont {Yerokhin}\ \emph {et~al.}(2022)\citenamefont
  {Yerokhin}, \citenamefont {Patk\'{o}\v{s}},\ and\ \citenamefont
  {Pachucki}}]{Yerokhin2022}%
  \BibitemOpen
  \bibfield  {author} {\bibinfo {author} {\bibfnamefont {V.~A.}\ \bibnamefont
  {Yerokhin}}, \bibinfo {author} {\bibfnamefont {V.}~\bibnamefont
  {Patk\'{o}\v{s}}},\ and\ \bibinfo {author} {\bibfnamefont {K.}~\bibnamefont
  {Pachucki}},\ }\bibfield  {title} {\bibinfo {title} {{QED} calculations of
  energy levels of heliumlike ions with $5 \leq {Z} \leq 30$},\ }\href
  {https://doi.org/10.1103/PhysRevA.106.022815} {\bibfield  {journal} {\bibinfo
   {journal} {Phys. Rev. A}\ }\textbf {\bibinfo {volume} {106}},\ \bibinfo
  {pages} {022815} (\bibinfo {year} {2022})}\BibitemShut {NoStop}%
\bibitem [{\citenamefont {Sodhi}\ and\ \citenamefont
  {Brion}(1984)}]{Sodhi1984a}%
  \BibitemOpen
  \bibfield  {author} {\bibinfo {author} {\bibfnamefont {R.~N.~S.}\
  \bibnamefont {Sodhi}}\ and\ \bibinfo {author} {\bibfnamefont {C.~E.}\
  \bibnamefont {Brion}},\ }\bibfield  {title} {\bibinfo {title} {Reference
  energies for inner shell electron energy-loss spectroscopy},\ }\href
  {https://doi.org/10.1016/0368-2048(84)80050-X} {\bibfield  {journal}
  {\bibinfo  {journal} {J. Elec. Spectrosc. Relat. Phenom.}\ }\textbf {\bibinfo
  {volume} {34}},\ \bibinfo {pages} {363} (\bibinfo {year} {1984})}\BibitemShut
  {NoStop}%
\bibitem [{\citenamefont {Stierhof}\ \emph {et~al.}(2022)\citenamefont
  {Stierhof}, \citenamefont {K\"{u}hn}, \citenamefont {Winter}, \citenamefont
  {Micke}, \citenamefont {Steinbr\"{u}gge}, \citenamefont {Shah}, \citenamefont
  {Hell}, \citenamefont {Bissinger}, \citenamefont {Hirsch}, \citenamefont
  {Ballhausen}, \citenamefont {Lang}, \citenamefont {Gr\"{a}fe}, \citenamefont
  {Wipf}, \citenamefont {Cumbee}, \citenamefont {Betancourt-Martinez},
  \citenamefont {Park}, \citenamefont {Niskanen}, \citenamefont {Chung},
  \citenamefont {Porter}, \citenamefont {St\"{o}hlker}, \citenamefont
  {Pfeifer}, \citenamefont {Brown}, \citenamefont {Bernitt}, \citenamefont
  {Hansmann}, \citenamefont {Wilms}, \citenamefont {L\'{o}pez-Urrutia},\ and\
  \citenamefont {Leutenegger}}]{Stierhof2022}%
  \BibitemOpen
  \bibfield  {author} {\bibinfo {author} {\bibfnamefont {J.}~\bibnamefont
  {Stierhof}}, \bibinfo {author} {\bibfnamefont {S.}~\bibnamefont {K\"{u}hn}},
  \bibinfo {author} {\bibfnamefont {M.}~\bibnamefont {Winter}}, \bibinfo
  {author} {\bibfnamefont {P.}~\bibnamefont {Micke}}, \bibinfo {author}
  {\bibfnamefont {R.}~\bibnamefont {Steinbr\"{u}gge}}, \bibinfo {author}
  {\bibfnamefont {C.}~\bibnamefont {Shah}}, \bibinfo {author} {\bibfnamefont
  {N.}~\bibnamefont {Hell}}, \bibinfo {author} {\bibfnamefont {M.}~\bibnamefont
  {Bissinger}}, \bibinfo {author} {\bibfnamefont {M.}~\bibnamefont {Hirsch}},
  \bibinfo {author} {\bibfnamefont {R.}~\bibnamefont {Ballhausen}}, \bibinfo
  {author} {\bibfnamefont {M.}~\bibnamefont {Lang}}, \bibinfo {author}
  {\bibfnamefont {C.}~\bibnamefont {Gr\"{a}fe}}, \bibinfo {author}
  {\bibfnamefont {S.}~\bibnamefont {Wipf}}, \bibinfo {author} {\bibfnamefont
  {R.}~\bibnamefont {Cumbee}}, \bibinfo {author} {\bibfnamefont {G.~L.}\
  \bibnamefont {Betancourt-Martinez}}, \bibinfo {author} {\bibfnamefont
  {S.}~\bibnamefont {Park}}, \bibinfo {author} {\bibfnamefont {J.}~\bibnamefont
  {Niskanen}}, \bibinfo {author} {\bibfnamefont {M.}~\bibnamefont {Chung}},
  \bibinfo {author} {\bibfnamefont {F.~S.}\ \bibnamefont {Porter}}, \bibinfo
  {author} {\bibfnamefont {T.}~\bibnamefont {St\"{o}hlker}}, \bibinfo {author}
  {\bibfnamefont {T.}~\bibnamefont {Pfeifer}}, \bibinfo {author} {\bibfnamefont
  {G.~V.}\ \bibnamefont {Brown}}, \bibinfo {author} {\bibfnamefont
  {S.}~\bibnamefont {Bernitt}}, \bibinfo {author} {\bibfnamefont
  {P.}~\bibnamefont {Hansmann}}, \bibinfo {author} {\bibfnamefont
  {J.}~\bibnamefont {Wilms}}, \bibinfo {author} {\bibfnamefont {J.~R.~C.}\
  \bibnamefont {L\'{o}pez-Urrutia}},\ and\ \bibinfo {author} {\bibfnamefont
  {M.~A.}\ \bibnamefont {Leutenegger}},\ }\bibfield  {title} {\bibinfo {title}
  {A new benchmark of soft x-ray transition energies of {N}e, {CO}$_2$, and
  {SF}$_6$: paving a pathway towards ppm accuracy},\ }\href
  {https://doi.org/10.1140/epjd/s10053-022-00355-0} {\bibfield  {journal}
  {\bibinfo  {journal} {Eur. Phys. J. D}\ }\textbf {\bibinfo {volume} {76}},\
  \bibinfo {pages} {38} (\bibinfo {year} {2022})}\BibitemShut {NoStop}%
\bibitem [{\citenamefont {Salomonson}\ and\ \citenamefont
  {{\"O}ster}(1989)}]{Salomonson1989a}%
  \BibitemOpen
  \bibfield  {author} {\bibinfo {author} {\bibfnamefont {S.}~\bibnamefont
  {Salomonson}}\ and\ \bibinfo {author} {\bibfnamefont {P.}~\bibnamefont
  {{\"O}ster}},\ }\bibfield  {title} {\bibinfo {title} {Relativistic all-order
  pair functions from a discretized single-particle {D}irac {H}amiltonian},\
  }\href {https://doi.org/10.1103/physreva.40.5548} {\bibfield  {journal}
  {\bibinfo  {journal} {Phys. Rev. A}\ }\textbf {\bibinfo {volume} {40}},\
  \bibinfo {pages} {5548 } (\bibinfo {year} {1989})}\BibitemShut {NoStop}%
\bibitem [{\citenamefont {Lindgren}(1974)}]{Lindgren1974}%
  \BibitemOpen
  \bibfield  {author} {\bibinfo {author} {\bibfnamefont {I.}~\bibnamefont
  {Lindgren}},\ }\bibfield  {title} {\bibinfo {title} {The
  {R}ayleigh-{S}chr{\"o}dinger perturbation and the linked-diagram theorem for
  a multi-configurational model space},\ }\href
  {https://doi.org/10.1088/0022-3700/7/18/010} {\bibfield  {journal} {\bibinfo
  {journal} {J. Phys. B: At. Mol. Opt. Phys.}\ }\textbf {\bibinfo {volume}
  {7}},\ \bibinfo {pages} {2441 } (\bibinfo {year} {1974})}\BibitemShut
  {NoStop}%
\bibitem [{\citenamefont {Lindroth}(1994)}]{Lindroth1994a}%
  \BibitemOpen
  \bibfield  {author} {\bibinfo {author} {\bibfnamefont {E.}~\bibnamefont
  {Lindroth}},\ }\bibfield  {title} {\bibinfo {title} {Calculation of doubly
  excited states of helium with a finite discrete spectrum},\ }\href
  {https://doi.org/10.1103/physreva.49.4473} {\bibfield  {journal} {\bibinfo
  {journal} {Phys. Rev. A}\ }\textbf {\bibinfo {volume} {49}},\ \bibinfo
  {pages} {4473 } (\bibinfo {year} {1994})}\BibitemShut {NoStop}%
\bibitem [{\citenamefont {Lestinsky}\ \emph {et~al.}(2008)\citenamefont
  {Lestinsky}, \citenamefont {Lindroth}, \citenamefont {Orlov}, \citenamefont
  {Schmidt}, \citenamefont {Schippers}, \citenamefont {B{\"o}hm}, \citenamefont
  {Brandau}, \citenamefont {Sprenger}, \citenamefont {Terekhov}, \citenamefont
  {M{\"u}ller},\ and\ \citenamefont {Wolf}}]{Lestinsky2008a}%
  \BibitemOpen
  \bibfield  {author} {\bibinfo {author} {\bibfnamefont {M.}~\bibnamefont
  {Lestinsky}}, \bibinfo {author} {\bibfnamefont {E.}~\bibnamefont {Lindroth}},
  \bibinfo {author} {\bibfnamefont {D.~A.}\ \bibnamefont {Orlov}}, \bibinfo
  {author} {\bibfnamefont {E.~W.}\ \bibnamefont {Schmidt}}, \bibinfo {author}
  {\bibfnamefont {S.}~\bibnamefont {Schippers}}, \bibinfo {author}
  {\bibfnamefont {S.}~\bibnamefont {B{\"o}hm}}, \bibinfo {author}
  {\bibfnamefont {C.}~\bibnamefont {Brandau}}, \bibinfo {author} {\bibfnamefont
  {F.}~\bibnamefont {Sprenger}}, \bibinfo {author} {\bibfnamefont {A.~S.}\
  \bibnamefont {Terekhov}}, \bibinfo {author} {\bibfnamefont {A.}~\bibnamefont
  {M{\"u}ller}},\ and\ \bibinfo {author} {\bibfnamefont {A.}~\bibnamefont
  {Wolf}},\ }\bibfield  {title} {\bibinfo {title} {Screened radiative
  corrections from hyperfine-split dielectronic resonances in lithiumlike
  scandium},\ }\href {https://doi.org/10.1103/physrevlett.100.033001}
  {\bibfield  {journal} {\bibinfo  {journal} {Phys. Rev. Lett.}\ }\textbf
  {\bibinfo {volume} {100}},\ \bibinfo {pages} {033001} (\bibinfo {year}
  {2008})}\BibitemShut {NoStop}%
\bibitem [{\citenamefont {Tokman}\ \emph {et~al.}(2002)\citenamefont {Tokman},
  \citenamefont {Ekl{\"o}w}, \citenamefont {Glans}, \citenamefont {Lindroth},
  \citenamefont {Schuch}, \citenamefont {Gwinner}, \citenamefont {Schwalm},
  \citenamefont {Wolf}, \citenamefont {Hoff\-knecht}, \citenamefont
  {M{\"u}ller},\ and\ \citenamefont {Schippers}}]{Tokman2002}%
  \BibitemOpen
  \bibfield  {author} {\bibinfo {author} {\bibfnamefont {M.}~\bibnamefont
  {Tokman}}, \bibinfo {author} {\bibfnamefont {N.}~\bibnamefont {Ekl{\"o}w}},
  \bibinfo {author} {\bibfnamefont {P.}~\bibnamefont {Glans}}, \bibinfo
  {author} {\bibfnamefont {E.}~\bibnamefont {Lindroth}}, \bibinfo {author}
  {\bibfnamefont {R.}~\bibnamefont {Schuch}}, \bibinfo {author} {\bibfnamefont
  {G.}~\bibnamefont {Gwinner}}, \bibinfo {author} {\bibfnamefont
  {D.}~\bibnamefont {Schwalm}}, \bibinfo {author} {\bibfnamefont
  {A.}~\bibnamefont {Wolf}}, \bibinfo {author} {\bibfnamefont {A.}~\bibnamefont
  {Hoff\-knecht}}, \bibinfo {author} {\bibfnamefont {A.}~\bibnamefont
  {M{\"u}ller}},\ and\ \bibinfo {author} {\bibfnamefont {S.}~\bibnamefont
  {Schippers}},\ }\bibfield  {title} {\bibinfo {title} {Dielectronic
  recombination resonances in {F}$^{6+}$},\ }\href
  {https://doi.org/10.1103/physreva.66.012703} {\bibfield  {journal} {\bibinfo
  {journal} {Phys. Rev. A}\ }\textbf {\bibinfo {volume} {66}},\ \bibinfo
  {pages} {012703} (\bibinfo {year} {2002})}\BibitemShut {NoStop}%
\bibitem [{\citenamefont {Lindroth}\ and\ \citenamefont
  {Argenti}(2012)}]{Lindroth2012a}%
  \BibitemOpen
  \bibfield  {author} {\bibinfo {author} {\bibfnamefont {E.}~\bibnamefont
  {Lindroth}}\ and\ \bibinfo {author} {\bibfnamefont {L.}~\bibnamefont
  {Argenti}},\ }\bibfield  {title} {\bibinfo {title} {Atomic resonance states
  and their role in charge-changing processes},\ }\href
  {https://doi.org/10.1016/B978-0-12-397009-1.00005-9} {\bibfield  {journal}
  {\bibinfo  {journal} {Adv. Quantum Chem.}\ }\textbf {\bibinfo {volume}
  {63}},\ \bibinfo {pages} {247 } (\bibinfo {year} {2012})}\BibitemShut
  {NoStop}%
\bibitem [{\citenamefont {Yerokhin}\ and\ \citenamefont
  {Shabaev}(2015)}]{Yerokhin2015}%
  \BibitemOpen
  \bibfield  {author} {\bibinfo {author} {\bibfnamefont {V.~A.}\ \bibnamefont
  {Yerokhin}}\ and\ \bibinfo {author} {\bibfnamefont {V.~M.}\ \bibnamefont
  {Shabaev}},\ }\bibfield  {title} {\bibinfo {title} {Lamb shift of $n = 1$ and
  $n = 2$ states of hydrogen-like atoms, $1 \leq {Z} \leq 110$},\ }\href
  {https://doi.org/10.1063/1.4927487} {\bibfield  {journal} {\bibinfo
  {journal} {J. Phys. Chem. Ref. Data}\ }\textbf {\bibinfo {volume} {44}},\
  \bibinfo {pages} {033103} (\bibinfo {year} {2015})}\BibitemShut {NoStop}%
\bibitem [{\citenamefont {Hughes}\ and\ \citenamefont
  {Eckart}(1930)}]{Hughes1930}%
  \BibitemOpen
  \bibfield  {author} {\bibinfo {author} {\bibfnamefont {D.~S.}\ \bibnamefont
  {Hughes}}\ and\ \bibinfo {author} {\bibfnamefont {C.}~\bibnamefont
  {Eckart}},\ }\bibfield  {title} {\bibinfo {title} {The effect of the motion
  of the nucleus on the spectra of {L}i {I} and {L}i {II}},\ }\href
  {https://doi.org/10.1103/PhysRev.36.694} {\bibfield  {journal} {\bibinfo
  {journal} {Phys. Rev.}\ }\textbf {\bibinfo {volume} {36}},\ \bibinfo {pages}
  {694 } (\bibinfo {year} {1930})}\BibitemShut {NoStop}%
\bibitem [{\citenamefont {Kheifets}\ and\ \citenamefont
  {Bray}(1996)}]{Kheifets1996}%
  \BibitemOpen
  \bibfield  {author} {\bibinfo {author} {\bibfnamefont {A.~S.}\ \bibnamefont
  {Kheifets}}\ and\ \bibinfo {author} {\bibfnamefont {I.}~\bibnamefont
  {Bray}},\ }\bibfield  {title} {\bibinfo {title} {Calculation of double
  photoionization of helium using the convergent close-coupling method},\
  }\href {https://doi.org/10.1103/PhysRevA.54.R995} {\bibfield  {journal}
  {\bibinfo  {journal} {Phys. Rev. A}\ }\textbf {\bibinfo {volume} {54}},\
  \bibinfo {pages} {R995 } (\bibinfo {year} {1996})}\BibitemShut {NoStop}%
\bibitem [{\citenamefont {Kheifets}\ and\ \citenamefont
  {Bray}(1998)}]{Kheifets1998}%
  \BibitemOpen
  \bibfield  {author} {\bibinfo {author} {\bibfnamefont {A.~S.}\ \bibnamefont
  {Kheifets}}\ and\ \bibinfo {author} {\bibfnamefont {I.}~\bibnamefont
  {Bray}},\ }\bibfield  {title} {\bibinfo {title} {Photoionization with
  excitation and double photoionization of the helium isoelectronic sequence},\
  }\href {https://doi.org/10.1103/physreva.58.4501} {\bibfield  {journal}
  {\bibinfo  {journal} {Phys. Rev. A}\ }\textbf {\bibinfo {volume} {58}},\
  \bibinfo {pages} {4501 } (\bibinfo {year} {1998})}\BibitemShut {NoStop}%
\bibitem [{\citenamefont {Bray}\ \emph {et~al.}(2002)\citenamefont {Bray},
  \citenamefont {Fursa}, \citenamefont {Kheifets},\ and\ \citenamefont
  {Stelbovics}}]{Bray2002a}%
  \BibitemOpen
  \bibfield  {author} {\bibinfo {author} {\bibfnamefont {I.}~\bibnamefont
  {Bray}}, \bibinfo {author} {\bibfnamefont {D.~V.}\ \bibnamefont {Fursa}},
  \bibinfo {author} {\bibfnamefont {A.~S.}\ \bibnamefont {Kheifets}},\ and\
  \bibinfo {author} {\bibfnamefont {A.~T.}\ \bibnamefont {Stelbovics}},\
  }\bibfield  {title} {\bibinfo {title} {Electrons and photons colliding with
  atoms: development and application of the convergent close-coupling method},\
  }\href {https://doi.org/10.1088/0953-4075/35/15/201} {\bibfield  {journal}
  {\bibinfo  {journal} {J. Phys. B: At. Mol. Opt. Phys.}\ }\textbf {\bibinfo
  {volume} {35}},\ \bibinfo {pages} {R117 } (\bibinfo {year}
  {2002})}\BibitemShut {NoStop}%
\bibitem [{\citenamefont {Bray}\ \emph {et~al.}(2017)\citenamefont {Bray},
  \citenamefont {Kheifets},\ and\ \citenamefont {Bray}}]{Bray2017}%
  \BibitemOpen
  \bibfield  {author} {\bibinfo {author} {\bibfnamefont {A.~W.}\ \bibnamefont
  {Bray}}, \bibinfo {author} {\bibfnamefont {A.~S.}\ \bibnamefont {Kheifets}},\
  and\ \bibinfo {author} {\bibfnamefont {I.}~\bibnamefont {Bray}},\ }\bibfield
  {title} {\bibinfo {title} {Calculation of atomic photoionization using the
  nonsingular convergent close-coupling method},\ }\href
  {https://doi.org/10.1103/PhysRevA.95.053405} {\bibfield  {journal} {\bibinfo
  {journal} {Phys. Rev. A}\ }\textbf {\bibinfo {volume} {95}},\ \bibinfo
  {pages} {053405} (\bibinfo {year} {2017})}\BibitemShut {NoStop}%
\bibitem [{\citenamefont {Cowan}(1981)}]{Cowan1981}%
  \BibitemOpen
  \bibfield  {author} {\bibinfo {author} {\bibfnamefont {R.~D.}\ \bibnamefont
  {Cowan}},\ }\href {https://doi.org/10.1525/9780520906150} {\emph {\bibinfo
  {title} {The Theory of Atomic Structure and Spectra}}}\ (\bibinfo
  {publisher} {University of California Press},\ \bibinfo {address}
  {Berkeley},\ \bibinfo {year} {1981})\BibitemShut {NoStop}%
\bibitem [{\citenamefont {Martins}(2001)}]{Martins2001}%
  \BibitemOpen
  \bibfield  {author} {\bibinfo {author} {\bibfnamefont {M.}~\bibnamefont
  {Martins}},\ }\bibfield  {title} {\bibinfo {title} {Photoionization of
  open-shell atoms: the chlorine 2p excitation},\ }\href
  {https://doi.org/10.1088/0953-4075/34/7/313} {\bibfield  {journal} {\bibinfo
  {journal} {J. Phys. B: At. Mol. Opt. Phys.}\ }\textbf {\bibinfo {volume}
  {34}},\ \bibinfo {pages} {1321 } (\bibinfo {year} {2001})}\BibitemShut
  {NoStop}%
\bibitem [{\citenamefont {Pines}(2016)}]{Pines2016}%
  \BibitemOpen
  \bibfield  {author} {\bibinfo {author} {\bibfnamefont {D.}~\bibnamefont
  {Pines}},\ }\bibfield  {title} {\bibinfo {title} {Emergent behavior in
  strongly correlated electron systems},\ }\href
  {https://doi.org/10.1088/0034-4885/79/9/092501} {\bibfield  {journal}
  {\bibinfo  {journal} {Rep. Prog. Phys.}\ }\textbf {\bibinfo {volume} {79}},\
  \bibinfo {pages} {092501} (\bibinfo {year} {2016})}\BibitemShut {NoStop}%
\bibitem [{\citenamefont {Co'}(2023)}]{Co2023}%
  \BibitemOpen
  \bibfield  {author} {\bibinfo {author} {\bibfnamefont {G.}~\bibnamefont
  {Co'}},\ }\bibfield  {title} {\bibinfo {title} {Introducing the random phase
  approximation theory},\ }\href {https://doi.org/10.3390/universe9030141}
  {\bibfield  {journal} {\bibinfo  {journal} {Universe}\ }\textbf {\bibinfo
  {volume} {9}},\ \bibinfo {pages} {141} (\bibinfo {year} {2023})}\BibitemShut
  {NoStop}%
\bibitem [{\citenamefont {Amusia}(1990)}]{Amusia1990}%
  \BibitemOpen
  \bibfield  {author} {\bibinfo {author} {\bibfnamefont {M.~Y.}\ \bibnamefont
  {Amusia}},\ }\bibinfo {title} {Atomic photoeffect},\ in\ \href
  {https://doi.org/10.1007/978-1-4757-9328-4} {\emph {\bibinfo {booktitle}
  {{Physics of Atoms and Molecules}}}},\ \bibinfo {editor} {edited by\ \bibinfo
  {editor} {\bibfnamefont {K.~T.}\ \bibnamefont {Taylor}}}\ (\bibinfo
  {publisher} {Plenum Press},\ \bibinfo {address} {New York},\ \bibinfo {year}
  {1990})\BibitemShut {NoStop}%
\bibitem [{\citenamefont {Amusia}(2004)}]{Amusia2004a}%
  \BibitemOpen
  \bibfield  {author} {\bibinfo {author} {\bibfnamefont {M.~Y.}\ \bibnamefont
  {Amusia}},\ }\bibfield  {title} {\bibinfo {title} {Random phase
  approximation: {F}rom giant to intra-doublet resonances},\ }\href
  {https://doi.org/10.1016/j.radphyschem.2003.12.014} {\bibfield  {journal}
  {\bibinfo  {journal} {Rad. Phys. Chem.}\ }\textbf {\bibinfo {volume} {70}},\
  \bibinfo {pages} {237 } (\bibinfo {year} {2004})}\BibitemShut {NoStop}%
\bibitem [{\citenamefont {Bell}\ and\ \citenamefont
  {Kingston}(1971)}]{Bell1971}%
  \BibitemOpen
  \bibfield  {author} {\bibinfo {author} {\bibfnamefont {K.~L.}\ \bibnamefont
  {Bell}}\ and\ \bibinfo {author} {\bibfnamefont {A.~E.}\ \bibnamefont
  {Kingston}},\ }\bibfield  {title} {\bibinfo {title} {Photoionization cross
  sections for the heliumisoelectronic series},\ }\href
  {https://doi.org/10.1088/0022-3700/4/10/016} {\bibfield  {journal} {\bibinfo
  {journal} {J. Phys. B: At. Mol. Opt. Phys.}\ }\textbf {\bibinfo {volume}
  {4}},\ \bibinfo {pages} {1308 } (\bibinfo {year} {1971})}\BibitemShut
  {NoStop}%
\bibitem [{\citenamefont {Reilman}\ and\ \citenamefont
  {Manson}(1979)}]{Reilman1979}%
  \BibitemOpen
  \bibfield  {author} {\bibinfo {author} {\bibfnamefont {R.~F.}\ \bibnamefont
  {Reilman}}\ and\ \bibinfo {author} {\bibfnamefont {S.~T.}\ \bibnamefont
  {Manson}},\ }\bibfield  {title} {\bibinfo {title} {Photoabsoprption cross
  sections for positive atomic ions with ${Z} \leq 30$},\ }\href
  {https://doi.org/10.1086/190605} {\bibfield  {journal} {\bibinfo  {journal}
  {Astrophys. J. Suppl. Ser.}\ }\textbf {\bibinfo {volume} {40}},\ \bibinfo
  {pages} {815 } (\bibinfo {year} {1979})}\BibitemShut {NoStop}%
\bibitem [{\citenamefont {Verner}\ \emph {et~al.}(1993)\citenamefont {Verner},
  \citenamefont {Yakovlev}, \citenamefont {Band},\ and\ \citenamefont
  {Trzhaskovskaya.}}]{Verner1993a}%
  \BibitemOpen
  \bibfield  {author} {\bibinfo {author} {\bibfnamefont {D.~A.}\ \bibnamefont
  {Verner}}, \bibinfo {author} {\bibfnamefont {D.~G.}\ \bibnamefont
  {Yakovlev}}, \bibinfo {author} {\bibfnamefont {I.~M.}\ \bibnamefont {Band}},\
  and\ \bibinfo {author} {\bibfnamefont {M.~B.}\ \bibnamefont
  {Trzhaskovskaya.}},\ }\bibfield  {title} {\bibinfo {title} {Subshell
  photoionization cross sections and ionization energies of atoms and ions from
  {H}e to {Z}n},\ }\href {https://doi.org/10.1006/adnd.1993.1022} {\bibfield
  {journal} {\bibinfo  {journal} {At. Data Nucl. Data Tables}\ }\textbf
  {\bibinfo {volume} {55}},\ \bibinfo {pages} {233} (\bibinfo {year}
  {1993})}\BibitemShut {NoStop}%
\bibitem [{\citenamefont {Mikhailov}\ \emph {et~al.}(2007)\citenamefont
  {Mikhailov}, \citenamefont {Nefiodov},\ and\ \citenamefont
  {Plunien}}]{Mikhailov2007}%
  \BibitemOpen
  \bibfield  {author} {\bibinfo {author} {\bibfnamefont {A.}~\bibnamefont
  {Mikhailov}}, \bibinfo {author} {\bibfnamefont {A.}~\bibnamefont
  {Nefiodov}},\ and\ \bibinfo {author} {\bibfnamefont {G.}~\bibnamefont
  {Plunien}},\ }\bibfield  {title} {\bibinfo {title} {Single photoeffect on
  helium-like ions in the non-relativistic region},\ }\href
  {https://doi.org/10.1016/j.physleta.2007.04.027} {\bibfield  {journal}
  {\bibinfo  {journal} {Physics Letters A}\ }\textbf {\bibinfo {volume}
  {368}},\ \bibinfo {pages} {391} (\bibinfo {year} {2007})}\BibitemShut
  {NoStop}%
\bibitem [{\citenamefont {Stobbe}(1930)}]{Stobbe1930}%
  \BibitemOpen
  \bibfield  {author} {\bibinfo {author} {\bibfnamefont {M.}~\bibnamefont
  {Stobbe}},\ }\bibfield  {title} {\bibinfo {title} {Zur {Q}uantenmechanik
  photoelektrischer {P}rozesse},\ }\href
  {https://doi.org/10.1002/andp.19303990604} {\bibfield  {journal} {\bibinfo
  {journal} {Ann. Phys. (Leipzig)}\ }\textbf {\bibinfo {volume} {7}},\ \bibinfo
  {pages} {661} (\bibinfo {year} {1930})}\BibitemShut {NoStop}%
\bibitem [{\citenamefont {Fano}(1961)}]{Fano1961}%
  \BibitemOpen
  \bibfield  {author} {\bibinfo {author} {\bibfnamefont {U.}~\bibnamefont
  {Fano}},\ }\bibfield  {title} {\bibinfo {title} {Effects of configuration
  interaction on intensities and phase shifts},\ }\href
  {https://doi.org/10.1103/physrev.124.1866} {\bibfield  {journal} {\bibinfo
  {journal} {Phys. Rev.}\ }\textbf {\bibinfo {volume} {124}},\ \bibinfo {pages}
  {1866} (\bibinfo {year} {1961})}\BibitemShut {NoStop}%
\bibitem [{\citenamefont {Herrick}\ and\ \citenamefont
  {Sinano\u{g}lu}(1975)}]{Herrick1975}%
  \BibitemOpen
  \bibfield  {author} {\bibinfo {author} {\bibfnamefont {D.~R.}\ \bibnamefont
  {Herrick}}\ and\ \bibinfo {author} {\bibfnamefont {O.}~\bibnamefont
  {Sinano\u{g}lu}},\ }\bibfield  {title} {\bibinfo {title} {Comparison of
  doubly-excited helium energy levels, isoelectronic series, autoionization
  lifetimes, and group-theoretical configuration-mixing predictions with
  large--configuration-interaction calculations and experimental spectra},\
  }\href {https://doi.org/10.1103/physreva.11.97} {\bibfield  {journal}
  {\bibinfo  {journal} {Phys. Rev. A}\ }\textbf {\bibinfo {volume} {11}},\
  \bibinfo {pages} {97 } (\bibinfo {year} {1975})}\BibitemShut {NoStop}%
\bibitem [{\citenamefont {Lin}(1983)}]{Lin1983}%
  \BibitemOpen
  \bibfield  {author} {\bibinfo {author} {\bibfnamefont {C.~D.}\ \bibnamefont
  {Lin}},\ }\bibfield  {title} {\bibinfo {title} {Classification of doubly
  excited states of two-electron atoms},\ }\href
  {https://doi.org/10.1103/physrevlett.51.1348} {\bibfield  {journal} {\bibinfo
   {journal} {Phys. Rev. Lett.}\ }\textbf {\bibinfo {volume} {51}},\ \bibinfo
  {pages} {1348 } (\bibinfo {year} {1983})}\BibitemShut {NoStop}%
\bibitem [{\citenamefont {Lin}(1984)}]{Lin1984}%
  \BibitemOpen
  \bibfield  {author} {\bibinfo {author} {\bibfnamefont {C.~D.}\ \bibnamefont
  {Lin}},\ }\bibfield  {title} {\bibinfo {title} {Classification and
  supermultiplet structure of doubly excited states},\ }\href
  {https://doi.org/10.1103/PhysRevA.29.1019} {\bibfield  {journal} {\bibinfo
  {journal} {Phys. Rev. A}\ }\textbf {\bibinfo {volume} {29}},\ \bibinfo
  {pages} {1019 } (\bibinfo {year} {1984})}\BibitemShut {NoStop}%
\bibitem [{\citenamefont {Lin}(1995)}]{Lin1995}%
  \BibitemOpen
  \bibfield  {author} {\bibinfo {author} {\bibfnamefont {C.~D.}\ \bibnamefont
  {Lin}},\ }\bibfield  {title} {\bibinfo {title} {Hyperspherical coordinate
  approach to atomic and other {C}oulombic three-body systems},\ }\href
  {https://doi.org/10.1016/0370-1573(94)00094-j} {\bibfield  {journal}
  {\bibinfo  {journal} {Phys. Rep.}\ }\textbf {\bibinfo {volume} {257}},\
  \bibinfo {pages} {1 } (\bibinfo {year} {1995})}\BibitemShut {NoStop}%
\bibitem [{\citenamefont {Domke}\ \emph {et~al.}(1991)\citenamefont {Domke},
  \citenamefont {Xue}, \citenamefont {Puschmann}, \citenamefont {Mandel},
  \citenamefont {Hudson}, \citenamefont {Shirley}, \citenamefont {Kaindl},
  \citenamefont {Greene}, \citenamefont {Sadeghpour},\ and\ \citenamefont
  {Petersen}}]{Domke1991a}%
  \BibitemOpen
  \bibfield  {author} {\bibinfo {author} {\bibfnamefont {M.}~\bibnamefont
  {Domke}}, \bibinfo {author} {\bibfnamefont {C.}~\bibnamefont {Xue}}, \bibinfo
  {author} {\bibfnamefont {A.}~\bibnamefont {Puschmann}}, \bibinfo {author}
  {\bibfnamefont {T.}~\bibnamefont {Mandel}}, \bibinfo {author} {\bibfnamefont
  {E.}~\bibnamefont {Hudson}}, \bibinfo {author} {\bibfnamefont {D.~A.}\
  \bibnamefont {Shirley}}, \bibinfo {author} {\bibfnamefont {G.}~\bibnamefont
  {Kaindl}}, \bibinfo {author} {\bibfnamefont {C.~H.}\ \bibnamefont {Greene}},
  \bibinfo {author} {\bibfnamefont {H.~R.}\ \bibnamefont {Sadeghpour}},\ and\
  \bibinfo {author} {\bibfnamefont {H.}~\bibnamefont {Petersen}},\ }\bibfield
  {title} {\bibinfo {title} {Extensive double-excitation states in atomic
  helium},\ }\href {https://doi.org/10.1103/physrevlett.66.1306} {\bibfield
  {journal} {\bibinfo  {journal} {Phys. Rev. Lett.}\ }\textbf {\bibinfo
  {volume} {66}},\ \bibinfo {pages} {1306 } (\bibinfo {year}
  {1991})}\BibitemShut {NoStop}%
\bibitem [{\citenamefont {Domke}\ \emph {et~al.}(1992)\citenamefont {Domke},
  \citenamefont {Remmers},\ and\ \citenamefont {Kaindl}}]{Domke1992a}%
  \BibitemOpen
  \bibfield  {author} {\bibinfo {author} {\bibfnamefont {M.}~\bibnamefont
  {Domke}}, \bibinfo {author} {\bibfnamefont {G.}~\bibnamefont {Remmers}},\
  and\ \bibinfo {author} {\bibfnamefont {G.}~\bibnamefont {Kaindl}},\
  }\bibfield  {title} {\bibinfo {title} {Observation of the $(2p,nd)~^1{P}^o$
  double-excitation {R}ydberg series of helium},\ }\href
  {https://doi.org/10.1103/physrevlett.69.1171} {\bibfield  {journal} {\bibinfo
   {journal} {Phys. Rev. Lett.}\ }\textbf {\bibinfo {volume} {69}},\ \bibinfo
  {pages} {1171 } (\bibinfo {year} {1992})}\BibitemShut {NoStop}%
\bibitem [{\citenamefont {Schulz}\ \emph {et~al.}(1996)\citenamefont {Schulz},
  \citenamefont {Kaindl}, \citenamefont {Domke}, \citenamefont {Bozek},
  \citenamefont {Heimann}, \citenamefont {Schlachter},\ and\ \citenamefont
  {Rost}}]{Schulz1996a}%
  \BibitemOpen
  \bibfield  {author} {\bibinfo {author} {\bibfnamefont {K.}~\bibnamefont
  {Schulz}}, \bibinfo {author} {\bibfnamefont {G.}~\bibnamefont {Kaindl}},
  \bibinfo {author} {\bibfnamefont {M.}~\bibnamefont {Domke}}, \bibinfo
  {author} {\bibfnamefont {J.~D.}\ \bibnamefont {Bozek}}, \bibinfo {author}
  {\bibfnamefont {P.~A.}\ \bibnamefont {Heimann}}, \bibinfo {author}
  {\bibfnamefont {A.~S.}\ \bibnamefont {Schlachter}},\ and\ \bibinfo {author}
  {\bibfnamefont {J.~M.}\ \bibnamefont {Rost}},\ }\bibfield  {title} {\bibinfo
  {title} {Observation of new {R}ydberg series and resonances in doubly excited
  helium at ultrahigh resolution},\ }\href
  {https://doi.org/10.1103/physrevlett.77.3086} {\bibfield  {journal} {\bibinfo
   {journal} {Phys. Rev. Lett.}\ }\textbf {\bibinfo {volume} {77}},\ \bibinfo
  {pages} {3086} (\bibinfo {year} {1996})}\BibitemShut {NoStop}%
\bibitem [{\citenamefont {P\"{u}ttner}\ \emph {et~al.}(2001)\citenamefont
  {P\"{u}ttner}, \citenamefont {Gr\'{e}maud}, \citenamefont {Delande},
  \citenamefont {Domke}, \citenamefont {Martins}, \citenamefont {Schlachter},\
  and\ \citenamefont {Kaindl}}]{Puettner2001}%
  \BibitemOpen
  \bibfield  {author} {\bibinfo {author} {\bibfnamefont {R.}~\bibnamefont
  {P\"{u}ttner}}, \bibinfo {author} {\bibfnamefont {B.}~\bibnamefont
  {Gr\'{e}maud}}, \bibinfo {author} {\bibfnamefont {D.}~\bibnamefont
  {Delande}}, \bibinfo {author} {\bibfnamefont {M.}~\bibnamefont {Domke}},
  \bibinfo {author} {\bibfnamefont {M.}~\bibnamefont {Martins}}, \bibinfo
  {author} {\bibfnamefont {A.~S.}\ \bibnamefont {Schlachter}},\ and\ \bibinfo
  {author} {\bibfnamefont {G.}~\bibnamefont {Kaindl}},\ }\bibfield  {title}
  {\bibinfo {title} {Statistical properties of inter-series mixing in helium:
  From integrability to chaos},\ }\href
  {https://doi.org/10.1103/physrevlett.86.3747} {\bibfield  {journal} {\bibinfo
   {journal} {Phys. Rev. Lett.}\ }\textbf {\bibinfo {volume} {86}},\ \bibinfo
  {pages} {3747 } (\bibinfo {year} {2001})}\BibitemShut {NoStop}%
\bibitem [{\citenamefont {Kennedy}\ and\ \citenamefont
  {Carroll}(1978)}]{Kennedy1978}%
  \BibitemOpen
  \bibfield  {author} {\bibinfo {author} {\bibfnamefont {E.~T.}\ \bibnamefont
  {Kennedy}}\ and\ \bibinfo {author} {\bibfnamefont {P.~K.}\ \bibnamefont
  {Carroll}},\ }\bibfield  {title} {\bibinfo {title} {Satellite lines of
  low-{Z} elements {(Li, Be, B)} observed in laser-produced plasmas},\ }\href
  {https://doi.org/10.1088/0022-3700/11/6/009} {\bibfield  {journal} {\bibinfo
  {journal} {J. Phys. B: At. Mol. Opt. Phys.}\ }\textbf {\bibinfo {volume}
  {11}},\ \bibinfo {pages} {965 } (\bibinfo {year} {1978})}\BibitemShut
  {NoStop}%
\bibitem [{\citenamefont {Gning}\ \emph {et~al.}(2015)\citenamefont {Gning},
  \citenamefont {Sow}, \citenamefont {Traor\'{e}}, \citenamefont {Dieng},
  \citenamefont {Diakhate}, \citenamefont {Biaye},\ and\ \citenamefont
  {Wagu\'{e}}}]{Gning2015}%
  \BibitemOpen
  \bibfield  {author} {\bibinfo {author} {\bibfnamefont {Y.}~\bibnamefont
  {Gning}}, \bibinfo {author} {\bibfnamefont {M.}~\bibnamefont {Sow}}, \bibinfo
  {author} {\bibfnamefont {A.}~\bibnamefont {Traor\'{e}}}, \bibinfo {author}
  {\bibfnamefont {M.}~\bibnamefont {Dieng}}, \bibinfo {author} {\bibfnamefont
  {B.}~\bibnamefont {Diakhate}}, \bibinfo {author} {\bibfnamefont
  {M.}~\bibnamefont {Biaye}},\ and\ \bibinfo {author} {\bibfnamefont
  {A.}~\bibnamefont {Wagu\'{e}}},\ }\bibfield  {title} {\bibinfo {title}
  {Calculations of resonances parameters for the ($(2s^2)~^1{S}^e,
  (2s2p)~^{1,3}{P}^o)$ and ($(3s^2)~^1{S}^e, (3s3p)~^{1,3}{P}^o)$ doubly
  excited states of helium-like ions with ${Z}\leq10$ using a complex rotation
  method implemented in {S}cilab},\ }\href
  {https://doi.org/10.1016/j.radphyschem.2014.06.015} {\bibfield  {journal}
  {\bibinfo  {journal} {Rad. Phys. Chem.}\ }\textbf {\bibinfo {volume} {106}},\
  \bibinfo {pages} {1 } (\bibinfo {year} {2015})}\BibitemShut {NoStop}%
\bibitem [{\citenamefont {Schippers}(2018)}]{Schippers2018a}%
  \BibitemOpen
  \bibfield  {author} {\bibinfo {author} {\bibfnamefont {S.}~\bibnamefont
  {Schippers}},\ }\bibfield  {title} {\bibinfo {title} {Analytical expression
  for the convolution of a {F}ano line profile with a {G}aussian},\ }\href
  {https://doi.org/10.1016/j.jqsrt.2018.08.003} {\bibfield  {journal} {\bibinfo
   {journal} {J. Quant. Spectrosc. Radiat. Transfer}\ }\textbf {\bibinfo
  {volume} {219}},\ \bibinfo {pages} {33 } (\bibinfo {year}
  {2018})}\BibitemShut {NoStop}%
\bibitem [{\citenamefont {Reininger}\ and\ \citenamefont {{de
  Castro}}(2005)}]{Reininger2005}%
  \BibitemOpen
  \bibfield  {author} {\bibinfo {author} {\bibfnamefont {R.}~\bibnamefont
  {Reininger}}\ and\ \bibinfo {author} {\bibfnamefont {A.~R.~B.}\ \bibnamefont
  {{de Castro}}},\ }\bibfield  {title} {\bibinfo {title} {High resolution,
  large spectral range, in variable-included-angle soft x-ray monochromators
  using a plane {VLS} grating},\ }\href
  {https://doi.org/10.1016/j.nima.2004.09.007} {\bibfield  {journal} {\bibinfo
  {journal} {Nucl. Instrum. Methods A}\ }\textbf {\bibinfo {volume} {538}},\
  \bibinfo {pages} {760 } (\bibinfo {year} {2005})}\BibitemShut {NoStop}%
\end{thebibliography}
\end{document}